\documentclass[manuscript,screen,nonacm]{acmart}
\setcopyright{none}
\usepackage{booktabs}
\usepackage{array}

\begin{document}

\title{Same Model, Different People: Numerical Execution Contracts and the Reproducibility of Fairness Audits}
\author{Erzheng Zhang}
\affiliation{%
  \institution{University of Toronto}
  \city{Toronto}
  \country{Canada}}
\email{erzheng.zhang@mail.utoronto.ca}

\author{Agnes Chu}
\affiliation{%
  \institution{McMaster University}
  \city{Hamilton}
  \country{Canada}}
\email{chua28@mcmaster.ca}

\begin{abstract}
Fairness audits treat a trained model as a fixed decision function, but the arithmetic that evaluates it is
part of that function. This study holds the model, records, threshold and audit code fixed and varies the
numerical execution contract in two tiers. Tier 1 changes only how products and sums are computed, with the
serialized weights byte-identical, and is the causal test of arithmetic. Tier 2 runs reduced-precision deployment
configurations (BF16 compute, FP8 and 8-bit weights), which can also round weights and change engines and kernels, so
its effects belong to the configuration. It audits three tabular model families
and two open-weight language models (4B and 24B parameters) on four public tasks, including hospital readmission,
on CPUs and GPUs from NVIDIA, AMD and Apple. At the frozen thresholds, strict FP32 changed none of the 4B model's 204{,}204 decisions on three
GPUs. Tier 1 changed decisions near the threshold, and which ones depended on the GPU and its software stack, but no
verdict under the prespecified, unweighted tests; with survey
weights, it moved one gap from 0.04999 to 0.0501 and 0.0503. Tier 2 configurations changed more decisions and some
diagnostic verdicts, but no four-fifths verdict. Serving the 24B model with vLLM FP8 selected 54\% of people rather than 45\% and passed an employment-status
classification audit by sex that its strict FP32 reference failed: its demographic-parity and equalized-odds
gaps fell from 0.0567 and 0.0515 to 0.0417 and 0.0394, against a diagnostic tolerance of 0.05. At the reference's selection
count FP8 failed demographic parity, and no reading gave its equalized-odds change a paired interval excluding zero. With survey
weights, strict FP32 already passed equalized odds, leaving a demographic-parity crossing whose interval includes
zero. The declared audit depends on the configuration; FP8 is not shown to be fairer. At the frozen thresholds, every pass-fail verdict a configuration changed had a sampling
interval containing its criterion. Each analysis is labeled prespecified before all runs, planned before a later arm, or post hoc. The
contribution is a systematic measurement of execution-contract dependence in fixed-model fairness audits, not a
finding that ordinary precision changes routinely reverse their conclusions, which Tier 1 contradicts. It proposes
an audit stability protocol, a domain-specific reporting specification in which the auditor declares contracts, groups, weights and
tests before scoring and records each verdict's range over them apart from its sampling uncertainty.

\end{abstract}


\maketitle

\section{Introduction}\label{sec:intro}
Fairness audits compute group statistics from thresholded decisions. Employment bias audits report selection rates
and impact ratios~\cite{nyc_aedt,cfr1607}. Health systems use risk-score cutoffs to select patients for
care-management programs, automatically or by referral to a clinician~\cite{obermeyer2019dissecting}. A thresholded
score becomes an action, so a small change in the score can change what happens to a person, and enough changed
decisions can change a group statistic.

Weights, inputs and a threshold do not fix the decision function. The arithmetic that evaluates the model also
fixes it: how inputs are rounded, the precision of each product and the order in which sums accumulate. A fixed
language model changes its outputs depending on numerical precision, batch size and
GPU~\cite{yuan2025nondeterminism,du2026greedy}. The default arithmetic differs by framework, version and
accelerator~\cite{tf_tf32,onnxruntime_cuda,pytorch_hip}. An audit report that names a model and a framework
therefore leaves part of the audited function unstated.

Work on arbitrariness in fairness has varied the model itself. It has retrained on other
hardware~\cite{nelaturu2024hardware}, pruned or quantized the model~\cite{tran2022pruning,marcuzzi2026quantization}
and varied training randomness~\cite{ganesh2023randomness,cooper2024arbitrariness}. The literature search for this study found no work that holds the serialized model,
records, threshold and audit code fixed, varies only the numerical execution, and measures changed decisions and
audit verdicts against a strict control.

Four public tasks supply the records: income, employment, mortgage lending and 30-day hospital readmission. Each is
scored by logistic regression (LR), a multilayer perceptron (MLP), gradient-boosted trees and an open-weight 4B language
model used as a zero-shot classifier. Tier 1 varies only the arithmetic. It compares FP64, strict FP32 and TF32 on
NVIDIA GPUs of three generations, PyTorch's TF32 setting on an AMD GPU and the relaxed default of MLX on an Apple
GPU. Tier 2 measures reduced-precision deployment configurations: BF16 compute, FP8 and 8-bit weights. A second language
model, Mistral Small 3.2 (24B), is scored under TF32 and the deployment configurations against strict FP32. Each
audit applies declared procedures as an auditor would, some deficient, such as a worst-pair test with
anti-conservative $p$-values. The study measures whether they give the same verdict under each contract, not whether
the verdicts are valid, with adjusted, exact and patient-clustered tests as checks (Section~\ref{sec:results-audit}).

Each analysis has one of three statuses: prespecified before any run, planned after
earlier results but before a later arm was scored (Tier 2 and the language-model and AMD arms), or post hoc and
exploratory (Appendix~\ref{app:notes}). The study measures execution-contract dependence in
fixed-model fairness audits; it does not show that ordinary precision changes routinely reverse their conclusions. It makes four contributions. First, it separates a causal claim about arithmetic from a
claim about deployments. Tier 1 arithmetic on GPUs from three vendors changed decisions and moved statistics but,
under the prespecified, unweighted tests, changed none of the 84 language-model and 144 tabular
verdicts at stake, and under TF32 none of the second model's 84. Of the 120 pass-fail and four-fifths verdicts, 11 lay within
first-order reach of Tier 1 flips, and none changed (Appendix~\ref{app:posthoc}). With survey weights TF32 moved one that
sat at its criterion (Appendix~\ref{app:split}). Tier 2 configurations changed verdicts in all three arms, though no four-fifths verdict. They also change
engines, kernels or weights, so their changes are not attributed to arithmetic alone.
Second, it measures individual-level arbitrariness with the model held fixed. Which people a contract flipped
depended on the GPU and its software stack. TF32 flipped
41 and 39 of the 4B model's decisions on two NVIDIA GPUs, with a pooled Jaccard index of 0.38 whose excess over chance
HMDA carries (0.31 to 0.67 by task). Third, it describes post hoc where deployment configurations changed verdicts. At the frozen thresholds,
among the 4B model and the tabular LR and MLP, they changed 6 of 17 pass-fail verdicts whose sampling interval
contained the criterion and 0 of the other 43. Only 2 of the 9 changed verdict and contract pairs had a paired interval
excluding zero. The 24B model, split by a rule and encounter-level intervals fixed in advance, showed 2 of 5 and 0 of
15, both in one employment-status classification audit whose demographic-parity pass ends at matched
selection. Fourth, it proposes an audit stability protocol, a domain-specific reporting specification in which the auditor
declares admissible contracts and records each verdict's range over them, with its cost measured on two GPUs.

\section{Background and Related Work}\label{sec:related}

\subsection{Numerical nondeterminism in inference}\label{sec:related-numerics}
Fixed large language model (LLM) inference diverges with batch size, GPU and precision, through finite-precision
arithmetic~\cite{yuan2025nondeterminism}. Du et al.~\cite{du2026greedy} compare BF16 with FP16 on identical models, prompts, decoding and hardware: whether the next token changed depended mainly on the
top-two logit margin relative to the perturbation, the closest mechanistic precedent for a thresholded decision.
These studies measure tokens, text and benchmark accuracy; none asks whether a person's thresholded decision
changes, whether changed decisions concentrate in protected groups, or whether an audit reaches another verdict.

\subsection{Hardware, compression and fairness}\label{sec:related-compression}
Training on different accelerators changes optimization, accuracy and group-fairness
disparities~\cite{nelaturu2024hardware}. Pruning can create or worsen disparate impact, which
Tran et al.~\cite{tran2022pruning} relate to group differences in gradient norms and in distance from the decision
boundary. Marcuzzi et al.~\cite{marcuzzi2026quantization} quantize LLMs and find slightly more stereotyping and unfairness
in generative tasks, most under aggressive compression. Linearizing activations for private inference lowers minority groups' accuracy more than
majority groups'~\cite{das2024linearization}.

In all of these studies the parameters or the architecture change, so they measure a different model. This study's
Tier 1 keeps both and changes only the arithmetic; its Tier 2 configurations, closer to that literature, are reported apart.

\subsection{Multiplicity and arbitrariness}\label{sec:related-multiplicity}
Predictive multiplicity describes near-equally accurate models that assign conflicting predictions to the same
individual~\cite{marx2020multiplicity}. Cooper et al.~\cite{cooper2024arbitrariness} study variance across training runs in fair binary classification, and show that nominal fairness conclusions can be misleading once individual arbitrariness is
considered. Ganesh et al.~\cite{ganesh2023randomness} find substantial variance in group fairness metrics across
training runs, traced to training randomness. In a multiverse analysis, design and evaluation choices give
one model very different fairness scores, which lets an actor ``hack'' a metric~\cite{simson2024multiverse}.

\subsection{Audit reliability and measurement}\label{sec:related-audit}
Jacobs and Wallach~\cite{jacobs2021measurement} frame fairness as a problem of measurement: unobservable constructs
are operationalized, and the operationalization must be checked for reliability and validity. Cherian and Cand\`es~\cite{cherian2024fairness}
treat fairness auditing as multiple hypothesis testing and bound disparities across groups simultaneously.
Non-determinism, Cooper, Frankle and De~Sa argue, puts machine-learning code outside the frame of code as law, because
one program yields a distribution of outcomes~\cite{cooper2022nondeterminism}. Liu et al.~\cite{liu2026trust} bound
the distribution shift under which a third-party audit's conclusion holds, and Broden et al.~\cite{broden2026conformal}
pair conformal prediction with a multiverse analysis of group-conditional coverage. Multiplicity varies the fitted
model, a multiverse the analyst's choices, and shift the records. This study fixes all of them and
varies only the execution contract, so its verdict changes measure execution-contract dependence. Its novelty rests
on that empirical axis, which these frameworks omit; its protocol is a domain-specific reporting specification.

\section{Numerical Execution Contracts}\label{sec:contracts}

\subsection{The decision function}\label{sec:contracts-function}
A thresholded model maps a record $x$ to a decision $f_c(x) = \mathbf{1}[s_c(x) \ge t]$, where $s_c(x)$ is the score
computed under execution contract $c$ and the threshold $t$ is fixed before scoring. A contract fixes, for every
operation, the format and rounding of its inputs, the precision of each product, the width and order of
accumulation, and the format of its output. No GPU run here identifies its accumulation order, so each GPU
contract is an empirically fingerprinted configuration: recorded settings and versions, checked where
Table~\ref{tab:contracts} says, but not fully specified. Under the MI350X's FP8, a decision also depends on the
record's batch. Two runs that share weights, inputs and $t$, but not the contract,
evaluate different functions. The references in this study are FP64, on a CPU for the tabular models and on a GPU for
the 4B model, and strict FP32 for the 24B model. FP64 is itself a contract, an empirical reference that the controls of
Section~\ref{sec:results-control} support at the prespecified operating points.

\subsection{When a decision flips}\label{sec:contracts-flip}
Let $\delta_c(x) = s_c(x) - s(x)$ be the score error of contract $c$ against the reference. Under the rule
$\mathbf{1}[s \ge t]$, a decision changes exactly when $s \ge t > s + \delta_c$ or $s < t \le s + \delta_c$; a score
that lands exactly on $t$ counts as positive, and Appendix~\ref{app:notes} reports the ties this decides. A flip
therefore requires $|s - t| \le |\delta_c|$, so the probability that a member of group $g$ flips is at most
$P(|s - t| \le |\delta_c| \mid g)$. Three things drive it: the size of $\delta_c$, the
density of $g$'s scores near $t$, and the sign of $\delta_c$.

For groups the sign of the error near $t$ matters most, not its mean. Error of both signs moves records across $t$
both ways; one-sided error moves them onto the groups on that side, as all 31 flips of the ACSEmployment MLP under
the M5's relaxed default went one way (Section~\ref{sec:results-who}).

\subsection{Two tiers of contracts}\label{sec:contracts-tiers}
Tier 1 contracts leave the serialized model byte-identical and change only how its products are computed
(Table~\ref{tab:contracts}). The relaxed default of MLX on the Apple M5 truncates both operands of a matrix product
with more than one output to 10 fraction bits, which the emulation of Section~\ref{sec:mechanism-rounding} reproduces. Apple's specification says only that
the mode may truncate the mantissa before the multiplication~\cite{apple2026msl}. TF32 rounds only the activations of
the language models, whose BF16 weights fit its 10 fraction bits. Tier 1 carries the causal claim
of this study: model, records, threshold and audit code are fixed, and only the arithmetic changes.

Tier 2 contracts are the reduced-precision ones deployments commonly run: BF16 compute, 8-bit weight
quantization and FP8. For the float32 tabular models, each rounds the weights, changing the represented model. The 4B model's BF16 contracts keep its stored BF16 weights but change activations and the
attention kernel, so they are Tier 2.

\subsection{Defaults are configuration}\label{sec:contracts-defaults}
Which contract runs depends on framework, version and accelerator, not on the type in which a model is declared.
TensorFlow and ONNX Runtime's CUDA provider enable TF32 by default on Ampere and later
GPUs~\cite{tf_tf32,onnxruntime_cuda}, vLLM's automatic data type serves a float32 checkpoint in
FP16~\cite{vllm_engine_args}, and PyTorch's NVIDIA TF32 controls, documented for the MI300 only~\cite{pytorch_hip},
select a different arithmetic on the MI350X (Table~\ref{tab:contracts}).
\label{sec:contracts-label}
A label such as \texttt{float32} names how numbers are stored, not the arithmetic of the kernel dispatched
(Section~\ref{sec:design-contracts} checks it).

\begin{table*}[t]
\caption{Execution contracts. Every contract ran twice, in separate processes (the 4B model's FP64 reference twice on one task and the 24B model's strict FP32 reference on two, once on the others).
Both language models' weights are stored in BF16, so their BF16 contracts keep them unchanged;
they also change the attention kernel (SDPA or vLLM's own), where Tier 1 uses eager attention.}
\label{tab:contracts}
\footnotesize
\begin{tabular}{@{}l c >{\raggedright\arraybackslash}p{0.31\textwidth} l >{\raggedright\arraybackslash}p{0.23\textwidth}@{}}
\toprule
Contract & Tier & What changes in the arithmetic & Checked by & Hardware \\
\midrule
FP64 (reference) & 1 & float64 throughout & -- & CPU; H100 (LLM); MI350X (control) \\
Strict FP32 & 1 & IEEE FP32 products & read-back, canary & CPU; RTX 4090, RTX A4000, H100, MI350X \\
TF32 & 1 & operands of matrix products rounded to 10 fraction bits & read-back, canary & RTX 4090, RTX A4000, H100 \\
TF32 setting, AMD & 1 & most products as three BF16 partial products of split operands (about 17 bits per
operand); an unidentified arithmetic in 1/8 of the output columns of the 4B model's large projections &
read-back, canary, probe & MI350X \\
M5 relaxed (MLX default) & 1 & both operands of matrix products truncated to 10 fraction bits & per-layer check & Apple M5 \\
M5 certified FP32 & 1 & products routed to an honest FP32 kernel & per-layer check & Apple M5 \\
Tree predictors & 1 & float32 comparisons and leaf sums & -- & CPU; RTX 4090, RTX A4000, H100 \\
\midrule
BF16 & 2 & inputs, every layer output and the tabular weights in BF16 & canary (H100: FP32 only) & M5 CPU and GPU; H100 (LLM); MI350X \\
INT8, dynamic & 2 & INT8 weights; activations quantized per batch & -- & Apple M5 CPU \\
8-bit and MXFP8 weights & 2 & weights quantized, FP32 activations & -- & Apple M5 \\
Serving: BF16, FP8 & 2 & an inference engine's BF16 defaults; FP8 weights and activations & -- & H100 (LLM) \\
FP8, PyTorch & 2 & FP8 weights and activations, per-tensor scales, FP32 accumulation & probe & MI350X \\
\bottomrule
\end{tabular}
\end{table*}

\section{Study Design}\label{sec:design}

\subsection{Data and groups}\label{sec:design-data}
Four public tasks supply the records (Table~\ref{tab:data}). ACSIncome and ACSEmployment follow the Folktables
definitions~\cite{ding2021retiring} on the 2018 one-year American Community Survey microdata for
California. HMDA uses California home-purchase applications from the 2018 public loan
records~\cite{cfpb2018hmda}, restricted to first-lien, owner-occupied, site-built applications that were approved
or denied. The tasks are testbeds whose positive outputs stand for hypothetical selections: ACSEmployment predicts
employment status, not a hiring decision, and HMDA reproduces past lending decisions.

Diabetes 130-US~\cite{strack2014impact} records inpatient encounters of patients with diabetes at 130
US hospitals from 1999 to 2008. The label is readmission within 30 days, the outcome that readmission risk models
predict~\cite{kansagara2011readmission}. The threshold flags the top 10\% of predicted risk, as a health system might
when prioritizing patients for additional support~\cite{obermeyer2019dissecting}. The task stands in for a
thresholded clinical risk score, not an evaluation of treatment or patient outcomes. Its split is by patient, so no
patient's encounters fall on both sides. For every task, a seeded split holds out 20\% of records for testing.

Groups are race and sex for ACS, derived race, ethnicity and sex for HMDA, and race, sex and age band for
Diabetes. HMDA's derived attributes include a Joint category for applications whose applicant and co-applicant
differ; it describes an application, not a protected characteristic of one person. Race and sex are model inputs in ACS, as in Folktables, and in Diabetes. HMDA uses none of its audited
attributes (race, ethnicity, sex) and no field that depends on the outcome, such as the interest rate or denial
reasons.

\begin{table}[t]
\caption{Tasks and frozen models. Positive: the test base rate. Test AUC is computed from FP64 reference scores.
Thresholds are frozen before test scoring: ACS uses 0.5, and from FP64 training scores HMDA approves the training
approval share and Diabetes flags the top 10\%. LLM: the 4B model's AUC and the share of test records it selects at the
prespecified threshold under FP64.}
\label{tab:data}
\small
\begin{tabular}{@{}l r r r r l r r@{}}
\toprule
Task & Train & Test & Features & Positive & AUC (LR / MLP / trees) & LLM AUC & LLM selects \\
\midrule
ACSIncome & 156{,}532 & 39{,}133 & 667 & 0.410 & 0.896 / 0.903 / 0.905 & 0.849 & 52.5\% \\
ACSEmployment & 303{,}054 & 75{,}763 & 74 & 0.454 & 0.870 / 0.902 / 0.904 & 0.821 & 0.094\% \\
HMDA & 277{,}478 & 69{,}370 & 538 & 0.902 & 0.887 / 0.922 / 0.919 & 0.650 & 90.4\% \\
Diabetes & 79{,}405 & 19{,}938 & 1{,}008 & 0.116 & 0.671 / 0.633 / 0.681 & 0.552 & 10.2\% \\
\bottomrule
\end{tabular}
\end{table}

\subsection{Models}\label{sec:design-models}
Three tabular families cover the computations an audit commonly meets: logistic regression, an MLP with hidden
layers of 256 and 128 units, and gradient-boosted trees (XGBoost, depth 6, 300 rounds). Each was trained once in
float32 with fixed seeds, serialized and loaded unchanged everywhere, and the SHA-256 of every weight and input array
is checked before each run. The models are adequate rather than tuned (Table~\ref{tab:data}).

The language-model arm uses an open-weight 4B instruction-tuned model, Qwen3-4B-Instruct, as a zero-shot
classifier on the same frozen test sets. Each record is rendered as text with a question naming two
label words. The score is read from the logits of each word's first token, $m(x) = z_{+} - z_{-}$ and
$s(x) = \sigma(m)$, computed in float64 per contract.

The 4B model ranks records well on the ACS tasks and weakly on HMDA and Diabetes (Table~\ref{tab:data}). Two of its operating points are poor: at the prespecified 0.5 it selected
0.094\% of ACSEmployment records against a base rate of 45.4\%, and the Diabetes top-10\% rule sits at a predicted
probability of 0.99977 on a task where it is near chance.
Its flip rates and verdict changes are read at these operating points; the tabular models and
the 24B model's employment threshold, near the base rate, supply more typical ones.

A second language model of another family, Mistral Small 3.2 (24B), was planned after the first arms' results, with its records, reference and analysis committed before any of its records were scored. It uses
the same prompts and label words. It scores 39{,}133 ACSIncome, 20{,}000 ACSEmployment, 20{,}000 HMDA and
10{,}000 Diabetes records per contract, all but ACSIncome seeded subsets. Its reference is strict
FP32, a reference the protocol can require when FP64 scoring needs several accelerators per job, as the 24B model's
weights do (Section~\ref{sec:protocol}). FP64
rescored 2{,}048 records each of ACSIncome, HMDA and, after the first draft, the ACSEmployment records nearest the threshold,
and changed none of their decisions (Appendix~\ref{app:notes}); the checks cover those records only. Its AUC was 0.879 and 0.794 on the ACS tasks, 0.667 on HMDA and 0.569 on Diabetes. It is a second model, not a scale test (Appendix~\ref{app:notes}).

\subsection{Thresholds}\label{sec:design-thresholds}
Operational thresholds were frozen before any test record was scored. ACS uses 0.5. From FP64 training scores, HMDA
approves the training approval share, which gives model-specific thresholds of 0.8130, 0.7557 and 0.8257
(Appendix~\ref{app:notes} gives a secondary point). Diabetes flags the top 10\% of those scores. That is a frozen numerical cutoff, not a capacity rule that flags a
fixed share of each deployment's scores, so a contract can move the flagged share (Section~\ref{sec:discussion}). The language model applies each threshold on the margin scale, $m \ge \mathrm{logit}(t)$, with its
quantiles taken from FP64 scores of a seeded sample of 10{,}000 training records. Flips are also counted over a
fixed grid of thresholds, but the prespecified point is the result. The 24B model takes its quantiles from strict FP32 scores of 2{,}048 training records per task and its
ACSEmployment threshold at the training employment rate, selecting 44.7\%.

\subsection{Contracts and hardware}\label{sec:design-contracts}
Table~\ref{tab:contracts} lists the contracts and hardware: CPUs, an RTX A4000, an RTX 4090 and an H100, an AMD
MI350X, and the CPU and GPU of an Apple M5 Max. The 4B language model ran on the H100, the RTX 4090 and the MI350X, and the 24B model on H100s, its float32
contracts split across two GPUs.

Before scoring, each Tier 1 run with matrix products read back its precision settings and evaluated a canary
matrix product against FP64. On the NVIDIA GPUs the canary separates TF32 from strict FP32 by more than two orders of magnitude in relative error. The canary checks the setting, not every product the model dispatches (Appendix~\ref{app:prospective}). On the M5, a per-layer check records which products run natively. On the MI350X, probes measured each setting's arithmetic, and untouched PyTorch
defaults computed strict FP32 (Table~\ref{tab:contracts}). vLLM reports no read-back or canary; a third run of the 24B model's vLLM contracts logged its kernels;
FP8 reproduced its margins bit for bit and vLLM BF16 moved 4 decisions (Appendix~\ref{app:audit-record}).

Tier 1 upcasts the language models' BF16 checkpoints exactly to float32 or float64, with explicit attention
products so the precision setting reaches every product. Their BF16 contracts run SDPA attention in
transformers or vLLM's own.

\subsection{Measures}\label{sec:design-measures}
Individual measures are flips against the reference, their rate by bins of $|s - t|$ on a log scale, and the
signed score perturbation. A group enters the statistics with a recorded label and at least 100 test records, and
the error rates with 30 positives and 30 negatives as well. A flip is reproducible when the same record flips in both runs of a contract. Audit measures are each eligible group's selection, true-positive and false-positive rates;
the demographic-parity gap, the highest minus the lowest selection rate; the equalized-odds gap, the larger of the
same ranges of true- and false-positive rates; and the impact ratio, the lowest over the highest selection rate.
Their 95\% intervals are percentiles of 1{,}000 unweighted bootstrap resamples of the test records, shared across
contracts, with eligible groups fixed and the extreme groups chosen again in each draw; they describe these
benchmark records, not the surveyed population. For Diabetes they draw the 13{,}998 test
patients, not the 19{,}938 encounters the plan drew, a post hoc change (Appendix~\ref{app:notes}).

Non-reproducibility is counted in six senses, all of which were fixed before any run. Senses 1 to 3 are pass or fail at a tolerance
of 0.05 (a diagnostic criterion, not a fairness standard), the significance of the worst-pair gap, and an equivalence test of the selection-rate gap at a margin of 0.05.
Sense 2 is a two-sided pooled two-proportion $z$-test between the groups with the highest and lowest rate, for
selection rates and for true-positive rates, chosen on the same records and not adjusted for that choice, so its $p$-values are anti-conservative. Sense 3 applies two one-sided tests to
the signed difference between the same two groups' selection rates, with unpooled Wald standard errors, and
declares equivalence when both reject at 0.05, which corresponds to a 90\% interval inside $\pm 0.05$; it is the verdict of this declared procedure, not a calibrated
test that every pair of groups is equivalent. Senses 2 and 3 treat a Diabetes patient's
encounters as independent. Adjusted and clustered tests are checks
(Appendix~\ref{app:posthoc}), and Table~\ref{tab:app-overview} sorts every result by status and by the checks it
survives. Senses 4 to 6 are the ranking of the model families by demographic-parity gap, equalized-odds gap and accuracy
(three tabular families, and four in the language model's arm, where the tabular families stay at their FP64
references),
the four-fifths rule, under which a group selected at less than 80\% of the most-selected group's rate indicates
adverse impact~\cite{cfr1607}, and a gap or impact ratio moving by more than 0.001. Senses 1 to 5 are verdicts, while sense 6 records movement. Three quantities are kept apart: the verdict under one contract on these test records, a deterministic point
verdict; its range over a declared set of contracts, the stability interval (Section~\ref{sec:protocol}); and how
likely resampling is to change it, which paired bootstrap draws estimate for the change in the statistic and for the
share of draws in which two contracts' verdicts differ. A verdict is one statistic of one model, task and
attribute, or one ranking. Per-model verdicts are counted for logistic regression and the MLP; the tree ensembles
enter only through the rankings (Appendix~\ref{app:notes}). It is counted once, however many contracts or runs changed it, and control contracts enter no denominator; this
rule was set post hoc (Appendix~\ref{app:notes}).

The study plan, with its expectations and a kill criterion, was committed to the study's version control before any
run; the anonymized artifact holds it with the dates of the study's commits (\url{https://osf.io/37p9y/?view_only=c60eb83f4f184aefa61f78a04d043093}), which are the authors'
own record rather than an external registration. Three of its flip-rate expectations and one of the MI350X plan's
were too high, and its kill criterion, a condition to proceed, was met only with Tier 2 or the language model
(Appendix~\ref{app:notes}, with the post hoc analyses).

\section{Results}\label{sec:results}
The results answer three questions: whether a decision changed (Sections~\ref{sec:results-control}
to~\ref{sec:results-who}), whether a statistic moved by more than 0.001 (Table~\ref{tab:senses}) and whether a
verdict changed (Sections~\ref{sec:results-audit} to~\ref{sec:results-24b}).

\subsection{The strict control}\label{sec:results-control}
Strict FP32 changed none of the 4B model's 204{,}204 decisions on any of three GPUs, and every contract run
twice returned the same decisions in both runs. vLLM BF16 alone did not repeat its scores bit for bit. For the tabular models, all 196 combinations of contract, dataset and
model, and 56 more on the MI350X, returned byte-identical scores in their two runs. No honest FP32 contract changed
a tabular decision or moved a score by more than $8.9\times10^{-7}$ from FP64. For the 4B and tabular arms, the disagreements below therefore come from the contract, not
run-to-run noise.

\subsection{Decisions under each contract}\label{sec:results-flips}
TF32 changed the 4B model's decisions on both NVIDIA GPU generations (Table~\ref{tab:flips}): 41 on the
H100 and 39 on the RTX 4090, at most 0.035\% of a dataset. The deployment configurations changed far more of them, up to 4.123\% of the Diabetes
decisions under FP8 serving (Table~\ref{tab:price} gives the totals). Exact ties at the threshold change no verdict (Appendix~\ref{app:notes}).

\begin{table}[t]
\caption{Decisions changed against FP64 at the prespecified threshold, run 1 (run 2 changed the same records in
every cell). LLM: the 4B classifier; BF16 and vLLM on the H100; MI350X: PyTorch's TF32 setting, BF16 and FP8 on
the AMD GPU. MLP: the tabular MLP, with the same TF32 counts on the three NVIDIA GPUs; the MI350X's TF32 setting changed none. Strict FP32 changed none for any model. 24B: the
second model on the H100, four tasks pooled, against its strict FP32 reference.}
\label{tab:flips}
\small
\setlength{\tabcolsep}{4pt}
\begin{tabular}{@{}l r r r r r r r@{}}
\toprule
 & \multicolumn{5}{c}{LLM} & \multicolumn{2}{c}{MLP} \\
\cmidrule(lr){2-6}\cmidrule(l){7-8}
Task ($n$) & TF32 H100 / 4090 & BF16 & vLLM BF16 & vLLM FP8 & MI350X TF32 / BF16 / FP8 & TF32 & M5 relaxed \\
\midrule
ACSIncome (39{,}133) & 9 / 8 & 176 & 166 & 976 & 0 / 191 / 927 & 3 & 1 \\
ACSEmployment (75{,}763) & 2 / 3 & 12 & 14 & 67 & 2 / 15 / 56 & 1 & 31 \\
HMDA (69{,}370) & 23 / 21 & 264 & 178 & 1{,}789 & 2 / 242 / 1{,}791 & 1 & 6 \\
Diabetes (19{,}938) & 7 / 7 & 134 & 90 & 822 & 0 / 106 / 629 & 0 & 6 \\
\midrule
24B, all tasks (89{,}133) & 23 / -- & 580 & 479 & 3{,}059 & -- & -- & -- \\
\bottomrule
\end{tabular}
\end{table}

Among the tabular families, only the MLP flipped under a Tier 1 contract (Table~\ref{tab:flips}). Tier 2 contracts
round the tabular weights, and they flipped 5 to 1{,}601 decisions of both dense families per dataset and model
(Appendix~\ref{app:tab-flips}).

\label{sec:results-margin}%
Every Tier 1 flip lay close to the threshold (Figure~\ref{fig:chain}A and B). For the language model, no TF32 flip had a logit margin above 0.0817. Every TF32 flip of the tabular MLP lay within
$2.1\times10^{-4}$ of the threshold in score, and every relaxed flip within $1.32\times10^{-3}$. The deployment contracts reached further, to a logit margin of 8.3577 under FP8.

\subsection{Who flips}\label{sec:results-who}
Which people flip depends on the GPU and its software stack (Appendix~\ref{app:versions}). Under TF32, the H100 and
the RTX 4090 flipped overlapping but different sets of the language model's decisions, with a Jaccard index of 0.38
pooled over the four tasks and 0.31 to 0.67 by task. Only HMDA's exceeded chance under every matching, and it carries
the pooled excess, so the pooled index does not show agreement on the other tasks (Appendix~\ref{app:robustness}). Under BF16 in one framework, the MI350X and the H100 had a pooled index of 0.31, and transformers and vLLM on one H100
0.19, closer to chance, so the engine moved who flips at least as much as the vendor.

This is individual-level arbitrariness in the sense of Cooper et al.~\cite{cooper2024arbitrariness}, with a single
fitted model. The tabular flips were more stable (Appendix~\ref{app:tab-flips}).

A contract with one-signed error sets the direction: men held 27 of the ACSEmployment MLP's 31 one-way relaxed flips, against
54\% of the records below the threshold down to the deepest flip (Figure~\ref{fig:chain}C), reported as mechanism, not a disparity.

\subsection{Audit statistics}\label{sec:results-audit}

\begin{table}[t]
\caption{Distinct audit verdicts (senses 1 to 5) and estimate movements (sense 6) changed against FP64 (for the 24B model, strict FP32), over those at stake, by sense. Each verdict is counted
once, a rule set post hoc, and every change held in both runs. Comp.: comparisons made (contract, run, task, model and attribute), not
changes; Appendix~\ref{app:tab-verdicts} names each change. In brackets: MI350X only. One flipped record moves
30 of the 90 tabular statistics (trees included) by more than sense 6's 0.001 (Section~\ref{sec:mechanism-granularity}).}
\label{tab:senses}
\footnotesize
\setlength{\tabcolsep}{4pt}
\begin{tabular}{@{}l >{\raggedright\arraybackslash}p{0.27\linewidth} r r r r r r r@{}}
\toprule
Arm & Non-control contracts & Comp. & 1 pass & 2 signif. & 3 equiv. & 4 rank & 5 4/5 & 6 moved \\
\midrule
LLM, Tier 1 & TF32 (H100, 4090); TF32 setting (MI350X) & 60 & 0 / 20 & 0 / 20 & 0 / 10 & 0 / 24 & 0 / 10 & 6 / 30 \\
LLM, Tier 2 & BF16, vLLM BF16, vLLM FP8 (H100); BF16, FP8 (MI350X) & 100 & 1 / 20 & 2 (1) / 20 & 0 / 10 & 2 / 24 & 0 / 10 & 26 (1) / 30 \\
Tabular, Tier 1 & TF32 (3 NVIDIA); TF32 setting (MI350X); M5 relaxed & 200 & 0 / 40 & 0 / 40 & 0 / 20 & 0 / 24 & 0 / 20 & 6 / 60 \\
Tabular, Tier 2 & CPU BF16, CPU INT8; MLX BF16, 8-bit, MXFP8; BF16, FP8 (MI350X) & 280 & 5 / 40 & 3 / 40 & 2 / 20 & 10 (4) / 24 & 0 / 20 & 54 (2) / 60 \\
\midrule
LLM 24B, Tier 1 & TF32 (H100) & 20 & 0 / 20 & 0 / 20 & 0 / 10 & 0 / 24 & 0 / 10 & 3 / 30 \\
LLM 24B, Tier 2 & BF16, vLLM BF16, vLLM FP8 (H100) & 60 & 2 / 20 & 0 / 20 & 0 / 10 & 2 / 24 & 0 / 10 & 28 / 30 \\
\bottomrule
\end{tabular}
\end{table}

Tier 1 contracts moved estimates but changed none of the 84 language-model and 144 tabular verdicts at
stake (Table~\ref{tab:senses}). With the survey's person weights, TF32 moved one ACS verdict that sat at its
criterion, the 4B model's ACSIncome demographic-parity gap by sex, from 0.04999 to 0.0501 and 0.0503 on the two NVIDIA GPUs (Appendix~\ref{app:split}). 

Across the 4B and tabular arms, deployment contracts changed 6 pass-fail, 5 significance, 2 equivalence and 12
ranking verdicts, some near ties and 3 by accuracy, which can move with the selection rate alone. Four of the 4B
model's five were on Diabetes, where it is near chance, and the fifth is the HMDA change below. No contract in any arm changed a four-fifths verdict, the one
criterion here drawn from regulation, an enforcement rule of thumb for employee selection~\cite{cfr1607}. The changed verdicts are diagnostic, at this study's 0.05
tolerance, not institutionally consequential, and that bounds the finding.

Table~\ref{tab:senses} thus counts procedural non-reproducibility: the declared tests, including a worst-pair
test that picks its pair on the records it tests, gave another verdict. Statistically defensible disagreement is
narrower. Of the five Tier 2 significance changes, two, the 4B model's Diabetes changes by sex, survived under pooled,
Wald and patient-clustered variances, the one test two groups allow. The three tabular ones, by race, recurred
under at most two and under no clustered, adjusted test. Exact tests in sparse Diabetes cells changed no Tier 1
verdict and moved each procedure's Tier 2 count by at most one. Only 2 of the 9 changed pass-fail pairs, both set by
HMDA's Joint category, had a paired interval excluding zero (Appendix~\ref{app:robustness}). Five of nine alternative tests, three of them adjusted, changed one of two sparse-cell 4B Tier 1 verdicts by
race, so the Tier 1 null
holds for the declared tests, not every defensible one (Appendices~\ref{app:posthoc} and~\ref{app:robustness}).

Under FP8 serving,
the 4B model passed the HMDA equalized-odds test by derived sex that it failed under FP64. Like the other two HMDA pass-fail
changes, by ethnicity, it involves a derived Joint category, here an application whose applicant and
co-applicant differ in sex, not one person's sex; none of the three remains without Joint. The gap lies between joint and female applications' false-positive rates, and between male and
female applicants both contracts pass (Appendices~\ref{app:hmda-groups}, \ref{app:tab-verdicts}).

\subsection{Verdict changes and sampling uncertainty}\label{sec:results-sampling}

In an exploratory, post hoc split, a deployment contract changed only pass-fail verdicts that sampling had left open
(Table~\ref{tab:twoway} in Appendix~\ref{app:posthoc}): 6 of the 17 whose FP64 interval contained 0.05 (three HMDA verdicts set by Joint, three
Diabetes ones), and none of the other 43. Their FP64
gaps lay 0.0024 to 0.0214 from 0.05. The contrast is
descriptive: the 17 verdicts nearest 0.05 also hold all six changes (Appendix~\ref{app:posthoc}). At an exploratory base-rate threshold for the 4B model on ACSEmployment, TF32 changed no verdict, but FP8 serving passed an equalized-odds test whose FP64
interval, $[0.065, 0.081]$, excluded 0.05, so the split holds at the frozen thresholds only
(Appendix~\ref{app:posthoc}).
For equivalence the split fails: INT8 ended an
equivalence whose FP64 interval lay inside the margin (Appendix~\ref{app:tab-ref}).

\begin{figure}[t]
\centering
\includegraphics[width=0.8\linewidth]{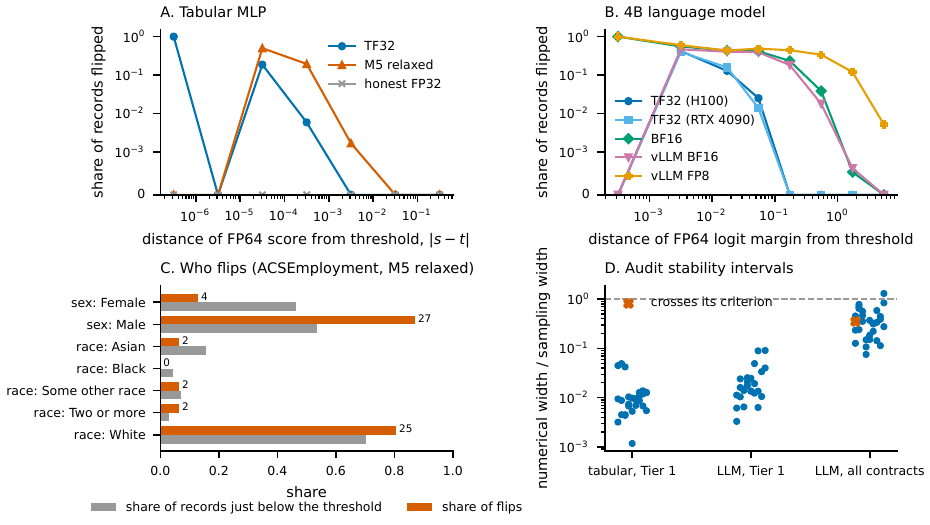}
\caption{From arithmetic to audit verdict. (A) Share of the tabular MLP's records whose decision flipped,
by distance of the FP64 score from the threshold, under TF32, the M5's relaxed default and the honest FP32
contracts. (B) The same for the 4B language model by distance of its logit margin from the threshold. (C) Under
the one-signed relaxed default, each group's share of the ACSEmployment MLP's 31 flipped records (not its flip rate)
against its share of the 71 records below the threshold, down to the deepest flip: 27 of 38 men and 4 of 33 women
flipped. (D) Each audit statistic's numerical spread over the contracts, the MI350X's excluded, relative
to its sampling interval; a cross marks an interval that crosses its criterion. Rates use reproducible flips; empty bins are not
drawn.}
\Description{Four panels. A and B are line charts, on logarithmic axes, of the share of records whose decision flipped
against the distance of the reference score from the threshold: TF32 and the relaxed default flip records only very
close to the threshold, while the deployment contracts reach much further. C is a bar chart comparing each sex and
race group's share of the flipped records with its share of the records below the threshold, down to the deepest
flip; men hold most of the flips although they are about half of those records. D is a dot plot of each audit statistic's spread over the
contracts divided by its sampling interval: small for the Tier 1 contracts and larger once the deployment contracts are
included, with one interval crossing its criterion.}
\label{fig:chain}
\end{figure}

\label{sec:results-intervals}
A stability interval is the range of a statistic over the declared contracts. Over the Tier 1 contracts, with or
without the MI350X's, every verdict of the prespecified tests was numerically robust. None of the 4B model's
60 intervals with a criterion crossed it, nor any tabular one, and the spread of each gap and impact ratio was at
most 0.091 of its sampling interval, 0.049 for the tabular models (Figure~\ref{fig:chain}D). Admitting the deployment contracts changed
this. Two 4B non-ranking verdicts were then not robust: the HMDA equalized-odds gap by derived sex, whose interval
$[0.0489, 0.0680]$ straddles 0.05, and the Diabetes true-positive gap by sex
(Appendix~\ref{app:tab-intervals} has the MI350X's and tabular intervals).

\begin{table}[t]
\caption{The 24B model's ACSEmployment audit by sex, run 1, 20{,}000 records. Post hoc: matched, each contract
selecting strict FP32's 8{,}948 records by its own margin, and weighted. Bold: within 0.05.
Change: FP8 minus strict FP32; 95\% intervals from paired bootstraps (raw, 10{,}000 draws; matched, 1{,}000 draws)
and replicate weights, whose variance and limits Appendix~\ref{app:split} gives.}
\label{tab:llm24-three}
\footnotesize
\setlength{\tabcolsep}{3.5pt}
\begin{tabular}{@{}l r r r r r r r r@{}}
\toprule
 & & \multicolumn{2}{c}{Raw} & \multicolumn{2}{c}{Matched selection} & \multicolumn{3}{c}{Survey-weighted} \\
\cmidrule(lr){3-4}\cmidrule(lr){5-6}\cmidrule(l){7-9}
Contract & Selected & DP & EO & DP & EO & Selected & DP & EO \\
\midrule
Strict FP32 & 0.447 & 0.0567 & 0.0515 & 0.0567 & 0.0515 & 0.429 & 0.0550 & \textbf{0.0353} \\
vLLM BF16 & 0.461 & 0.0516 & \textbf{0.0464} & 0.0581 & 0.0513 & 0.443 & 0.0515 & \textbf{0.0313} \\
vLLM FP8 & 0.544 & \textbf{0.0417} & \textbf{0.0394} & 0.0541 & \textbf{0.0481} & 0.533 & \textbf{0.0478} & \textbf{0.0271} \\
\midrule
Change & & $-0.0150$ & $-0.0120$ & $-0.0026$ & $-0.0033$ & & $-0.0072$ & $-0.0082$ \\
95\% interval & & $[-0.0238, -0.0064]$ & $[-0.0267, 0.0024]$ & $[-0.0094, 0.0044]$ & $[-0.0152, 0.0094]$ & & $[-0.018, 0.004]$ & $[-0.027, 0.011]$ \\
\bottomrule
\end{tabular}
\end{table}

\subsection{A second, larger model}\label{sec:results-24b}
Against its strict FP32 reference, TF32 changed 23 of the 24B model's 89{,}133 decisions, at most 0.065\% of a
dataset, and every contract repeated its decisions in runs 1 and 2, the reference on the two tasks where it ran twice,
ACSIncome and ACSEmployment; a third vLLM BF16 run moved 4 ACSEmployment decisions and no verdict. Table~\ref{tab:flips} gives the deployment
counts. TF32 changed no verdict, and none of its intervals with a criterion crossed
it. The deployment contracts changed 2 pass-fail and 2 ranking verdicts (Table~\ref{tab:senses}). The pass-fail
changes, men against women in an employment-status classification audit at an operating point near the base rate,
are this study's most readily interpreted diagnostic verdicts, and Table~\ref{tab:llm24-three} reads them three ways. FP8 serving passed both. At strict FP32's
selection count it failed demographic parity, and its equalized-odds change had a paired interval including zero: the
declared audit depends on the configuration; FP8 is not shown to be fairer. Strict FP32 failed both tolerances, inside sampling
intervals of $[0.043, 0.070]$ and $[0.033, 0.070]$.
As a chain, BF16 in transformers passed neither (0.0554 and 0.0520), a step that also moved from two GPUs and eager
attention to one and SDPA. vLLM's engine at the same BF16 passed equalized odds, and FP8 within vLLM, the controlled
step, passed demographic parity. Of the two crossings only FP8's had a paired interval excluding zero,
$[-0.0183, -0.0015]$ (Appendix~\ref{app:split}). FP8 serving's selection rate rose past the base rate of 45.4\%,
mostly in that step. With survey weights, strict FP32 already passed equalized odds. Of the table's six changes only the raw
demographic-parity one has an interval excluding zero. That crossing holds at the frozen threshold. Matched selection
takes its cutoffs from the scored test records, a diagnostic rather than a deployable policy. Of the five verdicts whose encounter-level interval contained 0.05, 2 changed,
and none of the other 15, by a rule fixed in advance (2 of 6 and 0 of 14 by patient).

\section{Mechanism}\label{sec:mechanism}

\subsection{What each contract rounds}\label{sec:mechanism-rounding}
A per-record model of the arithmetic explains the tabular flips under Tier 1 and under BF16. Language-model and serving flips are measured, not modeled. Emulating each contract's rounding in float64
reproduces the MLP's measured logit perturbations. The relaxed default's emulation truncates both
operands of every matrix product with more than one output and correlates with them at 0.99997 or more.
Round-to-nearest fits TF32, not the relaxed default.

At the prespecified thresholds the model reproduces the MLP's 16 Tier 1 flip sets, three of them empty, and the 16
distinct BF16 sets on the CPU and MLX; the MI350X's BF16 sets equal the CPU's. The model was built after the Tier 1 results were read, and its MLX
match fitted after the Tier 2 results, so these are reconstructions. Prospectively, for MLPs retrained with a new seed, its
flip sets were committed before any run: the M5's relaxed default flipped exactly those records on all four tasks and at
all 199 grid thresholds, and TF32 on an RTX 4090 and an H100 at every prespecified threshold. On an RTX A6000, which it had not
seen, the prediction failed for one model, whose perturbation a post hoc fit places in one hidden layer, not two
(Appendix~\ref{app:prospective}). The 32 sets under 8-bit and FP8 weights are not modeled; Appendix~\ref{app:fp16} tests an exploratory FP16 arm. Under Tier 1, logistic regression and trees never flipped (Appendix~\ref{app:unreached}).

\subsection{What a verdict needs}\label{sec:mechanism-verdicts}
\label{sec:mechanism-granularity}Small groups set a floor. One flipped record moves 30 of the 90 tabular statistics by more than sense 6's
tolerance of 0.001, because groups are small, positives are few or the selection is a top-10\% rule. Four tabular verdicts, one a tree
ensemble's, sit within one flip of their criterion, three of them among the five that 8-bit formats changed. Reproducibility tolerances should therefore scale with the cells behind each statistic.

\section{Audit Stability Protocol}\label{sec:protocol}

\subsection{Definition and steps}\label{sec:protocol-def}
Before scoring, the auditor, not the auditee, declares the set $C$ of admissible contracts: the deployment's, a
strict FP32 control and a reference (Table~\ref{tab:protocol}). The reference is FP64 or, for a model whose FP64 scoring needs several accelerators per job, strict FP32
checked against FP64 on a sample (Section~\ref{sec:design-models}). For an audit statistic $M$ with value $M_{c,r}$ in run $r$ of contract $c$, the
stability interval spans every run of every contract in $C$,
$I(M) = [\min_{c,r} M_{c,r}, \max_{c,r} M_{c,r}]$. A verdict is numerically robust if every run gives it
the same outcome under its declared rule (a gap passes at $M \le 0.05$, an impact ratio at $M \ge 0.8$, a test rejects
at $p < 0.05$, a ranking keeps its order with the other models' contracts as declared), and contract-dependent
otherwise. For a procedure that reselects groups under each contract, such as the worst-pair test, $I(M)$ describes
the procedure's outputs rather than one estimand.

\begin{table}[t]
\caption{Audit stability checklist. $^\dagger$: cannot currently be verified for a hosted engine. Appendix~\ref{app:audit-record} gives an example record.}
\label{tab:protocol}
\footnotesize
\begin{tabular}{@{}l p{0.78\linewidth}@{}}
\toprule
Step & Report \\
\midrule
1. Declare & by the auditor, before scoring: the admissible contracts (deployment, strict FP32 control, and an FP64
reference or a strict FP32 one checked against FP64), each with framework, version, accelerator and precision
settings$^\dagger$; the comparator, tie rule, groups, any weights, the tests and the claim made \\
2. Check & per run: the settings read back, a canary product against FP64 and the dispatched kernel names$^\dagger$, kept with the record \\
3. Score & the frozen model under every contract, twice, in separate processes, or $k$ declared times if scores do not repeat \\
4. Measure & flips against the reference per contract, their margin profile, the stability interval of every audited
statistic over all runs, with its ratio to the sampling interval where one is defined, and each decision's and verdict's share of runs \\
5. Classify & each verdict as numerically robust or contract-dependent and, for a gap or ratio, as settled or unsettled by its
sampling interval; $p$-values and rankings take no sampling status here \\
6. Resolve & for a contract-dependent verdict, audit the deployment contract or tighten the contract, and say
which \\
\bottomrule
\end{tabular}
\end{table}

\subsection{Three claims and who verifies them}\label{sec:protocol-fit}
The interval records the range over a finite, enumerated set $C$, not a confidence interval or a robustness
guarantee, and assures nothing outside $C$. An audit can make
three claims of different cost, and step 1 declares which. \emph{Reproducing one deployment}: $C$ holds its contract
and the two controls, with the contract record as evidence.
\emph{Portability across deployments}: $C$ enumerates every configuration the deployer may run, or a migration's
target, and the claim covers those alone; a contract-dependent verdict obliges the deployer to narrow $C$ or to have
each configuration audited.

\emph{Sampling uncertainty} is a third claim, separate from both: only the sampling interval speaks to other samples,
and neither interval describes a surveyed population without its weights, which changed 4 of the 96 ACS reference
verdicts (Appendix~\ref{app:split}). A verdict can be numerically robust yet statistically unsettled. The auditor, or its standard, fixes $C$; a set declared by the auditee alone certifies only its own choice. Verification of floating-point inference such as TAO~\cite{yao2026tao} accepts outputs within per-operator
tolerances, which leave thresholded decisions open near the threshold.

A contract is admissible when its record fixes what decides the arithmetic: the precision of weights, activations
and accumulation; framework, engine and kernel-library versions; batching and scheduler settings; and any
compilation or graph capture. No run here recorded its kernel-library version, and the vLLM runs did not record
batching or scheduler settings, so every record in this study is incomplete by this definition. A change to any of these, to the accelerator or to its driver reopens the audit, which
stands only if rescoring the audit records reproduces its decisions and verdicts.

The protocol separates three properties, each established here to a different degree. \emph{Within-contract
repeatability} is demonstrated for every contract run twice (Section~\ref{sec:results-control}), and two runs show
it only in part: a third vLLM BF16 run of the 24B model moved 4 decisions, no verdict (Section~\ref{sec:results-24b}).
A verdict that differs between runs is contract-dependent. \emph{Cross-contract
stability}, which the interval measures, is demonstrated for these contracts and records alone. vLLM FP8 repeated
its margins bit for bit yet passed tolerances strict FP32 failed: cross-contract dependence, not nondeterminism.

\emph{Independent contract verification} is partial: Tier 1 runs read back their settings and passed a canary, while
vLLM only repeated and logged its kernels. Its external form, each contract rerun with margin hashes compared
(Table~\ref{tab:app-audit-record}), is proposed, and a contract that cannot be verified, such as a hosted engine's
(Table~\ref{tab:protocol}), is recorded as unverified.

The protocol is a domain-specific reporting specification, a multiverse analysis~\cite{simson2024multiverse} over
execution contracts, not a new stability method or a reading of law. New York City's bias audits~\cite{nyc_aedt},
bank model-risk guidance~\cite{sr262} and medical-device change control~\cite{fda_pccp}, which counts new hardware as
a change, omit numerical precision.

\section{Measured Cost of the Contracts}\label{sec:price}
\begin{table}
\caption{Measured cost and decision changes of each execution contract for the 4B classifier, 204,204 decisions per
contract. GPU-hours per million decisions from run 1's scoring timers (every forward pass, not model loading; kernels
not profiled); the H100's FP64 tensor cores match the peak rate of its non-tensor FP32 units~\cite{nvidia_h100}. Dollars at \$3.49 per H100 hour on demand~\cite{runpod2026pricing}, read on 4 October 2026.
Strict/contract: strict FP32's GPU time on the same GPU over the contract's. vLLM ran with its default settings.}
\label{tab:price}
\begin{tabular}{llrrrr}
\toprule
Contract & GPU & GPU-hours per $10^6$ & \$ per $10^6$ & Strict/contract & Flips vs FP64 \\
\midrule
FP64 (reference) & H100 & 11.30 & 39.43 & 1.08 & -- \\
Strict FP32 & H100 & 12.17 & 42.49 & 1.00 & 0 \\
TF32 & H100 & 2.81 & 9.81 & 4.33 & 41 \\
BF16 (transformers) & H100 & 1.61 & 5.63 & 7.55 & 586 \\
BF16 (vLLM) & H100 & 3.71 & 12.93 & 3.29 & 448 \\
FP8 (vLLM) & H100 & 3.55 & 12.40 & 3.43 & 3{,}654 \\
Strict FP32 & RTX 4090 & 15.47 & -- & 1.00 & 0 \\
TF32 & RTX 4090 & 9.71 & -- & 1.59 & 39 \\

\bottomrule
\end{tabular}
\end{table}

A hosted interface removes the auditor's choice of contract and can change its model~\cite{chen2023chatgpt}.
Self-hosting restores the
choice but makes compute another resource in which auditors differ~\cite{birhane2024broken}. Table~\ref{tab:price} gives the measured cost of these
implementations on this workload, not an optimized frontier: engines, attention kernels and batching differ between
rows. On one H100, strict
FP32 cost 4.33 times the GPU time of TF32 per decision and 7.55 times that of BF16, a multiple that also includes a
change of attention kernel. The multiple depends on the hardware, 1.59 on an RTX 4090, so a cost quoted without the accelerator is
incomplete. The cost grows with the model: the 24B model's strict FP32 cost 6.33 times TF32's GPU time, both on two H100s, and 22.5 times BF16's
on one (115.68, 18.27 and 5.14 GPU-hours per $10^6$).

\section{Discussion}\label{sec:discussion}
The results draw one boundary. Tier 1 arithmetic changed which people received a decision but no verdict
under the prespecified, unweighted tests, and with survey weights only one that sat at its criterion.
Reduced-precision deployment configurations, which also change engines, kernels or weights, changed diagnostic
verdicts but no four-fifths one. That is a claim about configurations, not arithmetic: in the 24B employment-status
classification audit the engine and FP8 each changed one. Retraining with a new seed moves an audit more, changing 6 of the MLP's 60 verdicts
(Appendix~\ref{app:robustness}), but a contract changes the audited model's own decisions after the audit.

In one audit lifecycle, a model is audited under one contract, then its deployer, or a vendor hosting it, moves
to a new GPU, engine or precision, a configuration the deployer or vendor controls. The published audit summary gives
statistics, such as selection rates and impact ratios~\cite{nyc_aedt}, not the contract
(Section~\ref{sec:protocol-fit}). The recorded decision policy may then not reproduce, as the 24B employment-status
classification audit's did not under FP8 serving, and the report cannot show it. This study observed no deployment or hiring decision and claims no harm.

Resampling rarely tells these changes from none: at the frozen thresholds every changed pass-fail verdict lay
inside its sampling interval, and alternative tests and survey weights also moved verdicts near the criterion
(Sections~\ref{sec:results-audit} to~\ref{sec:results-24b}). Tests, weighting and group definitions therefore
belong in an audit's declaration beside the contract.

Instability, its concentration in groups, and unfairness are three distinct claims. This study makes the first, and
the second only as mechanism in one cell. Whether a changed decision is unfair depends on the domain and its doctrine, which numbers cannot settle.

Many bedside risk tools are integer point scores, such as CHA$_2$DS$_2$-VASc~\cite{lip2010chadsvasc} and the LACE
index and HOSPITAL score for readmission~\cite{vanwalraven2010lace,donze2013hospital}. Such a score sums small integers against an integer
cutoff, exactly in BF16 and every wider format, so no contract in those formats changes who crosses it. Machine-learned models intended to replace or extend such scores tend to
compute continuous scores, and their decisions inherit the contract. On the readmission task, under the frozen cutoff, FP8 serving changed
822 of the 4B model's 19{,}938 readmission decisions; in a separate analysis in which each deployment flags its own
top 10\%, it would still have flagged 232 different encounters and dropped 232. These counts show decision
sensitivity, not clinical harm: the 4B model ranks near chance, and no care decision was observed. A hospital that
validates a readmission model under one execution contract but deploys it under another could prioritize a different set of patients for transitional care. Fairness
evaluations in clinical machine learning already consider how a model is deployed, evaluated and
designed~\cite{rajkomar2018fairness}, and the execution contract belongs in that deployment record.

\section{Limitations}\label{sec:limitations}
The census and lending tasks cover one state and one year; the prespecified census results are unweighted, and the survey-weighted ones an exploratory, design-informed sensitivity check (Appendix~\ref{app:split}). The only health task is a public readmission dataset of
encounters of one disease (diabetes) from 1999 to 2008, with race recorded in coarse single-race categories that cannot represent people of more than one racial identity, and
no clinical deployment data were used. The models are small tabular models and
two open-weight language models used as zero-shot classifiers. The 4B
model ranks near chance on Diabetes and, at the prespecified threshold, selects almost no ACSEmployment record
(Section~\ref{sec:design-models}); the 24B model was scored on seeded subsets and on one GPU type. Hardware covers three NVIDIA generations, one MLX development build on Apple silicon, and one AMD MI350X, whose
TF32 setting is characterized by measurement and whose FP8 ran in PyTorch, not a serving engine; the FP16 arm used one CPU and one GPU type. Language-model Tier 1 runs used eager attention, not fused kernels. Only the reference scored training records, so no contract's threshold
was recalibrated on its own training scores. Flips are rare, so group concentration and cross-GPU Jaccard indices rest on small counts. The sampling split is post hoc for the 4B and tabular arms, and the 24B one adds two changes from one audit, so Section~\ref{sec:results-sampling} describes this study only.

\clearpage

\section*{Ethical Considerations}

\paragraph{Adverse impacts.}
The main risk is contract shopping. If a verdict depends on the execution contract, an auditee could choose the
contract under which an audit passes, and describing the effect also shows where to look. The protocol answers this
by having the auditor, not the auditee, declare the admissible contracts before scoring and report the whole
interval. A second risk is overreading small flip counts as group harm, so this study treats concentration as
mechanism. A third is the opposite misuse, dismissing a real disparity as numerical noise. A stability interval that
does not cross a criterion gives no reason to discount the verdict.

\paragraph{Health settings.}
In clinical use, a contract changes decisions for the patients whose scores sit nearest the cutoff, and a flipped
decision can decide who is offered a program, an alert or a follow-up. The cost of contract dependence therefore
falls on whoever the policy places at its boundary. A contract with one-signed error could also shift
which group gains or loses access, which this study did not test on the readmission task. There, the four-fifths rule and the 0.05 gap tolerance
are the auditor's declared rules, not clinical standards, and parity in who is flagged for transitional care is not
a clinical fairness criterion when need differs by group. A fixed top-10\% cutoff also treats patients just below it
as ineligible, although their needs may match those just above it. Clinicians and patients might not be aware of
these deployment choices, which is why this study asks that the contract be recorded with the model. Otherwise, a
change in which patients are flagged can go unnoticed.

\paragraph{Data.}
All data are public and de-identified: census microdata, public loan records and a public hospital readmission
dataset. No records were linked to other sources, no re-identification was attempted and no raw files are
redistributed. Race in the readmission data is recorded in coarse administrative categories, and group results
inherit that coding. Protected attributes were used to compute audit statistics, and as model inputs only where the
task definition includes them; the HMDA inputs include applicant age. There were no human participants, and credentialed clinical data were not
used. The language models ran locally on open weights, so no task record was sent to a hosted service; prompts
contain only each task's features.

\paragraph{Compute.}
Scoring the tabular models took minutes per contract on CPUs and GPUs, after training on a CPU; no other model was
trained. The 4B language-model arm used 12.97 H100 GPU-hours and 10.28 RTX 4090 GPU-hours, and its HMDA rescoring with
single-token labels 2.67 H100 GPU-hours. The 24B arm's 11 jobs held 41.77 H100
GPU-hours, its float32 jobs using two of the four GPUs they held. The AMD arm used 10.96 MI350X GPU-hours, 10.80 of
them for the language model. The checks added later (Appendix~\ref{app:notes}) are not counted here.

\paragraph{Scope.}
No fairness intervention was evaluated, and no claim is made that any deployed system, clinical or otherwise, is
unfair.

\section*{Generative AI Usage Statement}
Claude Code (Anthropic; Claude Opus 5.5) assisted with reviewing sections of the code and text; the authors
reviewed and revised all of it, and every number in the paper is checked against the result files by a script.
ChatGPT (OpenAI; Sol 5.6) was used for critique of drafts. The two language models this study audits,
Qwen3-4B-Instruct and Mistral Small 3.2, are research objects, not writing tools. The authors take full
responsibility for the content.

\bibliographystyle{ACM-Reference-Format}
\bibliography{refs}


\begin{thebibliography}{38}

%%% except for \shownote{} and \showURL{}.  The latter does
%%% not use final punctuation, in order to avoid confusing it with
%%% the Web address.

\ifx \showCODEN    \undefined \def \showCODEN     #1{\unskip}     \fi
\ifx \showISBNx    \undefined \def \showISBNx     #1{\unskip}     \fi
\ifx \showISBNxiii \undefined \def \showISBNxiii  #1{\unskip}     \fi
\ifx \showISSN     \undefined \def \showISSN      #1{\unskip}     \fi
\ifx \showLCCN     \undefined \def \showLCCN      #1{\unskip}     \fi
\ifx \shownote     \undefined \def \shownote      #1{#1}          \fi
\ifx \showarticletitle \undefined \def \showarticletitle #1{#1}   \fi
\ifx \showURL      \undefined \def \showURL       {\relax}        \fi
% invisible to TeX
\providecommand\bibfield[2]{#2}
\providecommand\bibinfo[2]{#2}
\providecommand\natexlab[1]{#1}
\providecommand\showeprint[2][]{arXiv:#2}
\makeatletter
\@ifundefined{NAT@parse@date}{}{\let\NAT@parse@date@orig\NAT@parse@date}
\@ifundefined{NAT@parse@date}{}{\def\NAT@parse@date#1#2#3#4#5#6@@{\NAT@parse@date@orig#1#2#3#4#5#6@@\def\NAT@tempyear{0000}\def\NAT@tempexlab{{?}}\ifx\NAT@year\NAT@tempyear\ifx\NAT@exlab\NAT@tempexlab\def\NAT@date{[n.\,d.]}\else\edef\NAT@date{[n.\,d.]\NAT@exlab}\fi\fi}}
\makeatother

\bibitem[{Apple Inc.}(2026)]%
        {apple2026msl}
\bibfield{author}{\bibinfo{person}{{Apple Inc.}}}
  \bibinfo{year}{2026}\natexlab{}.
\newblock \bibinfo{title}{Metal Shading Language Specification, Version 4.1}.
\newblock
\urldef\tempurl%
\url{https://developer.apple.com/metal/Metal-Shading-Language-Specification.pdf}
\showURL{%
Retrieved October 8, 2026 from \tempurl}
\newblock
\shownote{Dated 4 June 2026}.


\bibitem[Birhane et~al\mbox{.}(2024)]%
        {birhane2024broken}
\bibfield{author}{\bibinfo{person}{Abeba Birhane}, \bibinfo{person}{Ryan
  Steed}, \bibinfo{person}{Victor Ojewale}, \bibinfo{person}{Briana Vecchione},
  {and} \bibinfo{person}{Inioluwa~Deborah Raji}.}
  \bibinfo{year}{2024}\natexlab{}.
\newblock \showarticletitle{{AI} auditing: The Broken Bus on the Road to {AI}
  Accountability}. In \bibinfo{booktitle}{\emph{2024 IEEE Conference on Secure
  and Trustworthy Machine Learning (SaTML)}}. \bibinfo{publisher}{IEEE},
  \bibinfo{pages}{612--643}.
\newblock
\href{https://doi.org/10.1109/SaTML59370.2024.00037}{doi:\nolinkurl{10.1109/SaTML59370.2024.00037}}


\bibitem[{Board of Governors of the Federal Reserve System}(2026)]%
        {sr262}
\bibfield{author}{\bibinfo{person}{{Board of Governors of the Federal Reserve
  System}}.} \bibinfo{year}{2026}\natexlab{}.
\newblock \bibinfo{title}{{SR} 26-2: Revised Guidance on Model Risk
  Management}.
\newblock
\urldef\tempurl%
\url{https://www.federalreserve.gov/supervisionreg/srletters/SR2602.pdf}
\showURL{%
Retrieved October 8, 2026 from \tempurl}
\newblock
\shownote{17 April 2026}.


\bibitem[Broden et~al\mbox{.}(2026)]%
        {broden2026conformal}
\bibfield{author}{\bibinfo{person}{Julia Broden}, \bibinfo{person}{Jan Simson},
  {and} \bibinfo{person}{Christoph Kern}.} \bibinfo{year}{2026}\natexlab{}.
\newblock \showarticletitle{Conformal Prediction and the Many-Worlds of
  Fairness}. In \bibinfo{booktitle}{\emph{Proceedings of the 2026 ACM
  Conference on Fairness, Accountability, and Transparency}}
  \emph{(\bibinfo{series}{FAccT '26})}. \bibinfo{publisher}{ACM},
  \bibinfo{pages}{5981--6005}.
\newblock
\href{https://doi.org/10.1145/3805689.3812391}{doi:\nolinkurl{10.1145/3805689.3812391}}


\bibitem[Chen et~al\mbox{.}(2024)]%
        {chen2023chatgpt}
\bibfield{author}{\bibinfo{person}{Lingjiao Chen}, \bibinfo{person}{Matei
  Zaharia}, {and} \bibinfo{person}{James Zou}.}
  \bibinfo{year}{2024}\natexlab{}.
\newblock \showarticletitle{How Is {ChatGPT}'s Behavior Changing Over Time?}
\newblock \bibinfo{journal}{\emph{Harvard Data Science Review}}
  \bibinfo{volume}{6}, \bibinfo{number}{2} (\bibinfo{year}{2024}).
\newblock
\href{https://doi.org/10.1162/99608f92.5317da47}{doi:\nolinkurl{10.1162/99608f92.5317da47}}


\bibitem[Cherian and Cand{\`e}s(2024)]%
        {cherian2024fairness}
\bibfield{author}{\bibinfo{person}{John~J. Cherian} {and}
  \bibinfo{person}{Emmanuel~J. Cand{\`e}s}.} \bibinfo{year}{2024}\natexlab{}.
\newblock \showarticletitle{Statistical Inference for Fairness Auditing}.
\newblock \bibinfo{journal}{\emph{Journal of Machine Learning Research}}
  \bibinfo{volume}{25}, \bibinfo{number}{149} (\bibinfo{year}{2024}),
  \bibinfo{pages}{1--49}.
\newblock
\urldef\tempurl%
\url{https://jmlr.org/papers/v25/23-0739.html}
\showURL{%
\tempurl}


\bibitem[{Code of Federal Regulations}(2026)]%
        {cfr1607}
\bibfield{author}{\bibinfo{person}{{Code of Federal Regulations}}.}
  \bibinfo{year}{2026}\natexlab{}.
\newblock \bibinfo{title}{Uniform Guidelines on Employee Selection Procedures,
  29 {C.F.R.} \S 1607.4({D})}.
\newblock
\urldef\tempurl%
\url{https://www.ecfr.gov/current/title-29/subtitle-B/chapter-XIV/part-1607/section-1607.4}
\showURL{%
Retrieved October 8, 2026 from \tempurl}
\newblock
\shownote{Current text}.


\bibitem[{Consumer Financial Protection Bureau}(2026)]%
        {cfpb2018hmda}
\bibfield{author}{\bibinfo{person}{{Consumer Financial Protection Bureau}}.}
  \bibinfo{year}{2026}\natexlab{}.
\newblock \bibinfo{title}{{HMDA} Data Browser: 2018 Loan/Application Records}.
\newblock
\urldef\tempurl%
\url{https://ffiec.cfpb.gov/data-browser/}
\showURL{%
Retrieved October 4, 2026 from \tempurl}


\bibitem[Cooper et~al\mbox{.}(2022)]%
        {cooper2022nondeterminism}
\bibfield{author}{\bibinfo{person}{A.~Feder Cooper}, \bibinfo{person}{Jonathan
  Frankle}, {and} \bibinfo{person}{Christopher De~Sa}.}
  \bibinfo{year}{2022}\natexlab{}.
\newblock \showarticletitle{Non-Determinism and the Lawlessness of Machine
  Learning Code}. In \bibinfo{booktitle}{\emph{Proceedings of the 2022
  Symposium on Computer Science and Law}} \emph{(\bibinfo{series}{CSLAW '22})}.
  \bibinfo{publisher}{ACM}, \bibinfo{pages}{1--8}.
\newblock
\href{https://doi.org/10.1145/3511265.3550446}{doi:\nolinkurl{10.1145/3511265.3550446}}


\bibitem[Cooper et~al\mbox{.}(2024)]%
        {cooper2024arbitrariness}
\bibfield{author}{\bibinfo{person}{A.~Feder Cooper}, \bibinfo{person}{Katherine
  Lee}, \bibinfo{person}{Madiha~Zahrah Choksi}, \bibinfo{person}{Solon
  Barocas}, \bibinfo{person}{Christopher De~Sa}, \bibinfo{person}{James
  Grimmelmann}, \bibinfo{person}{Jon Kleinberg}, \bibinfo{person}{Siddhartha
  Sen}, {and} \bibinfo{person}{Baobao Zhang}.} \bibinfo{year}{2024}\natexlab{}.
\newblock \showarticletitle{Arbitrariness and Social Prediction: The
  Confounding Role of Variance in Fair Classification}. In
  \bibinfo{booktitle}{\emph{Proceedings of the AAAI Conference on Artificial
  Intelligence}}, Vol.~\bibinfo{volume}{38}. \bibinfo{publisher}{Association
  for the Advancement of Artificial Intelligence},
  \bibinfo{pages}{22004--22012}.
\newblock
\showeprint[arxiv]{2301.11562}
\href{https://doi.org/10.1609/aaai.v38i20.30203}{doi:\nolinkurl{10.1609/aaai.v38i20.30203}}


\bibitem[Das et~al\mbox{.}(2024)]%
        {das2024linearization}
\bibfield{author}{\bibinfo{person}{Saswat Das}, \bibinfo{person}{Marco
  Romanelli}, {and} \bibinfo{person}{Ferdinando Fioretto}.}
  \bibinfo{year}{2024}\natexlab{}.
\newblock \showarticletitle{Disparate Impact on Group Accuracy of Linearization
  for Private Inference}. In \bibinfo{booktitle}{\emph{Proceedings of the 41st
  International Conference on Machine Learning}}
  \emph{(\bibinfo{series}{Proceedings of Machine Learning Research},
  Vol.~\bibinfo{volume}{235})}. \bibinfo{publisher}{PMLR},
  \bibinfo{pages}{10168--10184}.
\newblock
\urldef\tempurl%
\url{https://proceedings.mlr.press/v235/das24d.html}
\showURL{%
\tempurl}


\bibitem[Ding et~al\mbox{.}(2021)]%
        {ding2021retiring}
\bibfield{author}{\bibinfo{person}{Frances Ding}, \bibinfo{person}{Moritz
  Hardt}, \bibinfo{person}{John Miller}, {and} \bibinfo{person}{Ludwig
  Schmidt}.} \bibinfo{year}{2021}\natexlab{}.
\newblock \showarticletitle{Retiring Adult: New Datasets for Fair Machine
  Learning}. In \bibinfo{booktitle}{\emph{Advances in Neural Information
  Processing Systems}}, Vol.~\bibinfo{volume}{34}. \bibinfo{publisher}{Curran
  Associates, Inc.}, \bibinfo{pages}{6478--6490}.
\newblock
\urldef\tempurl%
\url{https://proceedings.neurips.cc/paper_files/paper/2021/hash/32e54441e6382a7fbacbbbaf3c450059-Abstract.html}
\showURL{%
\tempurl}


\bibitem[Donz{\'e} et~al\mbox{.}(2013)]%
        {donze2013hospital}
\bibfield{author}{\bibinfo{person}{Jacques Donz{\'e}},
  \bibinfo{person}{Drahomir Aujesky}, \bibinfo{person}{Deborah Williams}, {and}
  \bibinfo{person}{Jeffrey~L. Schnipper}.} \bibinfo{year}{2013}\natexlab{}.
\newblock \showarticletitle{Potentially Avoidable 30-Day Hospital Readmissions
  in Medical Patients: Derivation and Validation of a Prediction Model}.
\newblock \bibinfo{journal}{\emph{JAMA Internal Medicine}}
  \bibinfo{volume}{173}, \bibinfo{number}{8} (\bibinfo{year}{2013}),
  \bibinfo{pages}{632--638}.
\newblock
\href{https://doi.org/10.1001/jamainternmed.2013.3023}{doi:\nolinkurl{10.1001/jamainternmed.2013.3023}}


\bibitem[Du et~al\mbox{.}(2026)]%
        {du2026greedy}
\bibfield{author}{\bibinfo{person}{Gaoyuan Du}, \bibinfo{person}{Anam~Nawaz
  Khan}, \bibinfo{person}{Rex Zhou}, \bibinfo{person}{Xiaoyang Liu},
  \bibinfo{person}{Deepayan Chakrabarti}, \bibinfo{person}{Fnu Suya}, {and}
  \bibinfo{person}{Xueping Li}.} \bibinfo{year}{2026}\natexlab{}.
\newblock \showarticletitle{Greedy Decoding Is Not Precision-Invariant:
  Cross-Precision Output Divergence in {LLM} Inference}.
\newblock \bibinfo{journal}{\emph{Transactions on Machine Learning Research}}
  (\bibinfo{year}{2026}).
\newblock
\showeprint[arxiv]{2609.26621}
\urldef\tempurl%
\url{https://openreview.net/forum?id=QDOKyg7a5e}
\showURL{%
\tempurl}


\bibitem[Ganesh et~al\mbox{.}(2023)]%
        {ganesh2023randomness}
\bibfield{author}{\bibinfo{person}{Prakhar Ganesh}, \bibinfo{person}{Hongyan
  Chang}, \bibinfo{person}{Martin Strobel}, {and} \bibinfo{person}{Reza
  Shokri}.} \bibinfo{year}{2023}\natexlab{}.
\newblock \showarticletitle{On the Impact of Machine Learning Randomness on
  Group Fairness}. In \bibinfo{booktitle}{\emph{Proceedings of the 2023 ACM
  Conference on Fairness, Accountability, and Transparency}}.
  \bibinfo{publisher}{Association for Computing Machinery},
  \bibinfo{address}{New York, NY, USA}, \bibinfo{pages}{1789--1800}.
\newblock
\href{https://doi.org/10.1145/3593013.3594116}{doi:\nolinkurl{10.1145/3593013.3594116}}


\bibitem[Jacobs and Wallach(2021)]%
        {jacobs2021measurement}
\bibfield{author}{\bibinfo{person}{Abigail~Z. Jacobs} {and}
  \bibinfo{person}{Hanna Wallach}.} \bibinfo{year}{2021}\natexlab{}.
\newblock \showarticletitle{Measurement and Fairness}. In
  \bibinfo{booktitle}{\emph{Proceedings of the 2021 ACM Conference on Fairness,
  Accountability, and Transparency}}. \bibinfo{publisher}{Association for
  Computing Machinery}, \bibinfo{address}{New York, NY, USA},
  \bibinfo{pages}{375--385}.
\newblock
\href{https://doi.org/10.1145/3442188.3445901}{doi:\nolinkurl{10.1145/3442188.3445901}}


\bibitem[Kansagara et~al\mbox{.}(2011)]%
        {kansagara2011readmission}
\bibfield{author}{\bibinfo{person}{Devan Kansagara}, \bibinfo{person}{Honora
  Englander}, \bibinfo{person}{Amanda Salanitro}, \bibinfo{person}{David
  Kagen}, \bibinfo{person}{Cecelia Theobald}, \bibinfo{person}{Michele
  Freeman}, {and} \bibinfo{person}{Sunil Kripalani}.}
  \bibinfo{year}{2011}\natexlab{}.
\newblock \showarticletitle{Risk Prediction Models for Hospital Readmission: A
  Systematic Review}.
\newblock \bibinfo{journal}{\emph{JAMA}} \bibinfo{volume}{306},
  \bibinfo{number}{15} (\bibinfo{year}{2011}), \bibinfo{pages}{1688--1698}.
\newblock
\href{https://doi.org/10.1001/jama.2011.1515}{doi:\nolinkurl{10.1001/jama.2011.1515}}


\bibitem[Lip et~al\mbox{.}(2010)]%
        {lip2010chadsvasc}
\bibfield{author}{\bibinfo{person}{Gregory Y.~H. Lip}, \bibinfo{person}{Robby
  Nieuwlaat}, \bibinfo{person}{Ron Pisters}, \bibinfo{person}{Deirdre~A. Lane},
  {and} \bibinfo{person}{Harry J. G.~M. Crijns}.}
  \bibinfo{year}{2010}\natexlab{}.
\newblock \showarticletitle{Refining Clinical Risk Stratification for
  Predicting Stroke and Thromboembolism in Atrial Fibrillation Using a Novel
  Risk Factor-Based Approach: The Euro Heart Survey on Atrial Fibrillation}.
\newblock \bibinfo{journal}{\emph{Chest}} \bibinfo{volume}{137},
  \bibinfo{number}{2} (\bibinfo{year}{2010}), \bibinfo{pages}{263--272}.
\newblock
\href{https://doi.org/10.1378/chest.09-1584}{doi:\nolinkurl{10.1378/chest.09-1584}}


\bibitem[Liu et~al\mbox{.}(2026)]%
        {liu2026trust}
\bibfield{author}{\bibinfo{person}{Yuanhao Liu}, \bibinfo{person}{Qi Cao},
  {and} \bibinfo{person}{Huawei Shen}.} \bibinfo{year}{2026}\natexlab{}.
\newblock \showarticletitle{When Can We Trust Fairness Audits? Identifying
  Reliability Boundaries of Third-party Audit Conclusions}. In
  \bibinfo{booktitle}{\emph{Proceedings of the Thirty-Fifth International Joint
  Conference on Artificial Intelligence, {IJCAI-26}}}.
  \bibinfo{publisher}{International Joint Conferences on Artificial
  Intelligence Organization}, \bibinfo{pages}{7319--7327}.
\newblock
\href{https://doi.org/10.24963/ijcai.2026/814}{doi:\nolinkurl{10.24963/ijcai.2026/814}}


\bibitem[Marcuzzi et~al\mbox{.}(2026)]%
        {marcuzzi2026quantization}
\bibfield{author}{\bibinfo{person}{Federico Marcuzzi}, \bibinfo{person}{Xuefei
  Ning}, \bibinfo{person}{Roy Schwartz}, {and} \bibinfo{person}{Iryna
  Gurevych}.} \bibinfo{year}{2026}\natexlab{}.
\newblock \showarticletitle{How Quantization Shapes Bias in Large Language
  Models}. In \bibinfo{booktitle}{\emph{Proceedings of the 19th Conference of
  the {E}uropean Chapter of the {A}ssociation for {C}omputational {L}inguistics
  (Volume 1: Long Papers)}}. \bibinfo{publisher}{Association for Computational
  Linguistics}, \bibinfo{address}{Stroudsburg, PA, USA},
  \bibinfo{pages}{363--404}.
\newblock
\showeprint[arxiv]{2508.18088}
\href{https://doi.org/10.18653/v1/2026.eacl-long.17}{doi:\nolinkurl{10.18653/v1/2026.eacl-long.17}}


\bibitem[Marx et~al\mbox{.}(2020)]%
        {marx2020multiplicity}
\bibfield{author}{\bibinfo{person}{Charles Marx}, \bibinfo{person}{Flavio
  Calmon}, {and} \bibinfo{person}{Berk Ustun}.}
  \bibinfo{year}{2020}\natexlab{}.
\newblock \showarticletitle{Predictive Multiplicity in Classification}. In
  \bibinfo{booktitle}{\emph{Proceedings of the International Conference on
  Machine Learning}} \emph{(\bibinfo{series}{Proceedings of Machine Learning
  Research}, Vol.~\bibinfo{volume}{119})}. \bibinfo{publisher}{PMLR},
  \bibinfo{pages}{6765--6774}.
\newblock
\urldef\tempurl%
\url{https://proceedings.mlr.press/v119/marx20a.html}
\showURL{%
\tempurl}


\bibitem[Nelaturu et~al\mbox{.}(2024)]%
        {nelaturu2024hardware}
\bibfield{author}{\bibinfo{person}{Sree~Harsha Nelaturu},
  \bibinfo{person}{Nishaanth~Kanna Ravichandran}, \bibinfo{person}{Cuong Tran},
  \bibinfo{person}{Sara Hooker}, {and} \bibinfo{person}{Ferdinando Fioretto}.}
  \bibinfo{year}{2024}\natexlab{}.
\newblock \showarticletitle{On The Fairness Impacts of Hardware Selection in
  Machine Learning}. In \bibinfo{booktitle}{\emph{Proceedings of the 41st
  International Conference on Machine Learning}}
  \emph{(\bibinfo{series}{Proceedings of Machine Learning Research},
  Vol.~\bibinfo{volume}{235})}. \bibinfo{publisher}{PMLR},
  \bibinfo{pages}{37486--37507}.
\newblock
\urldef\tempurl%
\url{https://proceedings.mlr.press/v235/nelaturu24a.html}
\showURL{%
\tempurl}


\bibitem[{New York City Department of Consumer and Worker Protection}(2023)]%
        {nyc_aedt}
\bibfield{author}{\bibinfo{person}{{New York City Department of Consumer and
  Worker Protection}}.} \bibinfo{year}{2023}\natexlab{}.
\newblock \bibinfo{title}{Automated Employment Decision Tools: Notice of
  Adoption of Final Rule, Subchapter {T} of Chapter 5 of Title 6 of the Rules
  of the City of New York}.
\newblock
\urldef\tempurl%
\url{https://rules.cityofnewyork.us/rule/automated-employment-decision-tools-updated/}
\showURL{%
Retrieved October 8, 2026 from \tempurl}
\newblock
\shownote{Effective 5 July 2023}.


\bibitem[{NVIDIA}(2026)]%
        {nvidia_h100}
\bibfield{author}{\bibinfo{person}{{NVIDIA}}.} \bibinfo{year}{2026}\natexlab{}.
\newblock \bibinfo{title}{{NVIDIA} {H100} Tensor Core {GPU}}.
\newblock
\urldef\tempurl%
\url{https://www.nvidia.com/en-us/data-center/h100/}
\showURL{%
Retrieved October 12, 2026 from \tempurl}
\newblock
\shownote{Product specifications}.


\bibitem[Obermeyer et~al\mbox{.}(2019)]%
        {obermeyer2019dissecting}
\bibfield{author}{\bibinfo{person}{Ziad Obermeyer}, \bibinfo{person}{Brian
  Powers}, \bibinfo{person}{Christine Vogeli}, {and} \bibinfo{person}{Sendhil
  Mullainathan}.} \bibinfo{year}{2019}\natexlab{}.
\newblock \showarticletitle{Dissecting Racial Bias in an Algorithm Used to
  Manage the Health of Populations}.
\newblock \bibinfo{journal}{\emph{Science}} \bibinfo{volume}{366},
  \bibinfo{number}{6464} (\bibinfo{year}{2019}), \bibinfo{pages}{447--453}.
\newblock
\href{https://doi.org/10.1126/science.aax2342}{doi:\nolinkurl{10.1126/science.aax2342}}


\bibitem[{ONNX Runtime}(2026)]%
        {onnxruntime_cuda}
\bibfield{author}{\bibinfo{person}{{ONNX Runtime}}.}
  \bibinfo{year}{2026}\natexlab{}.
\newblock \bibinfo{title}{{CUDA} Execution Provider}.
\newblock
\urldef\tempurl%
\url{https://onnxruntime.ai/docs/execution-providers/CUDA-ExecutionProvider.html}
\showURL{%
Retrieved October 8, 2026 from \tempurl}
\newblock
\shownote{Documentation}.


\bibitem[{PyTorch}(2026)]%
        {pytorch_hip}
\bibfield{author}{\bibinfo{person}{{PyTorch}}.}
  \bibinfo{year}{2026}\natexlab{}.
\newblock \bibinfo{title}{{HIP} ({ROCm}) Semantics: {TensorFloat-32} ({TF32})
  on {ROCm}}.
\newblock
\urldef\tempurl%
\url{https://docs.pytorch.org/docs/2.14/notes/hip.html#tensorfloat-32-tf32-on-rocm}
\showURL{%
Retrieved October 8, 2026 from \tempurl}
\newblock
\shownote{Documentation for version 2.14}.


\bibitem[Rajkomar et~al\mbox{.}(2018)]%
        {rajkomar2018fairness}
\bibfield{author}{\bibinfo{person}{Alvin Rajkomar}, \bibinfo{person}{Michaela
  Hardt}, \bibinfo{person}{Michael~D. Howell}, \bibinfo{person}{Greg Corrado},
  {and} \bibinfo{person}{Marshall~H. Chin}.} \bibinfo{year}{2018}\natexlab{}.
\newblock \showarticletitle{Ensuring Fairness in Machine Learning to Advance
  Health Equity}.
\newblock \bibinfo{journal}{\emph{Annals of Internal Medicine}}
  \bibinfo{volume}{169}, \bibinfo{number}{12} (\bibinfo{year}{2018}),
  \bibinfo{pages}{866--872}.
\newblock
\href{https://doi.org/10.7326/M18-1990}{doi:\nolinkurl{10.7326/M18-1990}}


\bibitem[{Runpod}(2026)]%
        {runpod2026pricing}
\bibfield{author}{\bibinfo{person}{{Runpod}}.} \bibinfo{year}{2026}\natexlab{}.
\newblock \bibinfo{title}{{GPU} Cloud Pricing}.
\newblock
\urldef\tempurl%
\url{https://www.runpod.io/pricing}
\showURL{%
Retrieved October 4, 2026 from \tempurl}
\newblock
\shownote{Secure Cloud on-demand rates}.


\bibitem[Simson et~al\mbox{.}(2024)]%
        {simson2024multiverse}
\bibfield{author}{\bibinfo{person}{Jan Simson}, \bibinfo{person}{Florian
  Pfisterer}, {and} \bibinfo{person}{Christoph Kern}.}
  \bibinfo{year}{2024}\natexlab{}.
\newblock \showarticletitle{One Model Many Scores: Using Multiverse Analysis to
  Prevent Fairness Hacking and Evaluate the Influence of Model Design
  Decisions}. In \bibinfo{booktitle}{\emph{Proceedings of the 2024 ACM
  Conference on Fairness, Accountability, and Transparency}}
  \emph{(\bibinfo{series}{FAccT '24})}. \bibinfo{publisher}{ACM},
  \bibinfo{pages}{1305--1320}.
\newblock
\href{https://doi.org/10.1145/3630106.3658974}{doi:\nolinkurl{10.1145/3630106.3658974}}


\bibitem[Strack et~al\mbox{.}(2014)]%
        {strack2014impact}
\bibfield{author}{\bibinfo{person}{Beata Strack}, \bibinfo{person}{Jonathan~P.
  DeShazo}, \bibinfo{person}{Chris Gennings}, \bibinfo{person}{Juan~L. Olmo},
  \bibinfo{person}{Sebastian Ventura}, \bibinfo{person}{Krzysztof~J. Cios},
  {and} \bibinfo{person}{John~N. Clore}.} \bibinfo{year}{2014}\natexlab{}.
\newblock \showarticletitle{Impact of {HbA1c} Measurement on Hospital
  Readmission Rates: Analysis of 70,000 Clinical Database Patient Records}.
\newblock \bibinfo{journal}{\emph{BioMed Research International}}
  \bibinfo{volume}{2014} (\bibinfo{year}{2014}), \bibinfo{pages}{781670}.
\newblock
\href{https://doi.org/10.1155/2014/781670}{doi:\nolinkurl{10.1155/2014/781670}}


\bibitem[{TensorFlow}(2026)]%
        {tf_tf32}
\bibfield{author}{\bibinfo{person}{{TensorFlow}}.}
  \bibinfo{year}{2026}\natexlab{}.
\newblock
  \bibinfo{title}{tf.config.experimental.enable\_tensor\_float\_32\_execution}.
\newblock
\urldef\tempurl%
\url{https://www.tensorflow.org/api_docs/python/tf/config/experimental/enable_tensor_float_32_execution}
\showURL{%
Retrieved October 8, 2026 from \tempurl}
\newblock
\shownote{Documentation for version 2.16.1}.


\bibitem[Tran et~al\mbox{.}(2022)]%
        {tran2022pruning}
\bibfield{author}{\bibinfo{person}{Cuong Tran}, \bibinfo{person}{Ferdinando
  Fioretto}, \bibinfo{person}{Jung-Eun Kim}, {and} \bibinfo{person}{Rakshit
  Naidu}.} \bibinfo{year}{2022}\natexlab{}.
\newblock \showarticletitle{Pruning Has a Disparate Impact on Model Accuracy}.
  In \bibinfo{booktitle}{\emph{Advances in Neural Information Processing
  Systems}}, Vol.~\bibinfo{volume}{35}. \bibinfo{publisher}{Neural Information
  Processing Systems Foundation, Inc.}, \bibinfo{address}{San Diego, CA, USA},
  \bibinfo{pages}{17652--17664}.
\newblock
\showeprint[arxiv]{2205.13574}
\href{https://doi.org/10.52202/068431-1283}{doi:\nolinkurl{10.52202/068431-1283}}


\bibitem[{U.S. Food and Drug Administration}(2025)]%
        {fda_pccp}
\bibfield{author}{\bibinfo{person}{{U.S. Food and Drug Administration}}.}
  \bibinfo{year}{2025}\natexlab{}.
\newblock \bibinfo{title}{Marketing Submission Recommendations for a
  Predetermined Change Control Plan for Artificial Intelligence-Enabled Device
  Software Functions: Guidance for Industry and {Food and Drug Administration}
  Staff}.
\newblock
\urldef\tempurl%
\url{https://www.fda.gov/regulatory-information/search-fda-guidance-documents/marketing-submission-recommendations-predetermined-change-control-plan-artificial-intelligence}
\showURL{%
Retrieved October 8, 2026 from \tempurl}
\newblock
\shownote{Issued 18 August 2025; originally issued 4 December 2024. Docket
  FDA-2022-D-2628}.


\bibitem[van Walraven et~al\mbox{.}(2010)]%
        {vanwalraven2010lace}
\bibfield{author}{\bibinfo{person}{Carl van Walraven},
  \bibinfo{person}{Irfan~A. Dhalla}, \bibinfo{person}{Chaim Bell},
  \bibinfo{person}{Edward Etchells}, \bibinfo{person}{Ian~G. Stiell},
  \bibinfo{person}{Kelly Zarnke}, \bibinfo{person}{Peter~C. Austin}, {and}
  \bibinfo{person}{Alan~J. Forster}.} \bibinfo{year}{2010}\natexlab{}.
\newblock \showarticletitle{Derivation and Validation of an Index to Predict
  Early Death or Unplanned Readmission after Discharge from Hospital to the
  Community}.
\newblock \bibinfo{journal}{\emph{Canadian Medical Association Journal}}
  \bibinfo{volume}{182}, \bibinfo{number}{6} (\bibinfo{year}{2010}),
  \bibinfo{pages}{551--557}.
\newblock
\href{https://doi.org/10.1503/cmaj.091117}{doi:\nolinkurl{10.1503/cmaj.091117}}


\bibitem[{vLLM}(2026)]%
        {vllm_engine_args}
\bibfield{author}{\bibinfo{person}{{vLLM}}.} \bibinfo{year}{2026}\natexlab{}.
\newblock \bibinfo{title}{Engine Arguments}.
\newblock
\urldef\tempurl%
\url{https://docs.vllm.ai/en/latest/configuration/engine_args/}
\showURL{%
Retrieved October 8, 2026 from \tempurl}
\newblock
\shownote{Documentation}.


\bibitem[Yao et~al\mbox{.}(2026)]%
        {yao2026tao}
\bibfield{author}{\bibinfo{person}{Jianzhu Yao}, \bibinfo{person}{Hongxu Su},
  \bibinfo{person}{Taobo Liao}, \bibinfo{person}{Zerui Cheng},
  \bibinfo{person}{Huan Zhang}, \bibinfo{person}{Xuechao Wang}, {and}
  \bibinfo{person}{Pramod Viswanath}.} \bibinfo{year}{2026}\natexlab{}.
\newblock \showarticletitle{{TAO}: Tolerance-Aware Optimistic Verification for
  Floating-Point Neural Networks}. In \bibinfo{booktitle}{\emph{Proceedings of
  the 21st European Conference on Computer Systems}}. \bibinfo{publisher}{ACM},
  \bibinfo{pages}{1515--1532}.
\newblock
\href{https://doi.org/10.1145/3767295.3803612}{doi:\nolinkurl{10.1145/3767295.3803612}}


\bibitem[Yuan et~al\mbox{.}(2025)]%
        {yuan2025nondeterminism}
\bibfield{author}{\bibinfo{person}{Jiayi Yuan}, \bibinfo{person}{Hao Li},
  \bibinfo{person}{Xinheng Ding}, \bibinfo{person}{Wenya Xie},
  \bibinfo{person}{Yu-Jhe Li}, \bibinfo{person}{Wentian Zhao},
  \bibinfo{person}{Kun Wan}, \bibinfo{person}{Jing Shi}, \bibinfo{person}{Xia
  Hu}, {and} \bibinfo{person}{Zirui Liu}.} \bibinfo{year}{2025}\natexlab{}.
\newblock \showarticletitle{Understanding and Mitigating Numerical Sources of
  Nondeterminism in {LLM} Inference}. In \bibinfo{booktitle}{\emph{Advances in
  Neural Information Processing Systems}}, Vol.~\bibinfo{volume}{38}.
  \bibinfo{publisher}{Neural Information Processing Systems Foundation, Inc.},
  \bibinfo{address}{San Diego, CA, USA}, \bibinfo{pages}{188045--188077}.
\newblock
\showeprint[arxiv]{2506.09501}
\href{https://doi.org/10.52202/085713-5653}{doi:\nolinkurl{10.52202/085713-5653}}


\end{thebibliography}

\appendix

\section{Supplementary Tables}\label{app:tables}

\subsection{Decision flips by contract and hardware}\label{app:tab-flips}

Tables~\ref{tab:app-tab1} to~\ref{tab:app-llm} give the flip counts behind Table~\ref{tab:flips} for every model,
contract and device. A flip is a test decision that differs from the declared reference at the prespecified threshold: FP64 for the
tabular and 4B arms, strict FP32 for the 24B arm.
Every run 2 changed the same records as its run 1, and every tabular run 2 returned byte-identical scores.
The tabular flips were more stable across devices than the language model's. The MLP's TF32 flips were the same
records on all three NVIDIA GPUs for ACSIncome and HMDA, but on ACSEmployment the RTX A4000 flipped one record and
the H100 and RTX 4090 another. Tabular BF16 flipped the same records on the MI350X as on the M5 Max CPU in all eight
cells, consistent with rounding to BF16, rather than accumulation, causing them.

\begin{table}[ht]
\caption{Tabular Tier 1: decisions changed against FP64 at the prespecified threshold (run 1; run 2 changed the same
records and returned identical scores in every cell), and the largest score change $\max|\Delta s|$. Strict: every
FP64 and strict, certified or CPU FP32 contract on every device, all without a flip. TF32 flips are given for the
RTX A4000, RTX 4090 and H100 in that order, and its $\max|\Delta s|$ is the largest over the three. The tree ensembles changed no decision; their largest score change
was $4.0\times10^{-7}$ with the CPU predictor and $8.0\times10^{-7}$ with the CUDA predictor.}
\label{tab:app-tab1}
\small
\begin{tabular}{@{}l l r r r r r r r@{}}
\toprule
 & & Strict & \multicolumn{2}{c}{TF32, NVIDIA} & \multicolumn{2}{c}{TF32 setting, MI350X} & \multicolumn{2}{c}{M5 relaxed} \\
\cmidrule(lr){3-3}\cmidrule(lr){4-5}\cmidrule(lr){6-7}\cmidrule(l){8-9}
Task & Model & $\max|\Delta s|$ & Flips & $\max|\Delta s|$ & Flips & $\max|\Delta s|$ & Flips & $\max|\Delta s|$ \\
\midrule
ACSIncome & LR & $1.4\times10^{-7}$ & 0 / 0 / 0 & $1.3\times10^{-7}$ & 0 & $5.3\times10^{-6}$ & 0 & $1.2\times10^{-7}$ \\
 & MLP & $3.2\times10^{-7}$ & 3 / 3 / 3 & $6.8\times10^{-4}$ & 0 & $1.5\times10^{-5}$ & 1 & $1.5\times10^{-3}$ \\
ACSEmployment & LR & $2.8\times10^{-7}$ & 0 / 0 / 0 & $2.8\times10^{-7}$ & 0 & $1.5\times10^{-5}$ & 0 & $2.4\times10^{-7}$ \\
 & MLP & $1.7\times10^{-7}$ & 1 / 1 / 1 & $6.6\times10^{-4}$ & 0 & $1.0\times10^{-5}$ & 31 & $1.6\times10^{-3}$ \\
HMDA & LR & $3.3\times10^{-7}$ & 0 / 0 / 0 & $3.3\times10^{-7}$ & 0 & $6.6\times10^{-6}$ & 0 & $1.6\times10^{-7}$ \\
 & MLP & $8.9\times10^{-7}$ & 1 / 1 / 1 & $1.2\times10^{-3}$ & 0 & $1.9\times10^{-5}$ & 6 & $2.5\times10^{-3}$ \\
Diabetes & LR & $1.6\times10^{-7}$ & 0 / 0 / 0 & $1.1\times10^{-7}$ & 0 & $3.9\times10^{-6}$ & 0 & $1.3\times10^{-7}$ \\
 & MLP & $2.5\times10^{-7}$ & 0 / 0 / 0 & $4.4\times10^{-4}$ & 0 & $5.2\times10^{-6}$ & 6 & $7.2\times10^{-4}$ \\
\bottomrule
\end{tabular}
\end{table}

\begin{table}[ht]
\caption{Tabular Tier 2: decisions changed against FP64 at the prespecified threshold (run 1; run 2 changed the same
records in every cell). CPU contracts ran on the CPU of an Apple M5 Max (BF16 with and without autocast gave
identical scores), MLX contracts on its GPU, and the last two on the MI350X.}
\label{tab:app-tab2}
\small
\begin{tabular}{@{}l l r r r r r r r@{}}
\toprule
Task & Model & CPU BF16 & CPU INT8 & MLX BF16 & MLX 8-bit & MLX MXFP8 & MI350X BF16 & MI350X FP8 \\
\midrule
ACSIncome & LR & 39 & 98 & 91 & 41 & 190 & 39 & 230 \\
 & MLP & 15 & 124 & 46 & 42 & 173 & 15 & 294 \\
ACSEmployment & LR & 185 & 266 & 49 & 300 & 957 & 185 & 1{,}601 \\
 & MLP & 96 & 480 & 150 & 59 & 152 & 96 & 860 \\
HMDA & LR & 74 & 296 & 61 & 105 & 584 & 74 & 536 \\
 & MLP & 15 & 323 & 15 & 53 & 172 & 15 & 514 \\
Diabetes & LR & 18 & 239 & 5 & 26 & 171 & 18 & 230 \\
 & MLP & 8 & 176 & 8 & 27 & 76 & 8 & 144 \\
\bottomrule
\end{tabular}
\end{table}

\begin{table}[ht]
\caption{The 4B language model at the prespecified threshold. (A) Decisions changed against FP64 in run 1; run 2
changed the same records in every cell. (B) The largest $|m|$ of a changed decision, where $m$ is the FP64 logit
margin from the threshold. BF16 is transformers with SDPA attention on each GPU; vLLM runs on the
H100; FP8 on the MI350X is PyTorch's scaled matrix product with per-tensor scales. Strict FP32 on all three GPUs
and FP64 on the MI350X changed no decision, and the MI350X's untouched defaults matched strict FP32 bit for bit
on the first shard of each task.}
\label{tab:app-llm}
\small
\begin{tabular}{@{}l r r r r r r@{}}
\toprule
 & TF32 & TF32 setting & BF16 & vLLM BF16 & vLLM FP8 & FP8 \\
Task ($n$) & H100 / 4090 & MI350X & H100 / MI350X & H100 & H100 & MI350X \\
\midrule
\multicolumn{7}{@{}l}{(A) Decisions changed} \\
ACSIncome (39{,}133) & 9 / 8 & 0 & 176 / 191 & 166 & 976 & 927 \\
ACSEmployment (75{,}763) & 2 / 3 & 2 & 12 / 15 & 14 & 67 & 56 \\
HMDA (69{,}370) & 23 / 21 & 2 & 264 / 242 & 178 & 1{,}789 & 1{,}791 \\
Diabetes (19{,}938) & 7 / 7 & 0 & 134 / 106 & 90 & 822 & 629 \\
\midrule
\multicolumn{7}{@{}l}{(B) Largest $|m|$ of a changed decision} \\
ACSIncome & 0.0339 / 0.0218 & -- & 0.551 / 0.847 & 0.484 & 3.91 & 6.74 \\
ACSEmployment & 0.0086 / 0.0174 & 0.0121 & 0.579 / 0.463 & 0.334 & 2.72 & 3.14 \\
HMDA & 0.0817 / 0.0771 & 0.00247 & 2.03 / 2.03 & 1.84 & 8.36 & 7.61 \\
Diabetes & 0.0175 / 0.0109 & -- & 0.623 / 0.435 & 0.518 & 2.94 & 2.65 \\
\midrule
\multicolumn{7}{@{}l}{(C) 24B model, decisions changed against strict FP32 on the H100} \\
ACSIncome (39{,}133) & 6 / -- & -- & 115 / -- & 88 & 403 & -- \\
ACSEmployment (20{,}000) & 13 / -- & -- & 380 / -- & 329 & 2{,}156 & -- \\
HMDA (20{,}000) & 1 / -- & -- & 22 / -- & 15 & 114 & -- \\
Diabetes (10{,}000) & 3 / -- & -- & 63 / -- & 47 & 386 & -- \\
\midrule
\multicolumn{7}{@{}l}{(D) 24B model, largest probability distance $|s - t|$ of a changed decision} \\
ACSIncome & $5.8\times10^{-4}$ / -- & -- & 0.0494 / -- & 0.0317 & 0.240 & -- \\
ACSEmployment & $3.2\times10^{-5}$ / -- & -- & $9.3\times10^{-4}$ / -- & $7.8\times10^{-4}$ & 0.00961 & -- \\
HMDA & $4.9\times10^{-4}$ / -- & -- & 0.0711 / -- & 0.0140 & 0.167 & -- \\
Diabetes & $6.7\times10^{-4}$ / -- & -- & 0.0163 / -- & 0.0163 & 0.0869 & -- \\
\bottomrule
\end{tabular}
\end{table}

\subsection{Contract checks}\label{app:tab-cert}

Before scoring, each Tier 1 run with matrix products read back its precision settings and compared a canary matrix
product with FP64 (Table~\ref{tab:app-cert}). On the NVIDIA GPUs, TF32 raised the canary error of a matrix product from at most
$1.3\times10^{-6}$ under strict FP32 to $3.0\times10^{-4}$, while a one-output product kept at most
$2.5\times10^{-7}$ under both. The vLLM runs recorded neither read-back nor canary, and the NVIDIA BF16 runs
recorded the canary in float32 only. On the MI350X, the TF32 setting gave $4.6\times10^{-6}$ and $4.3\times10^{-6}$ for the two shapes. On
the M5, the per-layer check found each one-output layer native and both hidden layers of every MLP
truncating, so certified FP32 routed those two layers to an honest kernel. Run 2 read back the same settings and
returned the same canary errors as run 1 in every case.

\begin{table}[p]
\caption{Contract checks per run. Read-back: PyTorch's TF32 flag and float32 matrix-product precision.
Canary: the largest relative error of a product against FP64, for a matrix product and for a one-output product.
The last column compares the product with its own rounded or dequantized operands
(BF16, 8-bit and MXFP8 weights) or with the FP64 product of the dequantized FP8 operands, so it reports the
remaining arithmetic error, including any further operand and output rounding. CPU contracts on a GPU's host ran on the CPU named. Run 2 gave the same values in every row.}
\label{tab:app-cert}
\footnotesize
\begin{tabular}{@{}l l l r r r@{}}
\toprule
Hardware & Contract & Read-back & Matrix product & One-output product & Against own operands \\
\midrule
\multicolumn{6}{@{}l}{Tabular models, Tier 1} \\
Apple M4 Pro CPU & FP64 & -- & 0 & 0 & -- \\
 & FP32 & -- & $1.4\times10^{-6}$ & $9.9\times10^{-8}$ & -- \\
RTX 4090 & CPU FP32 (Threadripper 1950X) & -- & $3.4\times10^{-7}$ & $3.4\times10^{-7}$ & -- \\
 & strict FP32 & off, highest & $1.3\times10^{-6}$ & $2.5\times10^{-7}$ & -- \\
 & TF32 & on, high & $3.0\times10^{-4}$ & $2.5\times10^{-7}$ & -- \\
RTX A4000 & CPU FP32 (Core i3-8100) & -- & $3.4\times10^{-7}$ & $2.5\times10^{-7}$ & -- \\
 & strict FP32 & off, highest & $8.6\times10^{-7}$ & $2.5\times10^{-7}$ & -- \\
 & TF32 & on, high & $3.0\times10^{-4}$ & $2.5\times10^{-7}$ & -- \\
AMD MI350X & CPU FP32 (EPYC 9755) & -- & $3.4\times10^{-7}$ & $3.4\times10^{-7}$ & -- \\
 & FP64 & off, highest & $2.4\times10^{-15}$ & $1.6\times10^{-15}$ & -- \\
 & strict FP32 & off, highest & $1.3\times10^{-6}$ & $1.8\times10^{-7}$ & -- \\
 & untouched defaults & off, highest & $1.3\times10^{-6}$ & $1.8\times10^{-7}$ & -- \\
 & TF32 setting & on, high & $4.6\times10^{-6}$ & $4.3\times10^{-6}$ & -- \\
H100 & CPU FP32 (EPYC 9654) & -- & $8.6\times10^{-7}$ & $2.2\times10^{-7}$ & -- \\
 & strict FP32 & off, highest & $4.2\times10^{-7}$ & $1.7\times10^{-7}$ & -- \\
 & TF32 & on, high & $3.0\times10^{-4}$ & $1.7\times10^{-7}$ & -- \\
Apple M5 Max GPU & \multicolumn{5}{l}{relaxed and certified FP32: per-layer check. LR layers and the MLP's
output layer run natively} \\
 & \multicolumn{5}{l}{with one output; both hidden MLP layers truncate natively and are routed to an honest kernel.} \\
\midrule
\multicolumn{6}{@{}l}{Tabular models, Tier 2} \\
Apple M5 Max CPU & FP32 & -- & $1.4\times10^{-6}$ & $9.9\times10^{-8}$ & -- \\
 & BF16 & -- & $4.0\times10^{-3}$ & $2.7\times10^{-3}$ & $3.0\times10^{-3}$ \\
 & INT8 dynamic & -- & $1.4\times10^{-2}$ & $1.2\times10^{-2}$ & -- \\
Apple M5 Max GPU (MLX) & BF16 & -- & $3.7\times10^{-3}$ & -- & $3.2\times10^{-3}$ \\
 & 8-bit weights & -- & $5.8\times10^{-3}$ & -- & $8.1\times10^{-4}$ \\
 & MXFP8 weights & -- & $3.0\times10^{-2}$ & -- & $4.4\times10^{-4}$ \\
AMD MI350X & BF16 & off, highest & $4.0\times10^{-3}$ & $2.7\times10^{-3}$ & -- \\
 & FP8 (E4M3, per-tensor scales) & off, highest & -- & -- & $1.6\times10^{-5}$ \\
\midrule
\multicolumn{6}{@{}l}{4B language model} \\
AMD MI350X & FP64 & off, highest & $2.4\times10^{-15}$ & $1.6\times10^{-15}$ & -- \\
 & strict FP32 & off, highest & $1.3\times10^{-6}$ & $1.8\times10^{-7}$ & -- \\
 & untouched defaults & off, highest & $1.3\times10^{-6}$ & $1.8\times10^{-7}$ & -- \\
 & TF32 setting & on, high & $4.6\times10^{-6}$ & $4.3\times10^{-6}$ & -- \\
 & BF16 & off, highest & $4.0\times10^{-3}$ & $2.7\times10^{-3}$ & -- \\
 & FP8 (E4M3, per-tensor scales) & off, highest & -- & -- & $1.6\times10^{-5}$ \\
H100 & FP64 & off, highest & $2.1\times10^{-15}$ & $3.9\times10^{-16}$ & -- \\
 & strict FP32 & off, highest & $4.2\times10^{-7}$ & $1.7\times10^{-7}$ & -- \\
 & TF32 & on, high & $3.0\times10^{-4}$ & $1.7\times10^{-7}$ & -- \\
 & BF16 & off, highest & \multicolumn{3}{l}{float32 canary only: $4.2\times10^{-7}$ and $1.7\times10^{-7}$} \\
 & vLLM BF16 and FP8 & \multicolumn{4}{l}{not recorded: the serving engine reports no read-back or canary} \\
RTX 4090 & strict FP32 & off, highest & $1.3\times10^{-6}$ & $2.5\times10^{-7}$ & -- \\
 & TF32 & on, high & $3.0\times10^{-4}$ & $2.5\times10^{-7}$ & -- \\
\multicolumn{6}{@{}l}{24B language model: on every H100 of each job and in both runs, the read-back and canary values of} \\
\multicolumn{6}{@{}l}{the 4B model's H100 rows above; vLLM recorded none.} \\
\bottomrule
\end{tabular}
\end{table}

\subsection{Audit verdicts counted once}\label{app:tab-verdicts}

Table~\ref{tab:senses} counts each changed verdict once, a rule set post hoc (Appendix~\ref{app:notes}).
Table~\ref{tab:app-verdicts} names them, with the
contracts that changed each in both runs. Three of the 12 rankings are accuracy rankings, which do not depend on
the attribute and are counted once per task. The 4B model's two ranking changes were both on Diabetes. Its accuracy ranking changed
because FP8 serving cut its selection rate from 0.102 to 0.062, moving its accuracy past the tabular families. It
measures selection volume at a fixed threshold rather than model quality. Both H100 BF16 contracts changed its
demographic-parity ranking by sex.

\begin{table}[p]
\caption{Every audit verdict a Tier 2 contract changed in senses 1 to 4, counted once. Contracts: CPU INT8 and
CPU BF16 on the M5 Max CPU; MLX BF16, 8-bit and MXFP8 weights on its GPU; vLLM on the H100; FP8 and BF16 on the
MI350X where named. Interval: whether the FP64 bootstrap 95\% interval of the statistic contains the 0.05
criterion (senses 1 and 3 only). No Tier 1 contract changed a verdict in these senses.}
\label{tab:app-verdicts}
\scriptsize
\setlength{\tabcolsep}{4pt}
\begin{tabular}{@{}l l l l >{\raggedright\arraybackslash}p{0.42\linewidth} l@{}}
\toprule
Sense & Task & Model & Verdict & Contracts & Interval \\
\midrule
1 & HMDA & LLM & EO gap by sex & vLLM FP8 (H100) & contains \\
1 & Diabetes & LR & DP gap by race & MLX 8-bit & contains \\
1 & Diabetes & MLP & DP gap by age & CPU INT8 & contains \\
1 & Diabetes & MLP & EO gap by race & MLX 8-bit, CPU INT8 & contains \\
1 & HMDA & LR & DP gap by ethnicity & MLX MXFP8 & contains \\
1 & HMDA & LR & EO gap by ethnicity & MLX 8-bit, CPU INT8, FP8 (MI350X) & contains \\
\addlinespace
2 & Diabetes & LLM & TPR gap by sex & vLLM FP8 (H100) & -- \\
2 & Diabetes & LLM & SR gap by sex & FP8 (MI350X) & -- \\
2 & Diabetes & LR & SR gap by race & MLX 8-bit & -- \\
2 & Diabetes & MLP & SR gap by race & CPU INT8 & -- \\
2 & HMDA & MLP & TPR gap by race & every Tier 2 contract: CPU BF16, CPU INT8, MLX BF16, 8-bit, MXFP8, BF16 and FP8 (MI350X) & -- \\
\addlinespace
3 & ACSEmployment & MLP & SR equivalence by sex & CPU INT8 & excludes \\
3 & Diabetes & LR & SR equivalence by age & CPU INT8 & contains \\
\addlinespace
4 & Diabetes & 4 families & accuracy & vLLM FP8 (H100) & -- \\
4 & Diabetes & 4 families & DP gap by sex & BF16 and vLLM BF16 (H100), FP8 (MI350X) & -- \\
4 & ACSEmployment & 3 families & EO gap by sex & CPU INT8 & -- \\
4 & ACSIncome & 3 families & accuracy & FP8 (MI350X) & -- \\
4 & ACSIncome & 3 families & EO gap by race & MLX MXFP8 & -- \\
4 & Diabetes & 3 families & accuracy & MLX MXFP8, CPU INT8, FP8 (MI350X) & -- \\
4 & Diabetes & 3 families & DP gap by age & CPU INT8, FP8 (MI350X) & -- \\
4 & Diabetes & 3 families & EO gap by age & FP8 (MI350X) & -- \\
4 & Diabetes & 3 families & DP gap by race & FP8 (MI350X) & -- \\
4 & Diabetes & 3 families & EO gap by race & MLX 8-bit, CPU INT8 & -- \\
4 & HMDA & 3 families & EO gap by race & MLX MXFP8, MLX 8-bit, CPU INT8, FP8 (MI350X) & -- \\
4 & HMDA & 3 families & EO gap by sex & FP8 (MI350X) & -- \\
\midrule
\multicolumn{6}{@{}l}{24B language model (intervals of its strict FP32 reference)} \\
1 & ACSEmployment & LLM 24B & DP gap by sex & vLLM FP8 (H100) & contains \\
1 & ACSEmployment & LLM 24B & EO gap by sex & vLLM BF16, vLLM FP8 (H100) & contains \\
4 & Diabetes & 4 families & DP gap by sex & BF16 (H100) & -- \\
4 & Diabetes & 4 families & EO gap by sex & BF16, vLLM FP8 (H100) & -- \\
\bottomrule
\end{tabular}
\end{table}

\subsection{Group sizes}\label{app:tab-groups}

Table~\ref{tab:app-groups} gives the denominators behind every group statistic. A group enters the selection-rate
statistics with at least 100 test records and a recorded label, and the error-rate statistics with at least 30
positives and 30 negatives as well. Error rates of the smallest eligible groups rest on few records: the Diabetes
true-positive rates by race compare groups with 43 and 31 readmitted encounters, from 38 and 24 patients, since
eligibility counts encounters. Joint, in HMDA, is an application whose applicant and co-applicant differ in the
attribute (Appendix~\ref{app:hmda-groups}).

\begin{table}[p]
\caption{Test records and positives per group, for the full test sets and the 24B model's seeded subsets (ACSIncome
in full). S: the group enters the selection-rate statistics; E: also the error-rate statistics; --: neither. AIAN:
American Indian or Alaska Native; NHPI: Native Hawaiian or Other Pacific Islander.}
\label{tab:app-groups}
\scriptsize
\setlength{\tabcolsep}{4pt}
\begin{tabular}{@{}l l l r r l r r l@{}}
\toprule
 & & & \multicolumn{3}{c}{Full test set} & \multicolumn{3}{c}{24B subset} \\
\cmidrule(lr){4-6}\cmidrule(l){7-9}
Task & Attribute & Group & Records & Positives & Enters & Records & Positives & Enters \\
\midrule
ACSIncome & race & AIAN/NHPI & 509 & 124 & S, E & 509 & 124 & S, E \\
 &  & Asian & 6{,}673 & 3{,}226 & S, E & 6{,}673 & 3{,}226 & S, E \\
 &  & Black & 1{,}716 & 566 & S, E & 1{,}716 & 566 & S, E \\
 &  & Some other race & 4{,}541 & 891 & S, E & 4{,}541 & 891 & S, E \\
 &  & Two or more & 1{,}569 & 535 & S, E & 1{,}569 & 535 & S, E \\
 &  & White & 24{,}125 & 10{,}688 & S, E & 24{,}125 & 10{,}688 & S, E \\
 & sex & Female & 18{,}532 & 6{,}466 & S, E & 18{,}532 & 6{,}466 & S, E \\
 &  & Male & 20{,}601 & 9{,}564 & S, E & 20{,}601 & 9{,}564 & S, E \\
\addlinespace
ACSEmployment & race & AIAN/NHPI & 972 & 414 & S, E & 248 & 94 & S, E \\
 &  & Asian & 11{,}926 & 5{,}869 & S, E & 3{,}168 & 1{,}573 & S, E \\
 &  & Black & 3{,}771 & 1{,}445 & S, E & 951 & 358 & S, E \\
 &  & Some other race & 8{,}821 & 4{,}028 & S, E & 2{,}361 & 1{,}072 & S, E \\
 &  & Two or more & 3{,}772 & 1{,}336 & S, E & 971 & 362 & S, E \\
 &  & White & 46{,}501 & 21{,}323 & S, E & 12{,}301 & 5{,}630 & S, E \\
 & sex & Female & 38{,}447 & 16{,}234 & S, E & 10{,}073 & 4{,}297 & S, E \\
 &  & Male & 37{,}316 & 18{,}181 & S, E & 9{,}927 & 4{,}792 & S, E \\
\addlinespace
HMDA & race & 2+ minority races & 131 & 123 & S & 46 & 44 & -- \\
 &  & AIAN & 352 & 304 & S, E & 92 & 77 & -- \\
 &  & Asian & 12{,}067 & 10{,}825 & S, E & 3{,}377 & 3{,}019 & S, E \\
 &  & Black & 2{,}510 & 2{,}157 & S, E & 763 & 651 & S, E \\
 &  & Free text only & 7 & 3 & -- & 1 & 1 & -- \\
 &  & Joint & 2{,}625 & 2{,}436 & S, E & 764 & 704 & S, E \\
 &  & NHPI & 299 & 267 & S, E & 81 & 72 & -- \\
 &  & Not available & 10{,}012 & 8{,}843 & -- & 2{,}900 & 2{,}574 & -- \\
 &  & White & 41{,}367 & 37{,}640 & S, E & 11{,}976 & 10{,}927 & S, E \\
 & ethnicity & Not available & 9{,}475 & 8{,}437 & -- & 2{,}757 & 2{,}463 & -- \\
 &  & Free text only & 11 & 11 & -- & 2 & 2 & -- \\
 &  & Hispanic or Latino & 13{,}776 & 12{,}257 & S, E & 4{,}018 & 3{,}583 & S, E \\
 &  & Joint & 2{,}798 & 2{,}582 & S, E & 817 & 746 & S, E \\
 &  & Not Hispanic or Latino & 43{,}310 & 39{,}311 & S, E & 12{,}406 & 11{,}275 & S, E \\
 & sex & Female & 13{,}169 & 11{,}782 & S, E & 3{,}745 & 3{,}360 & S, E \\
 &  & Joint & 29{,}982 & 27{,}462 & S, E & 8{,}732 & 8{,}005 & S, E \\
 &  & Male & 22{,}176 & 19{,}776 & S, E & 6{,}373 & 5{,}682 & S, E \\
 &  & Not available & 4{,}043 & 3{,}578 & -- & 1{,}150 & 1{,}022 & -- \\
\addlinespace
Diabetes & race & African American & 3{,}841 & 461 & S, E & 1{,}915 & 227 & S, E \\
 &  & Asian & 141 & 13 & S & 76 & 7 & -- \\
 &  & Caucasian & 14{,}761 & 1{,}731 & S, E & 7{,}410 & 867 & S, E \\
 &  & Hispanic & 448 & 43 & S, E & 232 & 23 & S \\
 &  & Other & 309 & 31 & S, E & 149 & 12 & S \\
 &  & Unknown & 438 & 35 & -- & 218 & 20 & -- \\
 & sex & Female & 10{,}637 & 1{,}225 & S, E & 5{,}292 & 600 & S, E \\
 &  & Male & 9{,}301 & 1{,}089 & S, E & 4{,}708 & 556 & S, E \\
 & age & 40-59 & 5{,}422 & 536 & S, E & 2{,}685 & 257 & S, E \\
 &  & 60-79 & 9{,}566 & 1{,}139 & S, E & 4{,}829 & 593 & S, E \\
 &  & 80+ & 3{,}643 & 477 & S, E & 1{,}852 & 230 & S, E \\
 &  & <40 & 1{,}307 & 162 & S, E & 634 & 76 & S, E \\

\bottomrule
\end{tabular}
\end{table}

\subsection{Reference audit values and sampling intervals}\label{app:tab-ref}

Table~\ref{tab:app-ref} gives each reference gap and impact ratio (FP64, or strict FP32 for the 24B model) with its
bootstrap 95\% interval, and the six verdicts the reference supports. A dagger marks an interval that contains its criterion, where sampling alone
leaves the verdict unsettled, and a bold letter marks a verdict that some contract changed. In a post hoc split
under Tier 2, 6 of the 17 pass-fail verdicts whose interval contained 0.05 changed, and none of the 43 whose interval
excluded it (Table~\ref{tab:twoway}). Within arms, the tabular LR and MLP split 5 of 14 against 0 of 26 and the
4B model 1 of 3 against 0 of 17, so the tabular arm carries the contrast; the verdicts share records, so no test is
attached.

Equivalence verdicts changed on both sides of that line. The one with a settled interval was the ACSEmployment MLP
by sex under INT8 dynamic quantization. Its FP64 gap of 0.0390 had an interval of [0.0316, 0.0460], inside the 0.05
margin, and its equivalence $p$-value rose from 0.0012 to 0.089. Significance verdicts have no sampling interval in
this study, so neither table can say whether they were settled.

\begin{table}[p]
\caption{Reference audit values (FP64; strict FP32 for the 24B model) with bootstrap 95\% intervals (1{,}000
resamples) and the reference $p$-value of the selection-rate gap. Verdicts, in the order of senses 1 (DP, EO), 2 (SR, TPR), 3 and 5: P pass, F fail, S
significant, E equivalent, N not. $^\dagger$: the interval contains the criterion (0.05 for the gaps, 0.8 for the impact ratio). Diabetes
intervals resample patients rather than encounters (Appendix~\ref{app:notes}).
Bold: a verdict that at least one contract changed in both runs. A $p$-value of 0 is below double precision.}
\label{tab:app-ref}
\scriptsize
\setlength{\tabcolsep}{3pt}
\begin{tabular}{@{}l l l l l r l@{}}
\toprule
Task & Attribute & DP gap [95\% CI] & EO gap [95\% CI] & Impact ratio [95\% CI] & $p$ (SR) & Verdicts \\
\midrule
\multicolumn{7}{@{}l}{Logistic regression} \\
ACSIncome & race & 0.3713 [0.3548, 0.3858] & 0.3575 [0.3219, 0.3933] & 0.2640 [0.2431, 0.2854] & 0 & F F S S N F \\
 & sex & 0.1375 [0.1282, 0.1472] & 0.0877 [0.0753, 0.1016] & 0.7102 [0.6924, 0.7272] & $6.5\times10^{-168}$ & F F S S N F \\
ACSEmployment & race & 0.1668 [0.1536, 0.1865] & 0.1386 [0.1204, 0.1579] & 0.7162 [0.6856, 0.7366] & $5.8\times10^{-72}$ & F F S S N F \\
 & sex & 0.0881 [0.0812, 0.0954] & 0.1150 [0.1076, 0.1228] & 0.8458 [0.8341, 0.8572] & $3.0\times10^{-130}$ & F F S S N P \\
HMDA & race & 0.0935 [0.0771, 0.1354] & 0.1430 [0.0839, 0.3446] & 0.9004 [0.8592, 0.9171] & 0.0035 & F F S S N P \\
 & ethnicity & 0.0473 [0.0364, 0.0584]$^\dagger$ & 0.0527 [0.0282, 0.1182]$^\dagger$ & 0.9486 [0.9371, 0.9602] & $1.6\times10^{-12}$ & \textbf{P} \textbf{F} S S N P \\
 & sex & 0.0331 [0.0280, 0.0391] & 0.0742 [0.0425, 0.1059]$^\dagger$ & 0.9640 [0.9576, 0.9696] & $2.4\times10^{-28}$ & P F S S E P \\
Diabetes & race & 0.0463 [0.0229, 0.1012]$^\dagger$ & 0.0757 [0.0437, 0.2824]$^\dagger$ & 0.5795 [0.1927, 0.7962] & 0.080 & \textbf{P} F \textbf{N} N N F \\
 & sex & 0.0060 [0.0002, 0.0212] & 0.0071 [0.0033, 0.0578]$^\dagger$ & 0.9449 [0.8191, 0.9980] & 0.17 & P P N N E P \\
 & age & 0.0414 [0.0239, 0.0641]$^\dagger$ & 0.0841 [0.0441, 0.2018]$^\dagger$ & 0.6659 [0.5209, 0.8009]$^\dagger$ & $9.8\times10^{-11}$ & P F S S \textbf{N} F \\
\midrule
\multicolumn{7}{@{}l}{MLP} \\
ACSIncome & race & 0.3203 [0.3043, 0.3362] & 0.2853 [0.2539, 0.3229] & 0.3302 [0.3067, 0.3541] & $1.7\times10^{-267}$ & F F S S N F \\
 & sex & 0.1355 [0.1256, 0.1449] & 0.0815 [0.0682, 0.0947] & 0.7106 [0.6927, 0.7292] & $7.3\times10^{-164}$ & F F S S N F \\
ACSEmployment & race & 0.1481 [0.1326, 0.1664] & 0.0911 [0.0748, 0.1110] & 0.7307 [0.6993, 0.7571] & $1.1\times10^{-56}$ & F F S S N F \\
 & sex & 0.0390 [0.0316, 0.0460] & 0.0616 [0.0555, 0.0697] & 0.9223 [0.9087, 0.9366] & $7.4\times10^{-27}$ & P F S S \textbf{E} P \\
HMDA & race & 0.0607 [0.0477, 0.1167]$^\dagger$ & 0.1468 [0.0629, 0.3118] & 0.9343 [0.8788, 0.9478] & 0.047 & F F S \textbf{N} N P \\
 & ethnicity & 0.0323 [0.0217, 0.0432] & 0.0127 [0.0098, 0.0796]$^\dagger$ & 0.9651 [0.9536, 0.9762] & $3.2\times10^{-7}$ & P P S S E P \\
 & sex & 0.0170 [0.0133, 0.0238] & 0.0378 [0.0139, 0.0715]$^\dagger$ & 0.9813 [0.9740, 0.9854] & $2.1\times10^{-8}$ & P P S N E P \\
Diabetes & race & 0.0361 [0.0231, 0.1048]$^\dagger$ & 0.0714 [0.0406, 0.2880]$^\dagger$ & 0.7062 [0.3217, 0.7925] & 0.012 & P \textbf{F} \textbf{S} N N F \\
 & sex & 0.0145 [0.0034, 0.0253] & 0.0148 [0.0072, 0.0475] & 0.8578 [0.7668, 0.9638]$^\dagger$ & $4.8\times10^{-4}$ & P P S N E P \\
 & age & 0.0476 [0.0224, 0.0786]$^\dagger$ & 0.1659 [0.0740, 0.2502] & 0.6273 [0.4882, 0.7828] & $3.5\times10^{-7}$ & \textbf{P} F S S N F \\
\midrule
\multicolumn{7}{@{}l}{Gradient-boosted trees} \\
ACSIncome & race & 0.3727 [0.3574, 0.3885] & 0.3610 [0.3258, 0.3985] & 0.2598 [0.2387, 0.2797] & 0 & F F S S N F \\
 & sex & 0.1341 [0.1242, 0.1433] & 0.0753 [0.0627, 0.0896] & 0.7186 [0.7018, 0.7364] & $2.2\times10^{-159}$ & F F S S N F \\
ACSEmployment & race & 0.1559 [0.1395, 0.1747] & 0.0969 [0.0800, 0.1145] & 0.7247 [0.6930, 0.7514] & $9.0\times10^{-63}$ & F F S S N F \\
 & sex & 0.0466 [0.0394, 0.0533]$^\dagger$ & 0.0663 [0.0594, 0.0739] & 0.9130 [0.9010, 0.9261] & $1.1\times10^{-37}$ & P F S S N P \\
HMDA & race & 0.0808 [0.0615, 0.1225] & 0.1108 [0.0633, 0.3017] & 0.9140 [0.8740, 0.9336] & 0.0089 & F F S N N P \\
 & ethnicity & 0.0401 [0.0291, 0.0510]$^\dagger$ & 0.0623 [0.0266, 0.1295]$^\dagger$ & 0.9565 [0.9452, 0.9682] & $6.9\times10^{-10}$ & P F S S E P \\
 & sex & 0.0277 [0.0243, 0.0342] & 0.0750 [0.0446, 0.1069]$^\dagger$ & 0.9699 [0.9629, 0.9735] & $1.2\times10^{-20}$ & P F S S E P \\
Diabetes & race & 0.0369 [0.0199, 0.0984]$^\dagger$ & 0.0453 [0.0356, 0.2611]$^\dagger$ & 0.6580 [0.2297, 0.8151]$^\dagger$ & 0.16 & P P N N N F \\
 & sex & 0.0020 [0.0003, 0.0175] & 0.0111 [0.0027, 0.0652]$^\dagger$ & 0.9810 [0.8468, 0.9974] & 0.64 & P P N N E P \\
 & age & 0.0372 [0.0126, 0.0788]$^\dagger$ & 0.1775 [0.0822, 0.2984] & 0.7204 [0.5230, 0.8869]$^\dagger$ & $7.1\times10^{-5}$ & P F S S N F \\
\midrule
\multicolumn{7}{@{}l}{4B language model} \\
ACSIncome & race & 0.4505 [0.4341, 0.4670] & 0.3656 [0.3405, 0.4233] & 0.3201 [0.3003, 0.3394] & 0 & F F S S N F \\
 & sex & 0.0590 [0.0494, 0.0688]$^\dagger$ & 0.0032 [0.0011, 0.0160] & 0.8934 [0.8765, 0.9097] & $2.0\times10^{-31}$ & F P S N N P \\
ACSEmployment & race & 0.0013 [0.0010, 0.0019] & 0.0020 [0.0015, 0.0031] & 0.0000 [0.0000, 0.0000] & 0.27 & P P N N E F \\
 & sex & 0.0013 [0.0008, 0.0017] & 0.0023 [0.0015, 0.0030] & 0.1974 [0.0861, 0.3499] & $1.2\times10^{-8}$ & P P S N E F \\
HMDA & race & 0.1085 [0.0701, 0.1581] & 0.1569 [0.0904, 0.3338] & 0.8854 [0.8364, 0.9249] & 0.0018 & F F S S N P \\
 & ethnicity & 0.0742 [0.0638, 0.0849] & 0.1262 [0.0825, 0.1843] & 0.9200 [0.9093, 0.9309] & $6.8\times10^{-26}$ & F F S S N P \\
 & sex & 0.0432 [0.0391, 0.0496] & 0.0672 [0.0423, 0.0966]$^\dagger$ & 0.9534 [0.9466, 0.9577] & $9.3\times10^{-49}$ & P \textbf{F} S S E P \\
Diabetes & race & 0.1064 [0.0688, 0.1428] & 0.1403 [0.0695, 0.3318] & 0.2857 [0.0638, 0.5219] & $4.3\times10^{-4}$ & F F S S N F \\
 & sex & 0.0141 [0.0040, 0.0249] & 0.0426 [0.0111, 0.0769]$^\dagger$ & 0.8703 [0.7801, 0.9618]$^\dagger$ & 0.0010 & P P \textbf{S} \textbf{S} E P \\
 & age & 0.1143 [0.0976, 0.1309] & 0.1161 [0.1031, 0.1543] & 0.1672 [0.0957, 0.2558] & $2.0\times10^{-30}$ & F F S S N F \\
\midrule
\multicolumn{7}{@{}l}{24B language model, strict FP32 reference} \\
ACSIncome & race & 0.4456 [0.4288, 0.4621] & 0.3845 [0.3528, 0.4202] & 0.2896 [0.2702, 0.3094] & 0 & F F S S N F \\
 & sex & 0.0798 [0.0703, 0.0891] & 0.0154 [0.0070, 0.0265] & 0.8503 [0.8344, 0.8679] & $5.3\times10^{-56}$ & F P S N N P \\
ACSEmployment & race & 0.3296 [0.3062, 0.3512] & 0.4065 [0.3707, 0.4385] & 0.4241 [0.3940, 0.4557] & $2.7\times10^{-132}$ & F F S S N F \\
 & sex & 0.0567 [0.0427, 0.0698]$^\dagger$ & 0.0515 [0.0327, 0.0700]$^\dagger$ & 0.8808 [0.8558, 0.9084] & $7.1\times10^{-16}$ & \textbf{F} \textbf{F} S S N P \\
HMDA & race & 0.1087 [0.0814, 0.1350] & 0.1221 [0.0940, 0.2247] & 0.8857 [0.8584, 0.9139] & $7.4\times10^{-27}$ & F F S S N P \\
 & ethnicity & 0.0942 [0.0830, 0.1079] & 0.1390 [0.0913, 0.2298] & 0.8995 [0.8851, 0.9114] & $2.5\times10^{-76}$ & F F S S N P \\
 & sex & 0.0316 [0.0207, 0.0431] & 0.0695 [0.0292, 0.1255]$^\dagger$ & 0.9659 [0.9537, 0.9776] & $4.1\times10^{-9}$ & P F S S E P \\
Diabetes & race & 0.0749 [0.0589, 0.1041] & 0.0754 [0.0600, 0.1533] & 0.5151 [0.3866, 0.5821] & $2.1\times10^{-23}$ & F F S S N F \\
 & sex & 0.0049 [0.0002, 0.0178] & 0.0106 [0.0039, 0.0559]$^\dagger$ & 0.9497 [0.8287, 0.9979] & 0.40 & P P N N E P \\
 & age & 0.0371 [0.0236, 0.0649]$^\dagger$ & 0.0804 [0.0371, 0.1834]$^\dagger$ & 0.6928 [0.4927, 0.7808] & 0.0081 & P F S S N F \\
\bottomrule
\end{tabular}
\end{table}

\subsection{Audit stability intervals}\label{app:tab-intervals}

The language model's 60 intervals with a criterion (its 70 less the 10 of accuracy) crossed none over Tier 1
contracts, with or without the MI350X (Table~\ref{tab:app-llm-intervals}). Over all contracts, 2 crossed on the
NVIDIA GPUs and 3 with the MI350X's contracts admitted. For the 4B model's Diabetes demographic-parity gap by age,
the numerical spread over all contracts was 1.31 times the sampling interval. None of the 180 tabular Tier 1 intervals with a criterion
crossed, and the widest spanned 0.049 of its sampling interval. Over the tabular Tier 2 contracts, 10 of 120 crossed,
1 of them over the BF16 contracts alone (Table~\ref{tab:app-tab2-intervals}). The MI350X's BF16 and FP8 contracts,
admitted beside FP64 and CPU FP32, made 2 of the 120 LR and MLP intervals with a criterion cross; both also crossed
under the M5 Max contracts. No MI350X Tier 2 contract scored the trees. None of the 24B model's 60 intervals with a
criterion crossed it over TF32, and 2 did over all its contracts, the ACSEmployment gaps by sex
(Table~\ref{tab:app-llm24-intervals}).

\begin{table}[p]
\caption{The 4B language model's audit stability intervals: the range of each statistic over the FP64 reference
and the admitted contracts and runs. Tier 1 admits strict FP32, TF32 and the MI350X's TF32 setting; ``all'' also
admits BF16, serving and FP8. $^\dagger$: the interval crosses its criterion (0.05 for gaps and $p$-values, 0.8
for the impact ratio), so the verdict is not numerically robust. TOST: the equivalence test of the selection-rate
gap.}
\label{tab:app-llm-intervals}
\scriptsize
\setlength{\tabcolsep}{4pt}
\begin{tabular}{@{}l l l l l l l@{}}
\toprule
Task & Attribute & Statistic & FP64 & Tier 1, all GPUs & All, NVIDIA & All, with MI350X \\
\midrule
ACSIncome & race & DP gap & 0.4505 & [0.4504, 0.4508] & [0.4504, 0.4546] & [0.4495, 0.4546] \\
 &  & EO gap & 0.3656 & [0.3656, 0.3656] & [0.3656, 0.3856] & [0.3630, 0.3856] \\
 &  & Impact ratio & 0.3201 & [0.3199, 0.3204] & [0.3021, 0.3204] & [0.3021, 0.3212] \\
 &  & $p$ (SR) & 0 & [0, 0] & [0, 0] & [0, 0] \\
 &  & $p$ (TPR) & $1.9\times10^{-143}$ & [$1.9\times10^{-143}$, $1.9\times10^{-143}$] & [$2.4\times10^{-152}$, $1.9\times10^{-143}$] & [$2.4\times10^{-152}$, $5.5\times10^{-143}$] \\
 &  & $p$ (TOST) & 1 & [1, 1] & [1, 1] & [1, 1] \\
 & sex & DP gap & 0.0590 & [0.0590, 0.0592] & [0.0573, 0.0602] & [0.0570, 0.0602] \\
 &  & EO gap & 0.0032 & [0.0029, 0.0032] & [0.0029, 0.0056] & [0.0029, 0.0056] \\
 &  & Impact ratio & 0.8934 & [0.8929, 0.8934] & [0.8886, 0.8961] & [0.8886, 0.8962] \\
 &  & $p$ (SR) & $2.0\times10^{-31}$ & [$1.0\times10^{-31}$, $2.0\times10^{-31}$] & [$1.3\times10^{-32}$, $1.0\times10^{-29}$] & [$1.3\times10^{-32}$, $1.7\times10^{-29}$] \\
 &  & $p$ (TPR) & 0.59 & [0.59, 0.63] & [0.35, 0.63] & [0.35, 0.63] \\
 &  & $p$ (TOST) & 0.96 & [0.96, 0.97] & [0.92, 0.98] & [0.92, 0.98] \\
\addlinespace
ACSEmployment & race & DP gap & 0.0013 & [0.0012, 0.0013] & [0.0008, 0.0013] & [0.0008, 0.0013] \\
 &  & EO gap & 0.0020 & [0.0020, 0.0020] & [0.0015, 0.0021] & [0.0014, 0.0021] \\
 &  & Impact ratio & 0.0000 & [0.0000, 0.0000] & [0.0000, 0.0000] & [0.0000, 0.0000] \\
 &  & $p$ (SR) & 0.27 & [0.27, 0.27] & [0.25, 0.37] & [0.25, 0.37] \\
 &  & $p$ (TPR) & 0.60 & [0.60, 0.60] & [0.60, 0.89] & [0.45, 0.89] \\
 &  & $p$ (TOST) & 0 & [0, 0] & [0, 0] & [0, 0] \\
 & sex & DP gap & 0.0013 & [0.0012, 0.0013] & [0.0011, 0.0013] & [0.0006, 0.0013] \\
 &  & EO gap & 0.0023 & [0.0022, 0.0023] & [0.0020, 0.0024] & [0.0010, 0.0024] \\
 &  & Impact ratio & 0.1974 & [0.1974, 0.2080] & [0.0237, 0.2080] & [0.0237, 0.3328] \\
 &  & $p$ (SR) & $1.2\times10^{-8}$ & [$1.2\times10^{-8}$, $4.7\times10^{-8}$] & [$3.6\times10^{-10}$, $4.7\times10^{-8}$] & [$3.6\times10^{-10}$, $5.4\times10^{-4}$] \\
 &  & $p$ (TPR) & 0.24 & [0.24, 0.47] & [0.24, 0.47] & [0.24, 0.47] \\
 &  & $p$ (TOST) & 0 & [0, 0] & [0, 0] & [0, 0] \\
\addlinespace
HMDA & race & DP gap & 0.1085 & [0.1085, 0.1085] & [0.1057, 0.1369] & [0.1057, 0.1369] \\
 &  & EO gap & 0.1569 & [0.1569, 0.1584] & [0.1537, 0.1721] & [0.1537, 0.1792] \\
 &  & Impact ratio & 0.8854 & [0.8854, 0.8854] & [0.8554, 0.8884] & [0.8554, 0.8884] \\
 &  & $p$ (SR) & 0.0018 & [0.0018, 0.0018] & [$2.0\times10^{-4}$, 0.0022] & [$2.0\times10^{-4}$, 0.0022] \\
 &  & $p$ (TPR) & 0.0013 & [0.0013, 0.0013] & [$2.8\times10^{-4}$, 0.0013] & [$2.2\times10^{-4}$, 0.0018] \\
 &  & $p$ (TOST) & 0.98 & [0.98, 0.98] & [0.98, 0.999] & [0.98, 0.999] \\
 & ethnicity & DP gap & 0.0742 & [0.0742, 0.0747] & [0.0731, 0.0797] & [0.0731, 0.0797] \\
 &  & EO gap & 0.1262 & [0.1255, 0.1262] & [0.1115, 0.1262] & [0.1115, 0.1262] \\
 &  & Impact ratio & 0.9200 & [0.9195, 0.9200] & [0.9128, 0.9212] & [0.9128, 0.9212] \\
 &  & $p$ (SR) & $6.8\times10^{-26}$ & [$3.7\times10^{-26}$, $6.8\times10^{-26}$] & [$9.3\times10^{-27}$, $3.5\times10^{-25}$] & [$9.3\times10^{-27}$, $3.5\times10^{-25}$] \\
 &  & $p$ (TPR) & $5.3\times10^{-20}$ & [$2.6\times10^{-20}$, $5.3\times10^{-20}$] & [$1.0\times10^{-21}$, $2.0\times10^{-19}$] & [$2.6\times10^{-110}$, $2.0\times10^{-19}$] \\
 &  & $p$ (TOST) & 1 & [1, 1] & [1, 1] & [1, 1] \\
 & sex & DP gap & 0.0432 & [0.0431, 0.0433] & [0.0345, 0.0433] & [0.0345, 0.0438] \\
 &  & EO gap & 0.0672 & [0.0670, 0.0680] & [0.0489, 0.0680]$^\dagger$ & [0.0489, 0.0680]$^\dagger$ \\
 &  & Impact ratio & 0.9534 & [0.9533, 0.9535] & [0.9533, 0.9620] & [0.9528, 0.9620] \\
 &  & $p$ (SR) & $9.3\times10^{-49}$ & [$4.4\times10^{-49}$, $1.1\times10^{-48}$] & [$4.4\times10^{-49}$, $3.2\times10^{-36}$] & [$2.3\times10^{-50}$, $3.2\times10^{-36}$] \\
 &  & $p$ (TPR) & $2.8\times10^{-54}$ & [$1.2\times10^{-54}$, $2.8\times10^{-54}$] & [$1.2\times10^{-54}$, $2.3\times10^{-30}$] & [$6.4\times10^{-55}$, $2.3\times10^{-30}$] \\
 &  & $p$ (TOST) & 0.016 & [0.015, 0.018] & [$1.4\times10^{-8}$, 0.018] & [$1.4\times10^{-8}$, 0.025] \\
\addlinespace
Diabetes & race & DP gap & 0.1064 & [0.1056, 0.1064] & [0.0645, 0.1066] & [0.0645, 0.1066] \\
 &  & EO gap & 0.1403 & [0.1176, 0.1414] & [0.1176, 0.1458] & [0.1176, 0.1458] \\
 &  & Impact ratio & 0.2857 & [0.2857, 0.2873] & [0.2852, 0.3547] & [0.2852, 0.4006] \\
 &  & $p$ (SR) & $4.3\times10^{-4}$ & [$4.3\times10^{-4}$, $4.6\times10^{-4}$] & [$4.2\times10^{-4}$, 0.011] & [$4.2\times10^{-4}$, 0.011] \\
 &  & $p$ (TPR) & 0.0050 & [0.0045, 0.018] & [$1.6\times10^{-4}$, 0.018] & [$1.6\times10^{-4}$, 0.018] \\
 &  & $p$ (TOST) & 0.999 & [0.999, 0.999] & [0.81, 0.999] & [0.81, 0.999] \\
 & sex & DP gap & 0.0141 & [0.0141, 0.0145] & [0.0097, 0.0164] & [0.0047, 0.0164] \\
 &  & EO gap & 0.0426 & [0.0426, 0.0435] & [0.0211, 0.0435] & [0.0211, 0.0440] \\
 &  & Impact ratio & 0.8703 & [0.8671, 0.8703] & [0.8496, 0.8703] & [0.8496, 0.9457] \\
 &  & $p$ (SR) & 0.0010 & [$7.6\times10^{-4}$, 0.0010] & [$1.3\times10^{-4}$, 0.0048] & [$1.3\times10^{-4}$, 0.24]$^\dagger$ \\
 &  & $p$ (TPR) & 0.0025 & [0.0020, 0.0025] & [0.0017, 0.059]$^\dagger$ & [0.0015, 0.059]$^\dagger$ \\
 &  & $p$ (TOST) & $5.9\times10^{-51}$ & [$1.4\times10^{-51}$, $5.9\times10^{-51}$] & [$7.1\times10^{-69}$, $1.9\times10^{-15}$] & [$7.1\times10^{-69}$, $1.9\times10^{-15}$] \\
 & age & DP gap & 0.1143 & [0.1135, 0.1143] & [0.0707, 0.1143] & [0.0707, 0.1143] \\
 &  & EO gap & 0.1161 & [0.1158, 0.1164] & [0.0931, 0.1164] & [0.0877, 0.1164] \\
 &  & Impact ratio & 0.1672 & [0.1672, 0.1682] & [0.0704, 0.1742] & [0.0704, 0.2163] \\
 &  & $p$ (SR) & $2.0\times10^{-30}$ & [$2.0\times10^{-30}$, $3.7\times10^{-30}$] & [$1.9\times10^{-30}$, $4.3\times10^{-21}$] & [$1.9\times10^{-30}$, $8.3\times10^{-18}$] \\
 &  & $p$ (TPR) & $1.7\times10^{-4}$ & [$1.7\times10^{-4}$, $1.9\times10^{-4}$] & [$5.1\times10^{-5}$, $3.2\times10^{-4}$] & [$5.1\times10^{-5}$, $7.5\times10^{-4}$] \\
 &  & $p$ (TOST) & 1 & [1, 1] & [1, 1] & [1, 1] \\
\midrule
\multicolumn{7}{@{}l}{24B language model: Table~\ref{tab:app-llm24-intervals}.} \\
\bottomrule
\end{tabular}
\end{table}

\begin{table}[p]
\caption{The 24B language model's audit stability intervals: the range of each statistic over its strict FP32
reference and the admitted contracts and runs, all on H100s. Tier 1 admits TF32; ``all'' also admits BF16 and
serving with BF16 and FP8. $^\dagger$: the interval crosses its criterion (0.05 for gaps and $p$-values, 0.8 for the
impact ratio). TOST: the equivalence test of the selection-rate gap.}
\label{tab:app-llm24-intervals}
\scriptsize
\setlength{\tabcolsep}{4pt}
\begin{tabular}{@{}l l l l l l@{}}
\toprule
Task & Attribute & Statistic & Strict FP32 & Tier 1 & All \\
\midrule
ACSIncome & race & DP gap & 0.4456 & [0.4456, 0.4456] & [0.4456, 0.4528] \\
 &  & EO gap & 0.3845 & [0.3845, 0.3845] & [0.3824, 0.3957] \\
 &  & Impact ratio & 0.2896 & [0.2896, 0.2896] & [0.2826, 0.2900] \\
 &  & $p$ (SR) & 0 & [0, 0] & [0, 0] \\
 &  & $p$ (TPR) & $1.2\times10^{-158}$ & [$1.2\times10^{-158}$, $1.2\times10^{-158}$] & [$1.1\times10^{-167}$, $4.0\times10^{-158}$] \\
 &  & $p$ (TOST) & 1 & [1, 1] & [1, 1] \\
 & sex & DP gap & 0.0798 & [0.0797, 0.0798] & [0.0742, 0.0799] \\
 &  & EO gap & 0.0154 & [0.0154, 0.0156] & [0.0103, 0.0159] \\
 &  & Impact ratio & 0.8503 & [0.8503, 0.8504] & [0.8503, 0.8603] \\
 &  & $p$ (SR) & $5.3\times10^{-56}$ & [$5.3\times10^{-56}$, $7.1\times10^{-56}$] & [$3.9\times10^{-56}$, $1.3\times10^{-48}$] \\
 &  & $p$ (TPR) & 0.21 & [0.21, 0.24] & [0.20, 0.79] \\
 &  & $p$ (TOST) & 1 & [1, 1] & [1, 1] \\
\addlinespace
ACSEmployment & race & DP gap & 0.3296 & [0.3285, 0.3296] & [0.2750, 0.3335] \\
 &  & EO gap & 0.4065 & [0.4065, 0.4065] & [0.2649, 0.4065] \\
 &  & Impact ratio & 0.4241 & [0.4241, 0.4253] & [0.4241, 0.5821] \\
 &  & $p$ (SR) & $2.7\times10^{-132}$ & [$2.7\times10^{-132}$, $1.8\times10^{-131}$] & [$2.5\times10^{-134}$, $1.3\times10^{-52}$] \\
 &  & $p$ (TPR) & $2.9\times10^{-103}$ & [$2.9\times10^{-103}$, $2.9\times10^{-103}$] & [$4.1\times10^{-106}$, $2.4\times10^{-55}$] \\
 &  & $p$ (TOST) & 1 & [1, 1] & [1, 1] \\
 & sex & DP gap & 0.0567 & [0.0567, 0.0568] & [0.0417, 0.0568]$^\dagger$ \\
 &  & EO gap & 0.0515 & [0.0515, 0.0516] & [0.0394, 0.0520]$^\dagger$ \\
 &  & Impact ratio & 0.8808 & [0.8806, 0.8808] & [0.8806, 0.9261] \\
 &  & $p$ (SR) & $7.1\times10^{-16}$ & [$6.4\times10^{-16}$, $7.1\times10^{-16}$] & [$6.4\times10^{-16}$, $3.1\times10^{-9}$] \\
 &  & $p$ (TPR) & $1.9\times10^{-7}$ & [$1.7\times10^{-7}$, $1.9\times10^{-7}$] & [$7.8\times10^{-8}$, $2.0\times10^{-6}$] \\
 &  & $p$ (TOST) & 0.83 & [0.83, 0.83] & [0.12, 0.83] \\
\addlinespace
HMDA & race & DP gap & 0.1087 & [0.1087, 0.1087] & [0.1083, 0.1160] \\
 &  & EO gap & 0.1221 & [0.1221, 0.1221] & [0.1221, 0.1399] \\
 &  & Impact ratio & 0.8857 & [0.8857, 0.8857] & [0.8780, 0.8863] \\
 &  & $p$ (SR) & $7.4\times10^{-27}$ & [$7.4\times10^{-27}$, $7.4\times10^{-27}$] & [$1.3\times10^{-29}$, $7.4\times10^{-27}$] \\
 &  & $p$ (TPR) & $1.5\times10^{-24}$ & [$1.5\times10^{-24}$, $1.5\times10^{-24}$] & [$2.1\times10^{-26}$, $1.9\times10^{-24}$] \\
 &  & $p$ (TOST) & 1 & [1, 1] & [1, 1] \\
 & ethnicity & DP gap & 0.0942 & [0.0941, 0.0942] & [0.0938, 0.0991] \\
 &  & EO gap & 0.1390 & [0.1390, 0.1390] & [0.1390, 0.1528] \\
 &  & Impact ratio & 0.8995 & [0.8995, 0.8996] & [0.8940, 0.8999] \\
 &  & $p$ (SR) & $2.5\times10^{-76}$ & [$2.5\times10^{-76}$, $3.7\times10^{-76}$] & [$4.7\times10^{-82}$, $3.7\times10^{-76}$] \\
 &  & $p$ (TPR) & $1.1\times10^{-71}$ & [$1.1\times10^{-71}$, $1.7\times10^{-71}$] & [$4.2\times10^{-76}$, $3.6\times10^{-71}$] \\
 &  & $p$ (TOST) & 1 & [1, 1] & [1, 1] \\
 & sex & DP gap & 0.0316 & [0.0316, 0.0319] & [0.0308, 0.0323] \\
 &  & EO gap & 0.0695 & [0.0695, 0.0695] & [0.0695, 0.0736] \\
 &  & Impact ratio & 0.9659 & [0.9656, 0.9659] & [0.9651, 0.9667] \\
 &  & $p$ (SR) & $4.1\times10^{-9}$ & [$3.1\times10^{-9}$, $4.1\times10^{-9}$] & [$1.7\times10^{-9}$, $1.7\times10^{-8}$] \\
 &  & $p$ (TPR) & $3.0\times10^{-6}$ & [$2.3\times10^{-6}$, $3.0\times10^{-6}$] & [$1.6\times10^{-6}$, $1.5\times10^{-5}$] \\
 &  & $p$ (TOST) & $6.8\times10^{-4}$ & [$6.8\times10^{-4}$, $8.1\times10^{-4}$] & [$4.5\times10^{-4}$, 0.0010] \\
\addlinespace
Diabetes & race & DP gap & 0.0749 & [0.0747, 0.0749] & [0.0730, 0.0756] \\
 &  & EO gap & 0.0754 & [0.0751, 0.0754] & [0.0751, 0.0823] \\
 &  & Impact ratio & 0.5151 & [0.5151, 0.5151] & [0.5151, 0.5986] \\
 &  & $p$ (SR) & $2.1\times10^{-23}$ & [$2.1\times10^{-23}$, $2.5\times10^{-23}$] & [$1.9\times10^{-23}$, $5.5\times10^{-18}$] \\
 &  & $p$ (TPR) & 0.0022 & [0.0022, 0.0022] & [0.0012, 0.0022] \\
 &  & $p$ (TOST) & 0.998 & [0.997, 0.998] & [0.99, 0.998] \\
 & sex & DP gap & 0.0049 & [0.0049, 0.0051] & [0.0003, 0.0055] \\
 &  & EO gap & 0.0106 & [0.0106, 0.0106] & [0.0067, 0.0171] \\
 &  & Impact ratio & 0.9497 & [0.9479, 0.9497] & [0.9479, 0.9966] \\
 &  & $p$ (SR) & 0.40 & [0.39, 0.40] & [0.39, 0.95] \\
 &  & $p$ (TPR) & 0.57 & [0.57, 0.57] & [0.37, 0.72] \\
 &  & $p$ (TOST) & $8.5\times10^{-15}$ & [$8.5\times10^{-15}$, $9.8\times10^{-15}$] & [$1.9\times10^{-17}$, $8.6\times10^{-12}$] \\
 & age & DP gap & 0.0371 & [0.0371, 0.0371] & [0.0351, 0.0416] \\
 &  & EO gap & 0.0804 & [0.0804, 0.0804] & [0.0804, 0.1351] \\
 &  & Impact ratio & 0.6928 & [0.6928, 0.6928] & [0.6633, 0.7304] \\
 &  & $p$ (SR) & 0.0081 & [0.0081, 0.0081] & [$3.2\times10^{-7}$, 0.012] \\
 &  & $p$ (TPR) & 0.0083 & [0.0083, 0.0083] & [0.0043, 0.0083] \\
 &  & $p$ (TOST) & 0.15 & [0.15, 0.15] & [0.12, 0.25] \\
\bottomrule
\end{tabular}
\end{table}

\begin{table}[ht]
\caption{Tabular Tier 2 statistics whose stability interval crossed its criterion over the M5 Max's Tier 2
contracts and controls. Beside it: the FP64 reference with its bootstrap 95\% interval (by patient for Diabetes), the interval over the BF16
contracts alone, and the interval over the MI350X's BF16 and FP8 contracts. $^\dagger$: crosses the criterion
(0.05).}
\label{tab:app-tab2-intervals}
\scriptsize
\setlength{\tabcolsep}{4pt}
\begin{tabular}{@{}l l l l l l l l@{}}
\toprule
Task & Model & Attribute & Statistic & FP64 [95\% CI] & All Tier 2, M5 Max & BF16 only & MI350X \\
\midrule
ACSEmployment & MLP & sex & $p$ (TOST) & 0.0012 & [$6.2\times10^{-4}$, 0.089]$^\dagger$ & [$6.2\times10^{-4}$, 0.0012] & [0.0010, 0.0013] \\
HMDA & LR & ethnicity & DP gap & 0.0473 [0.0364, 0.0584] & [0.0446, 0.0516]$^\dagger$ & [0.0473, 0.0479] & [0.0460, 0.0479] \\
HMDA & LR & ethnicity & EO gap & 0.0527 [0.0282, 0.1182] & [0.0415, 0.0573]$^\dagger$ & [0.0527, 0.0547] & [0.0418, 0.0547]$^\dagger$ \\
HMDA & MLP & race & $p$ (TPR) & 0.21 & [$7.7\times10^{-4}$, 0.21]$^\dagger$ & [0.0034, 0.21]$^\dagger$ & [$5.9\times10^{-4}$, 0.21]$^\dagger$ \\
Diabetes & LR & age & $p$ (TOST) & 0.098 & [0.020, 0.13]$^\dagger$ & [0.086, 0.11] & [0.098, 0.30] \\
Diabetes & LR & race & DP gap & 0.0463 [0.0229, 0.1012] & [0.0407, 0.0540]$^\dagger$ & [0.0461, 0.0473] & [0.0441, 0.0473] \\
Diabetes & LR & race & $p$ (SR) & 0.080 & [0.042, 0.14]$^\dagger$ & [0.075, 0.081] & [0.075, 0.11] \\
Diabetes & MLP & age & DP gap & 0.0476 [0.0224, 0.0786] & [0.0462, 0.0513]$^\dagger$ & [0.0476, 0.0482] & [0.0449, 0.0482] \\
Diabetes & MLP & race & EO gap & 0.0714 [0.0406, 0.2880] & [0.0326, 0.0714]$^\dagger$ & [0.0693, 0.0714] & [0.0693, 0.0773] \\
Diabetes & MLP & race & $p$ (SR) & 0.012 & [0.012, 0.18]$^\dagger$ & [0.012, 0.012] & [0.010, 0.012] \\
\bottomrule
\end{tabular}
\end{table}

\subsection{Threshold curves}\label{app:tab-threshold}

\begin{figure}[p]
\centering
\includegraphics[width=\linewidth]{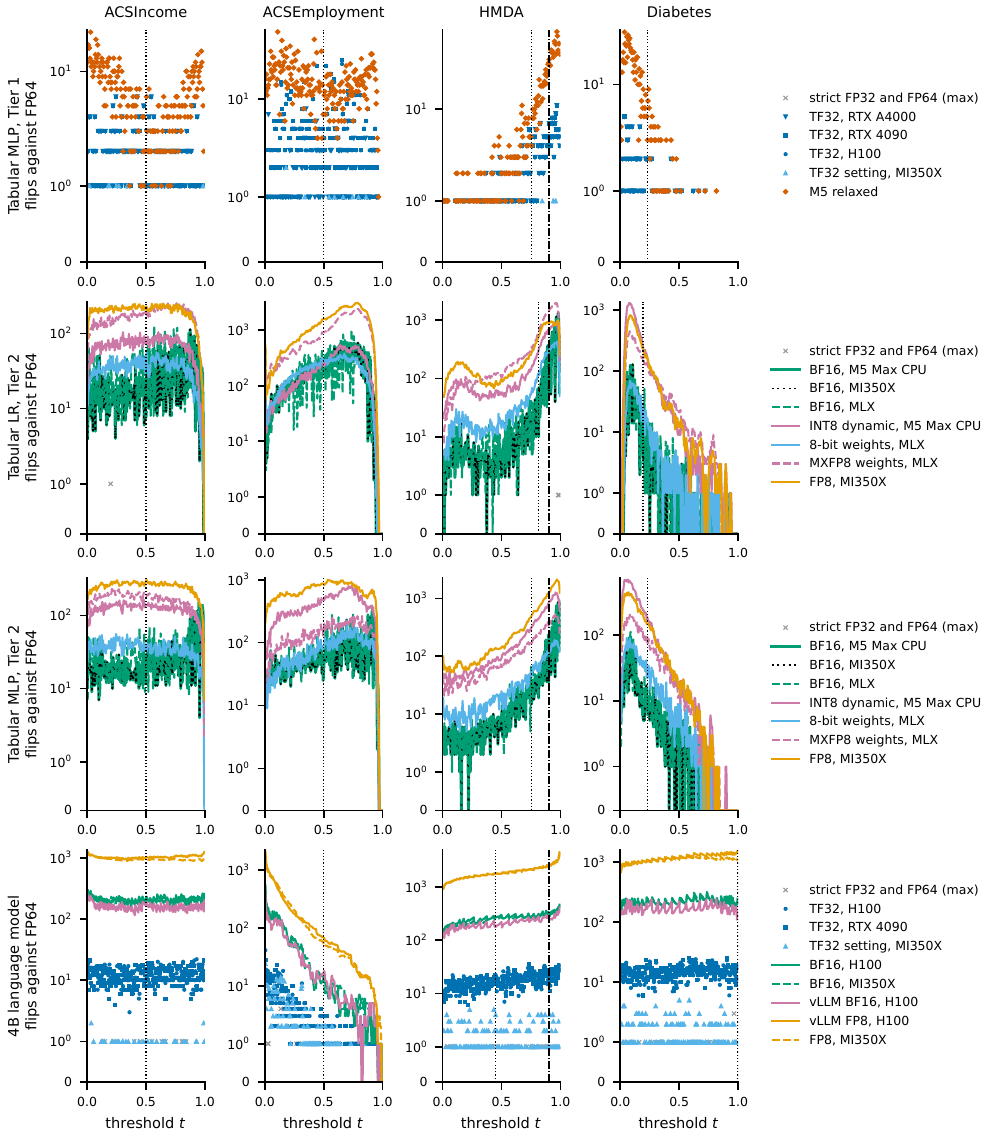}
\caption{Decisions changed against FP64 (run 1) at each threshold of the fixed probability grid, by task (columns)
and arm (rows); log scale above 1. Markers: Tier 1 contracts at the thresholds where they changed at least one
decision. Grey: the largest count over strict FP32 and FP64 contracts on every device. Dotted line: the
prespecified threshold; dash-dotted: the HMDA secondary point, the approval share read as a probability. The
language model's Diabetes threshold lies above the grid. The 24B model's curves are in
the artifact and its maxima in Table~\ref{tab:app-threshold}.}
\Description{A grid of line charts, one per task and arm, giving the number of decisions each contract changed
against FP64 at every threshold of a fixed probability grid, on a logarithmic scale, with markers for the Tier 1
contracts, a grey line for the strict controls, and vertical lines at the prespecified and secondary thresholds.}
\label{fig:app-threshold}
\end{figure}

Figure~\ref{fig:app-threshold} draws flips against FP64 over the fixed probability grid, the primary grid, and
Table~\ref{tab:app-threshold} gives the largest count on that grid and on the secondary grid of FP64 score
quantiles. The prespecified thresholds are the result, and these maxima were not searched for a headline. On the
probability grid, TF32 reached 25 tabular flips at one threshold and the M5 relaxed default 70. On the quantile
grid, output saturation near 1 made every float32 contract, strict included, disagree with FP64 on 1{,}516 HMDA MLP
records at the top quantile (Appendix~\ref{sec:mechanism-artifacts}).

\begin{table}[p]
\caption{Largest number of decisions changed against FP64 (run 1) at any threshold of the fixed probability grid
(0.005 to 0.995, primary) / of the FP64 score-quantile grid (secondary). Strict FP32 and FP64: the largest count
over those contracts on every device. The quantile grid places thresholds within an ulp of tied or saturated
scores, which every float32 contract then resolves differently from FP64 (Appendix~\ref{sec:mechanism-artifacts}).}
\label{tab:app-threshold}
\scriptsize
\begin{tabular}{@{}l r r r r@{}}
\toprule
Model and contract & ACSIncome & ACSEmployment & HMDA & Diabetes \\
\midrule
\multicolumn{5}{@{}l}{Tabular, Tier 1} \\
MLP, strict FP32 and FP64 & 0 / 2 & 1 / 35 & 1 / 1{,}516 & 0 / 1 \\
MLP, TF32, RTX A4000 & 6 / 7 & 8 / 35 & 11 / 1{,}516 & 5 / 6 \\
MLP, TF32, RTX 4090 & 6 / 7 & 25 / 40 & 11 / 1{,}516 & 5 / 6 \\
MLP, TF32, H100 & 6 / 7 & 23 / 40 & 11 / 1{,}516 & 5 / 6 \\
MLP, TF32 setting, MI350X & 1 / 1 & 2 / 33 & 1 / 1{,}516 & 0 / 2 \\
MLP, M5 relaxed & 22 / 24 & 49 / 57 & 70 / 1{,}297 & 31 / 39 \\
LR, every Tier 1 contract & 1 / 4 & 3 / 34 & 1 / 3 & 0 / 1 \\
Trees, CPU and CUDA predictors & 1 / 11 & 4 / 13 & 0 / 5 & 0 / 1 \\
\midrule
\multicolumn{5}{@{}l}{Tabular, Tier 2} \\
LR, strict FP32 and FP64 controls & 1 / 1 & 0 / 24 & 1 / 1 & 0 / 1 \\
LR, CPU BF16 & 113 / 113 & 591 / 714 & 1{,}344 / 2{,}957 & 128 / 162 \\
LR, CPU INT8 dynamic & 104 / 101 & 547 / 552 & 880 / 896 & 1{,}302 / 1{,}316 \\
LR, MLX BF16 & 120 / 124 & 1{,}016 / 774 & 1{,}280 / 2{,}964 & 131 / 195 \\
LR, MLX 8-bit weights & 56 / 53 & 381 / 385 & 433 / 439 & 69 / 82 \\
LR, MLX MXFP8 weights & 252 / 249 & 2{,}430 / 2{,}399 & 1{,}990 / 2{,}017 & 448 / 446 \\
LR, BF16, MI350X & 113 / 113 & 591 / 714 & 1{,}344 / 2{,}957 & 128 / 162 \\
LR, FP8, MI350X & 250 / 242 & 3{,}109 / 3{,}112 & 1{,}003 / 1{,}026 & 817 / 839 \\
MLP, strict FP32 and FP64 controls & 0 / 2 & 0 / 35 & 0 / 1{,}516 & 0 / 1 \\
MLP, CPU BF16 & 129 / 158 & 322 / 287 & 894 / 19{,}618 & 64 / 78 \\
MLP, CPU INT8 dynamic & 166 / 169 & 792 / 787 & 1{,}274 / 1{,}469 & 686 / 698 \\
MLP, MLX BF16 & 158 / 161 & 347 / 300 & 868 / 19{,}595 & 110 / 131 \\
MLP, MLX 8-bit weights & 56 / 57 & 202 / 255 & 192 / 1{,}290 & 90 / 91 \\
MLP, MLX MXFP8 weights & 231 / 227 & 266 / 281 & 540 / 1{,}285 & 202 / 202 \\
MLP, BF16, MI350X & 129 / 158 & 322 / 287 & 894 / 19{,}618 & 64 / 78 \\
MLP, FP8, MI350X & 307 / 302 & 1{,}004 / 998 & 2{,}131 / 2{,}162 & 442 / 457 \\
\midrule
\multicolumn{5}{@{}l}{4B language model} \\
strict FP32 and FP64 & 1 / 4 & 4 / 26 & 1 / 4 & 3 / 2 \\
TF32, H100 & 22 / 30 & 41 / 185 & 33 / 80 & 24 / 26 \\
TF32, RTX 4090 & 21 / 31 & 28 / 191 & 35 / 84 & 25 / 25 \\
TF32 setting, MI350X & 2 / 3 & 24 / 113 & 5 / 15 & 5 / 5 \\
BF16, H100 & 286 / 380 & 545 / 2{,}652 & 451 / 892 & 298 / 301 \\
BF16, MI350X & 304 / 396 & 444 / 2{,}363 & 485 / 898 & 316 / 319 \\
vLLM BF16, H100 & 248 / 390 & 394 / 2{,}046 & 410 / 692 & 242 / 252 \\
vLLM FP8, H100 & 1{,}244 / 1{,}534 & 2{,}087 / 10{,}346 & 4{,}282 / 6{,}054 & 1{,}490 / 1{,}558 \\
FP8, MI350X & 1{,}268 / 1{,}508 & 1{,}885 / 8{,}744 & 4{,}659 / 7{,}333 & 1{,}248 / 1{,}227 \\
\midrule
\multicolumn{5}{@{}l}{24B language model, against strict FP32} \\
TF32, H100 & 12 / 22 & 22 / 40 & 10 / 25 & 15 / 21 \\
BF16, H100 & 256 / 712 & 401 / 763 & 183 / 576 & 227 / 248 \\
vLLM BF16, H100 & 200 / 640 & 320 / 595 & 114 / 418 & 216 / 232 \\
vLLM FP8, H100 & 1{,}462 / 2{,}678 & 2{,}280 / 2{,}478 & 1{,}086 / 2{,}733 & 1{,}059 / 1{,}135 \\
\bottomrule
\end{tabular}
\end{table}

\subsection{Software versions}\label{app:versions}

Table~\ref{tab:app-versions} gives the software each platform ran, as recorded in the run metadata. The M5 GPU's
contracts ran on a 0.32.3 development build of MLX: official commit ce916dbbc plus one quantized-matmul kernel fix.
Rerun on the released MLX 0.32.3, which includes a version of this fix, the relaxed default, BF16 and the
8-bit and MXFP8 weight contracts returned byte-identical logits and scores for all eight models in both runs.

\begin{table}[ht]
\caption{Software by platform, from the run metadata of run 1 (run 2 recorded the same). No run recorded the
version of its BLAS library (cuBLAS or hipBLASLt), so none is given.}
\label{tab:app-versions}
\footnotesize
\begin{tabular}{@{}>{\raggedright\arraybackslash}p{0.2\linewidth} >{\raggedright\arraybackslash}p{0.2\linewidth}
>{\raggedright\arraybackslash}p{0.52\linewidth}@{}}
\toprule
Platform & Arm & Software \\
\midrule
Apple M4 Pro CPU & tabular reference & macOS 27.0.1; Python 3.12.13; PyTorch 2.14.1; XGBoost 3.2.0 \\
RTX 4090, RTX A4000 & tabular Tier 1 & Linux 7.0.0; Python 3.14.4; PyTorch 2.14.1+cu130 (CUDA 13.0 build);
cuDNN 9.24.0; NVIDIA driver 595.71.05; XGBoost 3.2.0 (CUDA 12.9) \\
H100 & tabular Tier 1 & Linux 5.14.0; Python 3.12.4; PyTorch 2.11.0 (CUDA 12.9 build); cuDNN 9.13.1; NVIDIA driver
580.173.02; XGBoost 3.2.0 (CUDA 12.9) \\
Apple M5 Max GPU & tabular Tiers 1 and 2 & macOS 27.0.1; Python 3.14.7; MLX 0.32.3.dev20260906, a development
build: official commit ce916dbbc plus one quantized-matmul kernel fix \\
Apple M5 Max CPU & tabular Tier 2 & macOS 27.0.1; Python 3.12.13; PyTorch 2.14.1 \\
AMD MI350X & tabular and language model & Linux 7.0.0; Python 3.12.15; PyTorch 2.14.1+rocm7.2; HIP 7.2.53211;
amdgpu driver 6.19.14; transformers 5.18.0 \\
H100 & language model & Linux 5.14.0; Python 3.12.4; PyTorch 2.11.0 (CUDA 12.9 build); cuDNN 9.13.1; NVIDIA
driver 580.173.02; transformers 5.18.0; vLLM 0.25.0 \\
RTX 4090 & language model & Linux 7.0.0; Python 3.14.4; PyTorch 2.14.1+cu130 (CUDA 13.0 build); cuDNN 9.24.0;
NVIDIA driver 595.71.05; transformers 5.18.0 \\
\bottomrule
\end{tabular}
\end{table}

\subsection{HMDA groups}\label{app:hmda-groups}

HMDA's derived sex, race and ethnicity each include a Joint category for applications whose applicant and
co-applicant differ in that attribute (29{,}982, 2{,}625 and 2{,}798 test applications). The study audits it as the derived fields
define it, but it describes an application, not a person, and under the declared groups every HMDA pass-fail change
in this study involved it. The 4B model's equalized-odds gap by sex is the false-positive-rate gap between joint and female applications,
under FP64 and FP8 and with either label set. Between male and female applicants alone, the gap is 0.028 under FP64
and 0.020 under FP8 serving with APPROVE/DENY, and 0.026 and 0.020 with YES/NO; all four pass. The logistic
regression's ethnicity pass-fail changes are set by joint against Hispanic or Latino applications, and the MLP's
race significance change by a contract making joint applications the highest true-positive rate.
Table~\ref{tab:app-hmda} recounts every HMDA verdict with each Joint category excluded. Tier 1 still changed none.
Tier 2 changed fewer, but not none: the MLP's race significance change survived under three contracts, and with
YES/NO labels, excluding Joint created an ethnicity pass-fail change under both BF16 contracts and a sex
significance change under FP8.

\begin{table}[ht]
\caption{Exploratory, post hoc: distinct HMDA verdicts changed against the reference (senses 1, 2, 3 and 5, and
the language models' rankings), with the derived attributes as declared and with each Joint category excluded.
With YES/NO labels each contract ran once (Appendix~\ref{app:notes}); in the other rows every non-reference
contract ran twice.}
\label{tab:app-hmda}
\footnotesize
\begin{tabular}{@{}l r r r r r@{}}
\toprule
 & \multicolumn{2}{c}{Declared} & \multicolumn{2}{c}{Joint excluded} & \\
\cmidrule(lr){2-3}\cmidrule(lr){4-5}
Arm & Tier 1 & Tier 2 & Tier 1 & Tier 2 & At stake \\
\midrule
Tabular LR and MLP & 0 & 3 & 0 & 1 & 36 \\
4B, APPROVE/DENY & 0 & 1 & 0 & 0 & 24 \\
4B, YES/NO & 0 & 3 & 0 & 2 & 24 \\
24B & 0 & 0 & 0 & 0 & 24 \\
\bottomrule
\end{tabular}
\end{table}

\subsection{Exploratory post hoc analyses}\label{app:posthoc}

Three analyses were designed after the first draft and written down before they ran. All three are
exploratory and post hoc. They read the per-record scores of the existing runs, use the pilot's audit code
unchanged, and change no count reported elsewhere in this paper.

\paragraph{Matched-perturbation null.} For each non-control contract, the null flips as many test decisions as
the contract did, chosen at random within strata of side of the threshold and FP64 margin bin, using the pilot's
own bins. Each draw therefore keeps the contract's flip count, direction and margin profile but picks random
people. Every verdict of senses 1 to 5 is recomputed and counted as in Table~\ref{tab:senses}, over 5{,}000 draws
per contract. Table~\ref{tab:app-null-contract} sets each contract against its null, and
Table~\ref{tab:app-null-arm} each arm and tier, with a finer and a coarser matching beside the primary one.

Tier 1 contracts changed no verdict of the prespecified tests, and their nulls at most one. The language model's Tier 1 null changed one in a
share 0.25 of draws, the near-tied Diabetes ranking by sex: on the RTX 4090, TF32 brought the 4B model's
demographic-parity gap by sex to within $2.0\times10^{-8}$ of the MLP's, against $3.4\times10^{-4}$ under FP64. The tabular Tier 2 contracts changed more
verdicts than their nulls, 20 against a mean of 14.3 (95\% range 11 to 18). A null count of 20 or more occurred in a
share 0.0024 of draws, and MLX 8-bit weights alone changed 7 verdicts against a mean of 2.74, an unadjusted share of
0.0138 that Holm's adjustment over the 12 Tier 2 contracts raises to 0.166. The language model's
Tier 2 contracts changed fewer, 5 against a mean of 7.34 (5 to 10). Both results hold under the finer and the coarser
matching. The tabular Tier 2 flips therefore fell where they move verdicts more often than this null's random flips
of the same size and margin profile. Each contract's null draws are independent of the other contracts', while the
observed flip sets overlap: the MI350X's tabular BF16 flips equal the M5 Max CPU's. Overlap shrinks the observed
union against independent draws, so this null favors a deficit and works against an excess. Under a null that keeps that
overlap (Appendix~\ref{app:robustness}), the tabular excess holds ($P = 0.0036$), but the language model's deficit
does not: 5 changes against a null mean of 7.23, with 5 or fewer in a share 0.105 of draws. Nor does the null show that errors target groups:
it holds group membership out of the matching, so it asks only whether these flips sit where verdicts are easy to
move.

The split by sampling interval points both ways. Among verdicts whose FP64 interval contained the criterion, the
tabular contracts changed 6 against a null mean of 4.49. The language model's contracts changed 1, its HMDA
equalized-odds verdict by sex, which no null draw changed; the null changed some unsettled language-model verdict
in a share 0.017 of draws. Among settled verdicts, INT8's equivalence change for the
ACSEmployment MLP by sex occurred in no null draw, and the null changed any settled tabular verdict in a share
0.0008 of draws. The null in turn changed the language model's HMDA equivalence
verdict by sex, settled by sampling, in nearly every draw, while no contract changed it.

\begin{table}[p]
\caption{Exploratory, post hoc: distinct verdicts of senses 1 to 5 each non-control contract changed, against a
matched-perturbation null (5{,}000 draws; primary matching by side of the threshold and FP64 margin bin). Flips:
run 1, summed over the contract's tasks and models. $P(\ge)$ and $P(\le)$: share of null draws with at least or
at most the observed count, unadjusted across contracts; 0 means no draw.}
\label{tab:app-null-contract}
\footnotesize
\begin{tabular}{@{}l r r r l r r@{}}
\toprule
Contract & Flips & Observed & Null mean & Null 95\% range & $P(\ge)$ & $P(\le)$ \\
\midrule
\multicolumn{7}{@{}l}{Language model, Tier 1} \\
TF32, H100 & 41 & 0 & 0.13 & [0, 1] & 1 & 0.8692 \\
TF32, RTX 4090 & 39 & 0 & 0.13 & [0, 1] & 1 & 0.866 \\
TF32 setting, MI350X & 4 & 0 & 0.00 & [0, 0] & 1 & 1 \\
\multicolumn{7}{@{}l}{Language model, Tier 2} \\
BF16, H100 & 586 & 1 & 0.94 & [0, 2] & 0.7224 & 0.7864 \\
BF16, MI350X & 554 & 0 & 0.84 & [0, 2] & 1 & 0.3308 \\
vLLM BF16, H100 & 448 & 1 & 0.69 & [0, 2] & 0.5912 & 0.9066 \\
vLLM FP8, H100 & 3{,}654 & 3 & 4.96 & [3, 8] & 0.9878 & 0.1144 \\
FP8, MI350X & 3{,}403 & 2 & 2.88 & [1, 5] & 0.886 & 0.4026 \\
\multicolumn{7}{@{}l}{Tabular, Tier 1} \\
TF32, RTX A4000 & 5 & 0 & 0.00 & [0, 0] & 1 & 1 \\
TF32, RTX 4090 & 5 & 0 & 0.00 & [0, 0] & 1 & 1 \\
TF32, H100 & 5 & 0 & 0.00 & [0, 0] & 1 & 1 \\
TF32 setting, MI350X & 0 & 0 & 0.00 & [0, 0] & 1 & 1 \\
M5 relaxed default & 44 & 0 & 0.05 & [0, 1] & 1 & 0.9474 \\
\multicolumn{7}{@{}l}{Tabular, Tier 2} \\
CPU BF16 & 450 & 1 & 0.51 & [0, 2] & 0.4626 & 0.959 \\
CPU INT8 dynamic & 2{,}002 & 12 & 8.69 & [5, 12] & 0.0636 & 0.9776 \\
MLX BF16 & 425 & 1 & 0.77 & [0, 3] & 0.564 & 0.8432 \\
MLX 8-bit weights & 653 & 7 & 2.74 & [0, 6] & 0.0138 & 0.9958 \\
MLX MXFP8 weights & 2{,}475 & 5 & 5.17 & [2, 9] & 0.6298 & 0.601 \\
BF16, MI350X & 450 & 1 & 0.51 & [0, 2] & 0.4642 & 0.9632 \\
FP8, MI350X & 4{,}409 & 9 & 6.90 & [4, 10] & 0.176 & 0.929 \\
\bottomrule
\end{tabular}
\end{table}

\begin{table}[p]
\caption{Exploratory, post hoc: distinct verdicts changed by any non-control contract of an arm and tier, against
the union over its contracts' null draws (5{,}000). Unsettled and settled: verdicts of senses 1, 3 and 5 whose FP64
bootstrap 95\% interval contains or excludes the criterion (by patient for Diabetes). Null columns: mean [95\% range] under the primary
matching, under strata of 25 records in margin order (finer), and under one band per side up to the largest
flipped margin (coarser). $P(\ge)$ and $P(\le)$: primary matching; 0 means no draw.}
\label{tab:app-null-arm}
\scriptsize
\setlength{\tabcolsep}{4pt}
\begin{tabular}{@{}l l r r l l l r r@{}}
\toprule
Arm & Verdicts & At stake & Observed & Primary & Finer & Coarser & $P(\ge)$ & $P(\le)$ \\
\midrule
LLM, Tier 1 & senses 1 to 5 & 84 & 0 & 0.25 [0, 1] & 0.25 [0, 1] & 0.05 [0, 1] & 1 & 0.7542 \\
 & unsettled & 5 & 0 & 0.00 [0, 0] & 0.00 [0, 0] & 0.00 [0, 0] & 1 & 1 \\
 & settled & 35 & 0 & 0.00 [0, 0] & 0.00 [0, 0] & 0.00 [0, 0] & 1 & 1 \\
LLM, Tier 2 & senses 1 to 5 & 84 & 5 & 7.34 [5, 10] & 7.29 [4, 10] & 7.28 [4, 11] & 0.986 & 0.0798 \\
 & unsettled & 5 & 1 & 0.02 [0, 0] & 0.01 [0, 0] & 0.02 [0, 0] & 0.017 & 1 \\
 & settled & 35 & 0 & 1.27 [1, 2] & 1.23 [1, 2] & 0.93 [0, 2] & 1 & 0.0002 \\
Tabular, Tier 1 & senses 1 to 5 & 144 & 0 & 0.05 [0, 1] & 0.18 [0, 1] & 0.00 [0, 0] & 1 & 0.9474 \\
 & unsettled & 22 & 0 & 0.00 [0, 0] & 0.00 [0, 0] & 0.00 [0, 0] & 1 & 1 \\
 & settled & 58 & 0 & 0.00 [0, 0] & 0.00 [0, 0] & 0.00 [0, 0] & 1 & 1 \\
Tabular, Tier 2 & senses 1 to 5 & 144 & 20 & 14.3 [11, 18] & 14.2 [11, 18] & 14.6 [11, 18] & 0.0024 & 0.9992 \\
 & unsettled & 22 & 6 & 4.49 [3, 6] & 4.50 [3, 6] & 4.29 [3, 6] & 0.0962 & 0.9946 \\
 & settled & 58 & 1 & 0.00 [0, 0] & 0.00 [0, 0] & 0.00 [0, 0] & 0.0008 & 1 \\
\bottomrule
\end{tabular}
\end{table}

\paragraph{Fixed pair for sense 2.} The significance test of sense 2 compares the groups with the highest and lowest
rate under each contract, so one record can change the pair being tested. With the FP64 reference's pair held
fixed, 3 of the 5 significance changes of Table~\ref{tab:senses} remain. Two are the language model's, whose pair
did not change. The third is the Diabetes LR selection-rate gap by race under MLX 8-bit weights, with $p$ 0.080 under
FP64 and 0.042 under the contract. The HMDA MLP true-positive gap by race, changed by all seven tabular Tier 2 contracts, kept $p$
between 0.20 and 0.54 with the pair fixed, against 0.21 under FP64. The Diabetes MLP selection-rate gap by race
kept $p = 0.049$ under INT8, against 0.012 under FP64, so it stayed significant.

\paragraph{Alternative tests for sense 2.} Because the prespecified test picks its pair on the decisions it tests
and does not adjust for that choice, its $p$-values are anti-conservative. Table~\ref{tab:app-sense2} recounts
the significance changes under alternatives applied alike to the reference and to every contract in both runs: the
reference's pair held fixed, unadjusted and Bonferroni-adjusted for the $K(K-1)/2$ pairs it was chosen from; every
pair under Holm's adjustment, significant if any pair is; and a test of equal rates across the $K$ eligible
groups. Each is computed with the prespecified pooled statistics (a pooled $z$ and a Pearson chi-square) and with
unpooled Wald statistics. The alternatives moved the changes rather than removing them: over Tier 2 they changed 3
to 7 verdicts each, not always the same ones, against 5 under the prespecified test. Five of the nine also changed
one Tier 1 verdict of the 4B model. Under the pooled Bonferroni test and the unpooled tests of the worst or the
fixed pair, it is the Diabetes true-positive gap by race between Hispanic and Caucasian encounters. TF32 on the
H100 flagged 10 rather than 11 of the 43 readmitted Hispanic encounters, which come from 38 patients, and 199 rather
than 200 of the 1{,}731 Caucasian ones: two of its 7 Diabetes flips. That moved the pooled $p$ from 0.0050 to
0.018 and its Bonferroni adjustment from 0.030 to 0.109; exact tests remove this change
(Appendix~\ref{app:robustness}). Under the unpooled Holm and omnibus tests it is the
ACSEmployment true-positive gap by race, at the operating point where the 4B model selects 0.094\% of records and
the Wald variance of a near-empty cell is near zero. A $p$-value that crosses 0.05 between two contracts marks a
changed verdict, not a significant difference between the contracts, which score the same records.

\begin{table}[ht]
\caption{Exploratory, post hoc: distinct significance verdicts (sense 2) changed against the reference, by test
procedure. Each procedure is applied alike to the reference and to every non-control contract, runs 1 and 2.}
\label{tab:app-sense2}
\footnotesize
\begin{tabular}{@{}l r r r r r r@{}}
\toprule
 & \multicolumn{2}{c}{4B LLM} & \multicolumn{2}{c}{Tabular} & \multicolumn{2}{c}{24B LLM} \\
\cmidrule(lr){2-3}\cmidrule(lr){4-5}\cmidrule(l){6-7}
Procedure & Tier 1 & Tier 2 & Tier 1 & Tier 2 & Tier 1 & Tier 2 \\
\midrule
\multicolumn{7}{@{}l}{Pooled statistics} \\
Each contract's own worst pair (prespecified) & 0 & 2 & 0 & 3 & 0 & 0 \\
Reference's pair, unadjusted & 0 & 2 & 0 & 1 & 0 & 0 \\
Reference's pair, Bonferroni & 1 & 4 & 0 & 1 & 0 & 2 \\
All pairs, Holm & 0 & 3 & 0 & 0 & 0 & 1 \\
Chi-square, equal rates & 0 & 4 & 0 & 0 & 0 & 0 \\
\multicolumn{7}{@{}l}{Unpooled Wald statistics} \\
Each contract's own worst pair & 1 & 4 & 0 & 3 & 0 & 0 \\
Reference's pair, unadjusted & 1 & 4 & 0 & 2 & 0 & 0 \\
Reference's pair, Bonferroni & 0 & 2 & 0 & 0 & 0 & 1 \\
All pairs, Holm & 1 & 4 & 0 & 1 & 0 & 0 \\
Wald chi-square, equal rates & 1 & 3 & 0 & 2 & 0 & 0 \\
\midrule
At stake & 20 & 20 & 40 & 40 & 20 & 20 \\
\bottomrule
\end{tabular}
\end{table}

\paragraph{Patients as clusters.} The prespecified tests also treat the encounters of one Diabetes patient as
independent, although 3{,}266 of the 13{,}998 test patients contribute more than one. Table~\ref{tab:app-cluster}
recomputes every Diabetes significance and equivalence verdict with a cluster-robust Wald covariance, in which each
group rate's influence is summed within patient, beside the same Wald statistics without clustering and the
prespecified ones. Against the same Wald statistics without clustering, clustering changed each count by at
most one; the choice between pooled and Wald statistics mattered more. With two groups the five procedures reduce
to one test, so each of the 4B model's two FP8 changes by sex is one test per statistic, and both held. Under vLLM
FP8, the true-positive gap between 1{,}225 female and 1{,}089 male readmitted encounters went from $p$ 0.0025 to
0.059 pooled and from 0.014 to 0.12 clustered; under the MI350X's FP8, the selection-rate gap went from 0.0010 to
0.24 and from 0.0080 to 0.32. The Tier 1 change of the race true-positive gap moved from the Bonferroni test to the
unadjusted ones, where its cluster-robust $p$ rose from 0.046 to 0.082 under TF32; no test that both clustered by
patient and adjusted for multiplicity changed it. The one Diabetes equivalence change, the
logistic regression's by age under INT8, did not survive.

\begin{table}[ht]
\caption{Exploratory, post hoc: Diabetes significance (sense 2) and equivalence (sense 3) verdicts changed, summed
over the 4B, tabular and 24B arms, under the prespecified statistics, unpooled Wald statistics with encounters
independent, and the same Wald statistics with patients as clusters.}
\label{tab:app-cluster}
\footnotesize
\begin{tabular}{@{}l r r r r r r@{}}
\toprule
 & \multicolumn{2}{c}{Prespecified} & \multicolumn{2}{c}{Wald} & \multicolumn{2}{c}{Wald, clustered} \\
\cmidrule(lr){2-3}\cmidrule(lr){4-5}\cmidrule(l){6-7}
Test & Tier 1 & Tier 2 & Tier 1 & Tier 2 & Tier 1 & Tier 2 \\
\midrule
Own worst pair & 0 & 4 & 1 & 5 & 1 & 5 \\
Reference's pair, unadjusted & 0 & 3 & 1 & 5 & 1 & 5 \\
Reference's pair, Bonferroni & 1 & 6 & 0 & 3 & 0 & 4 \\
All pairs, Holm & 0 & 3 & 0 & 3 & 0 & 4 \\
Equal rates (omnibus) & 0 & 2 & 0 & 3 & 0 & 4 \\
Equivalence (TOST) & 0 & 1 & 0 & 1 & 0 & 0 \\
\midrule
At stake, sense 2 & 24 & 24 & 24 & 24 & 24 & 24 \\
\bottomrule
\end{tabular}
\end{table}

\begin{table}[ht]
\caption{Exploratory, post hoc: Tier 2 pass-fail (sense 1) and equivalence (sense 3) verdicts at stake, pooled over the language model
and the tabular LR and MLP, split by whether the FP64 bootstrap 95\% interval of the statistic (resampled by patient for Diabetes) contains 0.05 and
by whether a non-control contract changed the verdict. For sense 3 the statistic is the selection-rate gap.}
\label{tab:twoway}
\small
\begin{tabular}{@{}l r r r r@{}}
\toprule
 & \multicolumn{2}{c}{Interval contains 0.05} & \multicolumn{2}{c}{Interval excludes 0.05} \\
\cmidrule(lr){2-3}\cmidrule(l){4-5}
Sense & Changed & Unchanged & Changed & Unchanged \\
\midrule
1, pass-fail (DP and EO gaps) & 6 & 11 & 0 & 43 \\
3, equivalence of selection rates & 1 & 6 & 1 & 22 \\
\bottomrule
\end{tabular}
\end{table}

\paragraph{Distance to the criterion.} The sampling split of Table~\ref{tab:twoway} is descriptive. Ranked by
distance to 0.05 alone, the four nearest of its 60 pass-fail verdicts all changed, and the 17 nearest hold all six
changes, the same 6 of 17 as the interval rule. Scaled by the half-width of the verdict's interval, the distance
needs only the 13 nearest. The 24B model's two changes are its two verdicts nearest 0.05. Both cutoffs were chosen
after the fact to capture the observed changes, so the comparison is descriptive and does not rank the two rules as
predictors. For sense 3, the split uses the
statistic's 95\% interval, wider than the 90\% interval that a TOST at 0.05 corresponds to, so that row is
descriptive too.

\paragraph{Paired bootstrap of the 24B employment changes.} The decisions on the scored records are fixed, so the
24B model's changed verdicts by sex on ACSEmployment are deterministic on those records. To ask how they would look
in other samples, each of 1{,}000 draws resamples the 20{,}000 records and scores every contract on the same draw.
Strict FP32 failed the demographic-parity tolerance in a share 0.85 of draws and the equalized-odds tolerance in
0.564. FP8 serving lowered the demographic-parity gap in every draw, by 0.015 (95\% interval 0.0064 to 0.024), and
its verdict differed from strict FP32's in a share 0.735 of draws; for equalized odds the share was 0.461. vLLM BF16
differed in 0.242 and 0.203, and TF32 in 0.011 and 0.027. The direction of FP8's demographic-parity change is stable
across samples; whether it crosses 0.05 is not, and its equalized-odds change includes zero
(Appendix~\ref{app:robustness}).

\paragraph{What the 24B employment changes rest on.} Table~\ref{tab:app-llm24-utility} gives each contract's decisions
for the 24B model on ACSEmployment. FP8 serving selected 54.4\% of records against strict FP32's 44.7\% and a base rate
of
45.4\%, raising women's selection rate by more than men's, and its AUC, accuracy and balanced accuracy rose rather
than fell. Selecting 8{,}948 records, the strict FP32 count, by each contract's own margin separates the operating
point from the ranking: FP8's demographic-parity gap is then 0.0541, failing, and its equalized-odds gap 0.0481,
passing, while vLLM BF16's gaps, 0.0581 and 0.0513, both fail. FP8's demographic-parity pass rests on the operating
point; its equalized-odds pass does not.

\begin{table}[ht]
\caption{Exploratory, post hoc: the 24B model on ACSEmployment by contract, run 1, 20{,}000 records. Sel.: share
selected; rates for women and men are selection / TPR / FPR; DP and EO: the audit gaps at the frozen threshold; last
column: the gaps when each contract selects 8{,}948 records, the strict FP32 count, by its own margin.}
\label{tab:app-llm24-utility}
\scriptsize
\setlength{\tabcolsep}{3pt}
\begin{tabular}{@{}l r r r r l l r r l@{}}
\toprule
Contract & Sel. & AUC & Acc. & Bal. acc. & Women & Men & DP & EO & DP / EO, matched \\
\midrule
Strict FP32 & 0.447 & 0.794 & 0.707 & 0.704 & 0.419 / 0.643 / 0.253 & 0.476 / 0.694 / 0.272 & 0.0567 & 0.0515 & 0.0567 / 0.0515 \\
TF32 & 0.448 & 0.794 & 0.708 & 0.705 & 0.419 / 0.643 / 0.253 & 0.476 / 0.695 / 0.272 & 0.0568 & 0.0516 & 0.0571 / 0.0521 \\
BF16 & 0.463 & 0.796 & 0.713 & 0.711 & 0.436 / 0.666 / 0.264 & 0.491 / 0.718 / 0.279 & 0.0554 & 0.0520 & 0.0619 / 0.0590 \\
vLLM BF16 & 0.461 & 0.794 & 0.710 & 0.708 & 0.435 / 0.663 / 0.265 & 0.487 / 0.710 / 0.279 & 0.0516 & 0.0464 & 0.0581 / 0.0513 \\
vLLM FP8 & 0.544 & 0.808 & 0.735 & 0.741 & 0.523 / 0.786 / 0.328 & 0.565 / 0.825 / 0.322 & 0.0417 & 0.0394 & 0.0541 / 0.0481 \\
\bottomrule
\end{tabular}
\end{table}

\paragraph{How firm the 24B employment changes are.} FP8 raised 84\% of the margins. Their mean signed change
was 0.41 logits (standard deviation 0.44; 0.46 within one logit of the threshold), so the change varied by record,
but the rank correlation with strict FP32 was 0.969, FP8 would reach strict FP32's 8{,}948 selections at a threshold
0.41 logits higher, and 2{,}044 of its 2{,}156 flips selected a record that strict FP32 had not. In paired bootstrap
draws of the matched-selection policy, which selects strict FP32's count anew in each draw, FP8 passed the
equalized-odds tolerance in a share 0.581 of draws and strict FP32 in 0.437; FP8 passed where strict FP32 failed in
0.196, and the paired change of the gap, FP8 minus strict FP32, had an interval of $[-0.0152, 0.0094]$, so the matched pass does not
separate the contracts. The threshold was set from 2{,}048 training records. Holding the full training set's
employment rate at 0.4566 and resampling their strict FP32 margins gave 95\% intervals of $[-5.89, -5.68]$ for $t_m$ and
$[0.426, 0.469]$ for strict FP32's selection rate; at all 1{,}000 resampled thresholds FP8 passed both tolerances,
while strict FP32 passed demographic parity at a share 0.010 of them and equalized odds at 0.222. Resampling
the training and test records together, FP8 passed demographic parity in 0.840 of draws against 0.168 for
strict FP32, and equalized odds in 0.870 against 0.449. The resampling describes these benchmark records; with
ACS survey weights, which a population estimate needs, the equalized-odds change disappears (Appendix~\ref{app:split}).

\paragraph{Distance in flips.} For each pass-fail and four-fifths verdict at stake under Tier 1, the distance of the
reference statistic to its criterion, divided by the largest change one record's decision causes (by central finite
differences), gives a first-order estimate of the flips needed to reach the criterion; it does not show that so
many feasible flips would reach it. Of the tabular models' 60 such verdicts, 3 were within one flip,
9 within five and 14 within ten; of the 4B model's 30, one was within two flips and 6 within ten; of the 24B
model's 30, none within five and 7 within ten. In 11 verdicts, some Tier 1 contract flipped at least
the estimated number of decisions in the same task and model, yet none of those verdicts changed; some of their
statistics moved toward the criterion without crossing it.

\paragraph{Fixed capacity on Diabetes.} The Diabetes threshold is a frozen cutoff, so a contract can change how many
encounters are flagged. A program with fixed capacity would instead flag the top 10\% of each deployment's own
scores, 1{,}994 encounters, breaking ties at the boundary by record order. Under that rule the 4B model's FP8 serving changed 464 decisions, 232 encounters flagged
and 232 dropped, against 822 under the frozen cutoff, and the MI350X's FP8 changed 556 against 629. TF32 changed 8
on the H100 and the RTX 4090, against 7, and no tabular Tier 1 contract changed any. The 232 newly flagged encounters involve
229 distinct patients and the 232 dropped ones 225. Ties at the boundary decide some BF16 selections: 181 of the 4B model's
encounters share the 1{,}994th score under BF16 on the H100, and over 200 random orders of tied scores its count
ranged from 112 to 146, while FP8's stayed at 464.

\paragraph{Direction of the relaxed flips.} Under the M5's relaxed default, the ACSEmployment MLP's scores within
0.005 of the threshold rose by $3.6\times10^{-4}$ on average for men and $1.5\times10^{-4}$ for women. A recount
removes the part of each shift linked to sex: every record keeps its FP64 score and its own measured shift, less its
sex's mean shift there plus the pooled mean. It flips 16 men and 7 women rather than 27 and 4. The recount changes
every record's shift, so the total changes too, from 31 to 23.

\paragraph{Base-rate threshold for the 4B model on ACSEmployment.} At $m \ge 0$ the 4B model selects 0.094\% of
ACSEmployment test records against a base rate of 45.4\%. The 24B arm's rule instead sets $t_m$ at the quantile of
its strict FP32 training margins that selects the training employment rate. The 4B arm had scored training samples
only for HMDA and Diabetes, so after the first draft a sample of 10{,}000 ACSEmployment training records, drawn by the same
seeded rule, was scored under FP64; the quantile of its margins that selects the training employment rate, 45.7\%, gives
$t_m = -9.85$ and an FP64 selection rate of 45.6\% on the test records. Table~\ref{tab:app-baserate} gives each
contract's flips and changed verdicts at that threshold. A threshold taken from the test margins instead, as an
earlier draft did ($t_m = -9.86$), changed the same verdicts.

TF32 changed 122, 135 and 48 decisions on the H100, RTX 4090 and MI350X, and no verdict in senses 1 to 5. Strict
FP32 changed 2 decisions on each GPU, both records whose FP64 margin equals $t_m$ exactly. No BF16 contract changed a
verdict. FP8 serving changed 10{,}214 decisions and four verdicts by sex. It passed the equalized-odds test that FP64
failed, with a gap of 0.0458 against 0.0736 and an FP64 interval of $[0.0653, 0.0811]$. That is the only pass-fail
change in this study whose FP64 interval excluded the criterion. Over Tier 1 contracts every stability interval kept
its verdict.

\begin{table}[ht]
\caption{Exploratory, post hoc: the 4B model on ACSEmployment at a base-rate threshold set from 10{,}000 training
records ($t_m = -9.85$, FP64 selection rate 45.6\%). Flips against FP64 in run 1 (run 2 changed the same records),
and the verdicts of senses 1 to 5 that changed. All changes are by sex.}
\label{tab:app-baserate}
\footnotesize
\begin{tabular}{@{}l r l@{}}
\toprule
Contract & Flips & Verdicts changed \\
\midrule
Strict FP32, H100 / RTX 4090 / MI350X & 2 / 2 / 2 & none \\
TF32, H100 / RTX 4090; TF32 setting, MI350X & 122 / 135 / 48 & none \\
BF16, H100 / MI350X & 1{,}791 / 1{,}673 & none \\
vLLM BF16, H100 & 1{,}243 & none \\
vLLM FP8, H100 & 10{,}214 & EO pass-fail; SR and TPR significance; EO ranking \\
FP8, MI350X & 7{,}558 & SR significance \\
\bottomrule
\end{tabular}
\end{table}

\paragraph{Distance to the criterion by linearization.} Moved from Section~\ref{sec:mechanism-verdicts} for the page
limit. For each verdict, a local linearization estimates the uniform score shift that would carry its statistic to the
criterion, from each cell's FP64 score density within 0.005 of the threshold and the change per flipped record.
TF32's mean shift reached at most 0.004 of it. Ranked by that estimate, which needs no deployment run, the five
tabular pass-fail verdicts that 8-bit formats changed were among the six nearest of 90, the other a tree ensemble's
that no Tier 2 contract scored. The ranking was committed after the Tier 2 results.

\subsection{Post hoc robustness checks}\label{app:robustness}
The eight checks below were run post hoc, after the first draft, on CPU from the committed per-record
outputs. Each first reproduced the committed number it builds on. They are exploratory and change no count reported
elsewhere. The prespecified analyses stay primary; these checks qualify how some of them read.

\paragraph{Chance overlap of flip sets.} A null draws each GPU's TF32 flips at random among the 4B model's records in
the same task, side of the threshold and logit-margin bin, keeping each GPU's count per stratum. It expects 13.7
records shared by the H100 and the RTX 4090 rather than the observed 22. That is a Jaccard index of 0.21 (95\% range
0.14 to 0.29) rather than 0.38, and 22 or more shared records occurred with probability 0.00015. Six stratifications,
from one band per side to blocks of 5 records in margin order, put chance at 0.14 to 0.23 and the observed shared records at
1.5 to 2.3 times the expected count ($P \le 0.0011$).

HMDA carries the excess, 12 shared records against a chance index of 0.17 ($P = 0.0008$), and keeps it under
all six stratifications ($P \le 0.0032$). Under the same bins, the
other three tasks are each consistent with chance ($P$ 0.12, 0.33 and 0.58), though blocks of 25 records make
ACSEmployment and Diabetes significant ($P$ 0.010 and 0.019). The by-task range of Section~\ref{sec:results-who}
therefore depends on the matching. Under the bin and block strata, cross-vendor BF16 (0.31) and FP8 (0.29) shared 1.2
to 1.7 times the chance overlap, and transformers and vLLM BF16 on one H100 (0.19) shared 1.1 to 1.2 times it. The
tabular MLP's TF32 flips were the same 5 records on the H100 and the RTX 4090, against a chance index of 0.32
($P = 0.003$).

\paragraph{Paired changes behind the changed verdicts.} For each of the 9 verdict and contract pairs behind the 6
changed pass-fail verdicts of Table~\ref{tab:twoway}, 10{,}000 paired draws give the change in the gap, contract
minus FP64, on the same resampled units (patients for Diabetes). The percentile interval of the change excludes zero for 2
of the 9. One is the 4B model's HMDA equalized-odds gap by sex under vLLM FP8, $-0.0183$ $[-0.0297, -0.0058]$. The
other is the logistic regression's HMDA demographic-parity gap by ethnicity under MLX MXFP8, 0.0042 $[0.0006,
0.0077]$. BCa intervals exclude zero for 4 more, though up to 19\% of draws show no change there, which strains BCa.
The two verdicts differed in 4\% to 41\% of draws. For most pairs, the change cannot be told from zero in other
samples, so the changed verdict describes these test records.

For the 24B model, with a separate bootstrap seed from Table~\ref{tab:llm24-three}, FP8 serving's
demographic-parity change, $-0.0150$ $[-0.0237, -0.0062]$, excludes zero, and the two
verdicts differ in 71\% of draws (74\% of the 1{,}000 draws above). Its equalized-odds changes include zero:
$-0.0050$ $[-0.0111, 0.0010]$ under vLLM BF16 and $-0.0120$ $[-0.0267, 0.0027]$ under FP8. Of the two equivalence
changes, INT8 moved the ACSEmployment MLP's gap by sex by 0.0062 $[0.0050, 0.0073]$, while the Diabetes one includes
zero.

\paragraph{A null that keeps the contracts' overlap.} Unlike the matched-perturbation null of
Appendix~\ref{app:posthoc}, one random permutation within each stratum here moves every contract's flips at once
(5{,}000 draws). Each contract's own null is unchanged, and every overlap between contracts is the
observed one. The language model's Tier 2 union then has a null mean of 7.23 (95\% range 5 to 10). Five or fewer
changes occurred in a share 0.105 of draws, against 0.0798 with independent draws (0.132 under the finer matching),
so its 5 changes do not separate from this null. The tabular Tier 2 union has a null mean of 14.04 (10 to 18), and 20
or more occurred in a share 0.0036 (0.0012 under the finer matching), so the tabular excess holds. Holm's adjustment
over the 12 Tier 2 contracts raises MLX 8-bit weights' share of 0.0138 to 0.166.

\paragraph{Bootstrap size and BCa.} Recomputed from the pilot's seeds, the committed $B = 1{,}000$ intervals match to
within $5\times10^{-11}$. At $B = 10{,}000$, no percentile interval of the 60 pass-fail verdicts or of the
24B model's 20 changes whether it contains 0.05. BCa intervals move 4 of the 60, two
each way and none a changed verdict, and 2 of the 20. Both splits stay as reported: 6 of 17 against 0 of 43, and 2 of
6 against 0 of 14 by patient. One endpoint lay within two Monte Carlo standard errors of 0.05 at $B = 1{,}000$, the
MLP's upper end for the Diabetes equalized-odds gap by sex, 0.0475; it is 0.0471 at $B = 10{,}000$. Two of the 30
four-fifths intervals, whose verdicts no contract changed, change status.

\paragraph{Flip slope per GPU.} The pilot's pooled fit over both NVIDIA GPUs, 80 flips in 408{,}408 decisions,
reproduces a fall of 3.93 in the log-odds of a TF32 flip per tenfold increase in logit margin. Fitted per GPU, the log-odds fall by 3.90 on the H100
(95\% Wald interval 3.38 to 4.43) and by 3.95 on the RTX 4090 (3.40 to 4.51). Pooled with each record's two rows
clustered, the interval is 3.45 to 4.41. The logistic form predicts 3.6 to 4.1 flips per GPU beyond 0.1 logit, where
none occurred, so flips fall off faster than the fit.

\paragraph{The 24B model's distances in logits.} Table~\ref{tab:app-llm} (D) gives the largest distance of a changed
decision in probability. In logits, $|m - t_m|$, the largest per task were 0.0023 to 0.0105 under TF32, 0.145 to
0.444 under BF16, 0.097 to 0.295 under vLLM BF16 and 0.92 to 1.43 under FP8 serving. FP8's ACSEmployment 0.00961 in
probability is 1.43 logits. Every TF32, BF16 and vLLM BF16 change lay within one logit, as did 98.8\% to 100\% of
FP8's per task.

\paragraph{Exact tests in sparse Diabetes cells.} In place of the pooled $z$-test of Table~\ref{tab:app-cluster},
which is asymptotic in small cells, Fisher's exact test or its mid-$p$ version, under the same five procedures and
three arms, changes no Tier 1 verdict. Tier 2 changes 4, 4, 6, 3 and 2
verdicts under Fisher's test and 4, 4, 5, 2 and 2 under mid-$p$, against 4, 3, 6, 3 and 2. The one Tier 1 change
disappears: the 4B model's true-positive gap by race between Hispanic (11 of 43 flagged under FP64) and Caucasian (200
of 1{,}731) encounters. Its Bonferroni-adjusted $p$ under FP64 already exceeds 0.05 under Fisher's (0.080), mid-$p$
(0.062) and Boschloo's (0.093) tests, against 0.030 under the $z$-test, so TF32 changes nothing. The ACSEmployment Tier 1 change of Table~\ref{tab:app-sense2}, under unpooled Holm and omnibus tests, was not
retested.

\paragraph{Retraining with a new seed.} The MLP retrained with a new seed for the prospective test
(Appendix~\ref{app:prospective}) was scored in FP64, its threshold set by the same frozen rule from its own training
scores. Against the original MLP it changed 2{,}330, 2{,}594, 2{,}503 and 1{,}587 decisions on ACSIncome,
ACSEmployment, HMDA and Diabetes, 6.0\%, 3.4\%, 3.6\% and 8.0\% of each test set. Tier 1 contracts changed at most 31
of the MLP's decisions per dataset, and Tier 2 contracts 8 to 860. The reseed changed 6 of the MLP's 60 verdicts of
senses 1, 2, 3 and 5 (four significance and two equivalence verdicts, no pass-fail). It moved 28 of its 30 sense-6
statistics and changed 7 rankings in each arm.

With only the MLP perturbed, Tier 1 contracts changed no verdict or ranking and moved 6 statistics. Tier 2 contracts
changed 5 verdicts (2 pass-fail) and 8 and 7 rankings in the language-model and tabular arms, and moved 24
statistics. The reseed moved gaps by up to 0.059 and impact ratios by up to 0.11, at most 1.1 times the FP64
sampling-interval width. Tier 1 moved them by at most 0.0022, and Tier 2 by at most 0.039 and 0.043. The reseed's
change exceeded the largest Tier 2 change in 21 of the 30 statistics.

\subsection{Engine against precision, survey weights and fixed capacity}\label{app:split}
These checks were run after the first draft, from committed outputs and the raw ACS file; none was
planned, and each is exploratory.

\paragraph{Engine against precision.} Every Tier 2 change of the 24B model compares strict FP32 in transformers with a
contract that changes precision and, under vLLM, the engine as well. The saved logits order its contracts as a chain:
strict FP32, BF16 in transformers, vLLM BF16 and vLLM FP8, whose steps add up to the total change. The first step is a
configuration change within the framework: BF16, one H100 rather than two and SDPA rather than eager attention. The
second changes the engine at fixed precision on one H100, and the third, the controlled comparison, changes precision
within one engine and GPU. Table~\ref{tab:app-split} gives each link
of the ACSEmployment audit by sex and each step's paired change over 10{,}000 bootstrap draws of the test records.
BF16 in transformers flipped 380 decisions and changed no pass-fail verdict. vLLM's engine at BF16 flipped 357 and passed
equalized odds, a change whose paired interval includes zero and whose verdict differed across the step in 0.23 of
draws; its demographic-parity change, which crossed no criterion, had an interval whose upper end,
$-7.7\times10^{-6}$, is within Monte Carlo error of zero. FP8 within vLLM flipped 1{,}967, passed demographic parity with a paired interval that excludes zero, and
raised the selection rate by 8.3 points. Along the same chain no other task or attribute changed a pass-fail verdict.
The chain is one path: strict FP32 under vLLM and FP8 under transformers were not run, nor BF16 on two H100s with
eager attention. It therefore cannot separate precision from GPU count and attention in its first step, or an
interaction of engine and precision, which a full engine-by-precision design on the GPU would need.

\begin{table}[ht]
\caption{Exploratory, post hoc: the 24B model's ACSEmployment audit by sex along a chain of its contracts (run 1,
20{,}000 records). Steps are paired changes, later link minus earlier, with 95\% percentile intervals over 10{,}000
bootstrap draws of the records; flips are decisions that differ across the step. The first step also changes
two H100s to one and eager attention to SDPA (Table~\ref{tab:app-audit-record}).}
\label{tab:app-split}
\footnotesize
\begin{tabular}{@{}l r r r@{}}
\toprule
Link & Selection rate & DP gap & EO gap \\
\midrule
Strict FP32, transformers & 0.447 & 0.0567 & 0.0515 \\
BF16, transformers & 0.463 & 0.0554 & 0.0520 \\
BF16, vLLM & 0.461 & 0.0516 & 0.0464 \\
FP8, vLLM & 0.544 & 0.0417 & 0.0394 \\
\midrule
Step (flips) & & DP change & EO change \\
\midrule
Configuration, transformers (380) & & $-0.0014$ $[-0.0051, 0.0025]$ & $0.0005$ $[-0.0061, 0.0070]$ \\
Engine at BF16 (357) & & $-0.0037$ $[-0.0074, -7.7\times10^{-6}]$ & $-0.0056$ $[-0.0118, 0.0006]$ \\
Precision, vLLM (1{,}967) & & $-0.0099$ $[-0.0183, -0.0015]$ & $-0.0070$ $[-0.0209, 0.0068]$ \\
Total (2{,}156) & & $-0.0150$ $[-0.0238, -0.0064]$ & $-0.0120$ $[-0.0267, 0.0024]$ \\
\bottomrule
\end{tabular}
\end{table}

\paragraph{Survey weights.} The ACS statistics weight every sampled person equally, so they describe the test records.
The ACS PUMS person weights describe California's 2018 population within each task's Folktables filters. Each weighted rate is the weighted count of
favorable decisions over the weighted count of people among the records scored. Its variance is the Census Bureau's
successive-difference one. The statistic is recomputed under each of the 80 replicate weights on the same records,
and the variance is 4/80 times the sum of the replicate values' squared deviations from the full-sample value. An
interval is the estimate plus or minus 1.96 standard errors. A paired change differences two contracts' statistics
replicate by replicate, and each gap keeps the worst pair chosen under the full-sample weight. The replicate weights
are applied unchanged to the records scored, the seeded test split or, for the 24B model, a 20{,}000-record subset of it.
Neither draw is modeled as a stage of the design, nor are training and the threshold. The intervals therefore condition
on the records scored and the trained model, and are not shown to cover the variance those steps add. The weighted
analysis is therefore a design-informed sensitivity check, not a population-level adjudication. The weights were read from the raw file the pilot built both ACS tasks from,
with folktables' row filters and the pilot's seeded split reapplied; the mapping reproduced every test label and every
sex and race code. Senses 1, 2, 3 and 5 were recomputed with weights, groups keeping their unweighted eligibility
and senses 2 and 3 using the replicate-weight standard error of the weighted pair difference. Weighting changed 4 of
the 96 ACS reference verdicts at stake: the 4B model's ACSIncome demographic-parity gap by sex went from 0.0590 to
0.04999, inside the tolerance; the MLP's ACSEmployment gap by sex from 0.0390 to 0.0540, ending its pass and its
equivalence; and the 24B model's ACSEmployment equalized-odds gap by sex from 0.0515 to 0.0353, a pass. Contracts
changed 3 distinct ACS verdicts unweighted and 4 weighted. The 24B demographic-parity change remained, from 0.0550
(interval $[0.041, 0.069]$) to 0.0478 under FP8, with a change of $-0.0072$ and a replicate-weight interval of $[-0.018, 0.004]$; its
equalized-odds change did not, since the reference already passed.
Weighted, the 24B model's selected shares were 0.429, 0.443 and 0.533 under strict FP32, vLLM BF16 and vLLM FP8
(Table~\ref{tab:llm24-three}), each below its raw share. The 4B model's ACSIncome gap by sex, 0.04999 under
FP64, crossed the tolerance under four contracts: 0.05009 and 0.05031 under TF32 on the H100 and the
RTX 4090, and 0.05002 and 0.05006 under vLLM BF16 and FP8. This is the one weighted Tier 1 change, a verdict at its
criterion.
Paired over the replicate weights, TF32 changed the gap by 0.00011 on the H100, with an interval of
$[-0.00017, 0.00038]$, and by 0.00033 on the RTX 4090, with $[0.00003, 0.00062]$, which barely excludes zero. Both
changes are small against the reference gap's own standard error of 0.0062: the crossing is deterministic, not
evidence of a wider disparity.
The other two weighted changes are 4B ACSEmployment significance verdicts by race under Tier 2, where the
model selects 0.094\% of records and the replicate variance of a near-empty cell is unstable. INT8's equivalence change
of the MLP disappears, since the weighted reference is no longer equivalent.

\paragraph{Fixed capacity on Diabetes, audited.} Appendix~\ref{app:posthoc} counts the decisions a top-10\% policy
changes. Auditing those decisions, against the reference's own top 10\%, Tier 1 changed no Diabetes verdict under
either policy. Tier 2 changed 8 distinct per-model Diabetes verdicts (senses 1, 2, 3 and 5; rankings not recomputed) under the
frozen cutoff and 6 at fixed capacity, 4 of them the same. The pass-fail changes fell from 3 to 1, the logistic
regression's demographic-parity gap by race. Of the
changes at fixed capacity, the 4B model's true-positive significance by race and the 24B model's equivalence by age
appear under that policy only. The fixed-capacity decisions reproduce the counts of Appendix~\ref{app:posthoc}.

\subsection{What each result rests on}\label{app:overview}
Table~\ref{tab:app-overview} sorts the main results by whether their analysis was prespecified and by the checks
of Appendices~\ref{app:posthoc} to~\ref{app:split} that they survive or fail. A prespecified result keeps that status
when a check fails it; the check qualifies how it reads.

\begin{table}[ht]
\caption{The main results by status and by the checks they survive and fail. Prespecified: the analysis in the
study plan; the verdict-counting rule was set post hoc (Appendix~\ref{app:notes}). Exploratory checks are those of
Appendices~\ref{app:posthoc} to~\ref{app:split}.}
\label{tab:app-overview}
\footnotesize
\begin{tabular}{@{}>{\raggedright\arraybackslash}p{0.27\linewidth} >{\raggedright\arraybackslash}p{0.17\linewidth}
>{\raggedright\arraybackslash}p{0.24\linewidth} >{\raggedright\arraybackslash}p{0.24\linewidth}@{}}
\toprule
Result & Status & Survives & Fails or untested \\
\midrule
Strict FP32 changed no 4B decision at the prespecified thresholds; every contract repeated its decisions & Prespecified & three GPUs, both runs &
vLLM BF16 did not repeat its scores bit for bit \\
Tier 1 flipped decisions near the threshold (TF32: 41 and 39 of the 4B model's) & Prespecified & both runs;
predicted in advance for new MLPs on the M5, an RTX 4090 and an H100 & the prediction for one MLP on an RTX A6000 \\
Tier 1 changed no verdict (84, 144 and 84 at stake) & Prespecified, unweighted tests & exact tests; tests both clustered and
adjusted (Diabetes); fixed capacity & five of nine alternative tests (one sparse-cell verdict each); survey weights (one
verdict at its criterion) \\
Which records flip depends on the GPU and stack (pooled Jaccard 0.38) & Prespecified measure; chance level post hoc &
above-chance agreement on HMDA, under all six matchings & chance agreement on the other three tasks \\
Tier 2 changed verdicts (Table~\ref{tab:senses}) & Prespecified senses & both runs & 7 of 9 pass-fail pairs have
paired intervals including zero \\
Tier 2 changed only pass-fail verdicts whose interval contains 0.05 (6 of 17, 0 of 43) & Exploratory; 24B arm by a
rule fixed in advance (2 of 5, 0 of 15) & $B = 10{,}000$, BCa, resampling by patient & a base-rate threshold (one
change outside) \\
24B FP8 passed both employment tolerances by sex & Prespecified verdicts; chain, matched and weighted readings
exploratory & the raw demographic-parity change excludes zero & matched selection (demographic parity fails; the
equalized-odds change's interval includes zero); weights (the reference passes equalized odds; the demographic-parity change's interval includes zero) \\
Five Tier 2 significance changes & Prespecified test & the 4B model's two by sex, under pooled, Wald and
patient-clustered variances, the one test two groups allow & the three tabular ones by race, under at least four of the six \\
Three HMDA pass-fail changes & Prespecified groups & -- & each involves the derived Joint category; none remains
without it \\
Per-record model of the arithmetic & Post hoc reconstruction; one prospective test & as in the second row & 8-bit
and FP8 weights, not modeled \\
Strict control's measured cost (4.33 times TF32 on one H100) & Measured & two GPU types & kernels not profiled \\
\bottomrule
\end{tabular}
\end{table}

\section{Implementation Notes}\label{app:notes}

\paragraph{Analysis status.} Each analysis has one of three statuses, set by the order of commits in the study's
version control. Prespecified before any run: the study plan, with its contracts, the six senses, the tolerance, the
expectations and the kill criterion, and the tabular Tier 1 constants and harness. Planned after earlier results but
before a later arm was scored: the 4B arm's prompts, label tokens and code and the Tier 2 constants, each committed
after the tabular Tier 1 results and before its own scoring; the 24B arm, with its records, reference, threshold rule
and analysis; and the MI350X arm, with its contracts and expectations. Post hoc: the analyses under ``Analyses added
after results were seen'' below, including the FP16 arm, all of them exploratory.

\paragraph{Runs.}
\begin{itemize}
\item On an RTX A4500 node, large float32 matrix products returned NaN under strict FP32, while the same code
passed on the RTX A4000 and the RTX 4090. That run was excluded as a node fault, and the RTX A4000 covers Ampere.
\item Language-model batches in transformers hold records of equal token length without padding, with the same
batch plan for every contract. vLLM schedules its own batches over requests sent in record order. Under FP64 the 4B model's top next token was the first token of a candidate label for every record. Under strict FP32
the 24B model's top next token was a label for every ACS and HMDA record and 98.5\% of Diabetes records (moved
from Section~\ref{sec:design-models} for the page limit).
\item For the tabular models, BF16 compute ran on the CPU and GPU of an Apple M5 Max, INT8 on its CPU, and 8-bit
and MXFP8 weights on its GPU; BF16 and FP8 also ran on the AMD MI350X. NVIDIA BF16 and FP8 compute were not run for them, and the MXFP8 contract,
which quantizes weights only, stands in for neither. The tree ensembles have no BF16 or 8-bit path, so every
Tier 2 ranking takes them from their float32 CPU predictor, whose decisions equal FP64's. The language model's BF16
and FP8 contracts ran on the H100 and the MI350X.
\item Every contract ran twice, except the language model's FP64 reference, which ran twice on ACSIncome and once
on the other three tasks. Its two ACSIncome runs were bit-identical. On the H100, run 2 of each contract timed within 2\%
of run 1 wherever it repeated every task (Table~\ref{tab:price}). Both vLLM runs sent requests in the same
order, and vLLM BF16 alone did not repeat its scores bit for bit, though its decisions repeated. vLLM ran with its
weights in BF16, FP8 quantization applied online for the FP8 contract, a KV cache in the model's type, prefix
caching, chunked prefill and CUDA graphs; its matrix-product kernels and attention backend were not recorded. For a
BF16 checkpoint quantized online, vLLM 0.25.0's source fixes one FP8 scale per weight tensor at load and, on a GPU
with CUTLASS FP8 kernels such as the H100, a scale per token computed from the activations of each call; FP8 on the
MI350X scales activations per tensor instead, one scale over the whole batch of each call, so a record's FP8 score
there depends on the other records in its batch (fixed, and the same in every run).
\item On the MI350X, FP64 changed no language-model decision, its margins within $3.8\times10^{-12}$ of the H100's,
though not bit-identical.
\item On the MI350X, the language model's strict FP32 used 1.38 times the GPU-hours of the TF32 setting over whole
jobs, model loading included, so the ratio is not comparable with those of Table~\ref{tab:price}, which exclude loading.
\item The 24B model (Mistral Small 3.2 Instruct, 2506) deviates from the 4B arm by design: a strict FP32 reference,
which FP64 checked on 2{,}048 records each of ACSIncome and HMDA (no decision changed; the largest margin difference
was $1.6\times10^{-4}$, against an expected bound of 0.01) and, after the first draft, on the 2{,}048 ACSEmployment records
nearest the threshold, all within 0.23 logits of it (no decision changed; largest difference $1.6\times10^{-4}$, with 2
records closer to the threshold than that). The ACSEmployment check ran FP64 on a CPU, because the GPU partitions
were full, and compared it with the strict FP32 margins of the test run, scored in other batches; seeded subsets of 20{,}000, 20{,}000 and 10{,}000 records
for ACSEmployment, HMDA and Diabetes; the first token of each label word; and an ACSEmployment threshold at the
training employment rate. Strict FP32 ran a second time on ACSIncome and, after the first draft, on
ACSEmployment, bit-identically both times. Later, both vLLM
contracts ran a third time with the engine's log kept, on the same weights (the Mistral-format file restored at the
same revision, its SHA-256 checked against the hub) and settings: the engine selected FlashAttention, and CUTLASS
kernels for FP8's products, with per-tensor weight scales and per-token activation scales; FP8 reproduced its margins bit for bit on every task, and vLLM BF16 changed 4
ACSEmployment decisions, with margins up to 0.375 apart, but no verdict (Appendix~\ref{app:audit-record}). Of its 16 written
flip-rate expectations, 11 held. Its deployment contracts exceeded the expected range on ACSEmployment, where every
flip lay within 0.01 of the threshold on the probability scale. In logits, 26 of FP8's 2{,}156 flips there lay 1 to 3
from the threshold ($t_m = -5.78$), and none further. vLLM fell below the expected range on HMDA. Its most likely next token
was a label token for 96.8\% to 98.9\% of Diabetes records, against an expected 99\%, and for every other record.
\item The first training pass crashed in XGBoost after writing two ACSIncome models (PyTorch and XGBoost each
bundle an OpenMP runtime on macOS). Those files were discarded, and every model was trained in one complete pass
before anything was scored.
\end{itemize}

\paragraph{Effects of where a threshold sits.}\label{sec:mechanism-artifacts}
Two effects depend on where a threshold sits rather than on the arithmetic, and both are kept out of every count in
the paper. A float32 output cannot fall between $1-6\times10^{-8}$ and 1, and 2{,}050 HMDA MLP test records lie
within $10^{-7}$ of 1, so a threshold there separates FP64 from every float32 contract, strict included. A quantile threshold can also
fall within an ulp of identical scores, which any perturbation then flips together. A third effect is not kept out.
At the ACS threshold, a logit margin of 0, the label logits tie exactly under BF16 and FP8 for 5 to 149 records per
contract and task for the 4B model and 189 to 218 on ACSIncome for the 24B model, and $m \ge 0$ decides them. FP64,
strict FP32 and TF32 produce no such tie. Counting the ties as negative instead changes the flip counts, for the 4B
model's BF16 on ACSIncome from 176 to 241, but no verdict of either model in any of the five senses, rankings
included.

Table~\ref{tab:app-ties} splits each tying contract's flips into ties and untied records. Most of the 24B model's
BF16 flips on ACSIncome are ties: 98 of 115 under transformers and 80 of 88 under vLLM.

\begin{table}[ht]
\caption{Ties at the ACS threshold ($m = 0$) and the flips they decide, run 1. Under the paper's rule $m \ge 0$ a tie
is positive, so it flips where the reference rejects; under $m > 0$ it is negative and flips where the reference
accepts. Untied flips are the same under both rules. No reference produces a tie, and no contract ties on HMDA or
Diabetes.}
\label{tab:app-ties}
\footnotesize
\begin{tabular}{@{}l l r r r r r r@{}}
\toprule
 & & & \multicolumn{2}{c}{Flips, $m \ge 0$} & \multicolumn{2}{c}{Flips, $m > 0$} & \\
\cmidrule(lr){4-5}\cmidrule(lr){6-7}
Model, task & Contract & Ties & All & Ties & All & Ties & Untied \\
\midrule
4B, ACSIncome & BF16, H100 & 147 & 176 & 41 & 241 & 106 & 135 \\
 & vLLM BF16, H100 & 149 & 166 & 60 & 195 & 89 & 106 \\
 & vLLM FP8, H100 & 145 & 976 & 47 & 1{,}027 & 98 & 929 \\
 & BF16, MI350X & 128 & 191 & 39 & 241 & 89 & 152 \\
 & FP8, MI350X & 117 & 927 & 60 & 924 & 57 & 867 \\
\addlinespace
4B, ACSEmployment & BF16, H100 & 19 & 12 & 6 & 19 & 13 & 6 \\
 & vLLM BF16, H100 & 17 & 14 & 5 & 21 & 12 & 9 \\
 & vLLM FP8, H100 & 5 & 67 & 2 & 68 & 3 & 65 \\
 & BF16, MI350X & 11 & 15 & 5 & 16 & 6 & 10 \\
 & FP8, MI350X & 10 & 56 & 6 & 54 & 4 & 50 \\
\addlinespace
24B, ACSIncome & BF16, H100 & 189 & 115 & 98 & 108 & 91 & 17 \\
 & vLLM BF16, H100 & 190 & 88 & 80 & 118 & 110 & 8 \\
 & vLLM FP8, H100 & 218 & 403 & 98 & 425 & 120 & 305 \\
\bottomrule
\end{tabular}
\end{table}

\paragraph{Thresholds and labels.}
\begin{itemize}
\item The HMDA threshold approves the training approval share. A threshold of 0.9025 on the probability is a
marked secondary point of the threshold curves.
\item The language model's label words are HIGH/LOW for ACSIncome, APPROVE/DENY for HMDA, and YES/NO for
ACSEmployment and Diabetes. APPROVE and DENY span several tokens, so the first token of each is read. After the
first draft, HMDA was rescored once on the H100 with its question asking for YES or NO, single tokens,
under FP64, TF32, BF16, vLLM BF16 and vLLM FP8, each once. The threshold was set by the same rule from FP64 scores
of the same training sample, so FP64 selected 90.5\% of test records against 90.4\% with APPROVE/DENY. AUC was
0.647 against 0.650, and the four contracts flipped 24,
255, 216 and 1{,}822 decisions against 23, 264, 178 and 1{,}789. FP8 again passed the equalized-odds test by sex
that FP64 failed, with a gap of 0.0353 against 0.0557 and an FP64 interval of $[0.0314, 0.0837]$, and it also
changed the demographic-parity and equalized-odds rankings by sex. No other contract changed a verdict. The gap
is set by joint applications, whose applicant and co-applicant differ in sex, against female ones
(Appendix~\ref{app:hmda-groups}).
\item The 4B language model uses $m \ge 0$ on ACSIncome and ACSEmployment. Its HMDA and Diabetes thresholds come from
FP64 scores of a seeded sample of 10{,}000 training records, because a forward pass of the model costs far more than
one of a tabular model.
\end{itemize}

\paragraph{Analyses added after results were seen.} These are exploratory.
\begin{itemize}
\item The AMD MI350X arm was added after the other arms' results. Its contracts and expectations were written
before any AMD job, and a probe established what each setting computes before scoring. Two planned contracts
changed with the probe. A variant of the TF32 setting with hipBLASLt's TF32 environment variable was dropped, its
outputs bit-identical to the setting's. The language model's untouched PyTorch defaults, which the probe found
equal to strict FP32, were scored on the first shard of each task only, where they matched strict FP32 bit for
bit. The analysis code ran unchanged and reproduced every earlier contract's results. The probe outputs behind the
MI350X rows of Table~\ref{tab:contracts} are in the artifact.
\item The fixed probability grid for threshold curves was made primary after the quantile grid showed the
saturation and tie effects of Appendix~\ref{sec:mechanism-artifacts}. Both grids are reported.
\item The Tier 2 diagnostics were written after the Tier 2 results were seen: the decomposition of logistic
regression's BF16 flips, the logit granularity, the threshold grid under both decision readings and the HMDA
worst-pair breakdown. They explain mechanisms and change no count or verdict.
\item The per-record model of Section~\ref{sec:mechanism} was built after the Tier 1 results. Its CPU BF16
predictions and its ranking of verdicts were committed after the Tier 2 results. That they were made before those
results were read rests on a commit message, so the repository cannot show the order. Its description of MLX
BF16 arithmetic was fitted afterwards, including a search over sigmoid implementations.
\item The matched-perturbation null, the fixed-pair rerun of sense 2 and the base-rate threshold for the 4B
model's ACSEmployment decisions (Appendix~\ref{app:posthoc}) were designed after the first draft and written
down before they ran.
\item Analyses added after the first draft read the per-record outputs or committed results: the tie counts
above, the group sizes of Table~\ref{tab:app-groups}, the alternative and clustered tests for sense 2, the distance
to the criterion, the paired bootstrap, utility and matched selection of the 24B employment changes, the distance of
Tier 1 verdicts in flips and the fixed-capacity Diabetes policy (Appendix~\ref{app:posthoc}), the HMDA group attribution (Appendix~\ref{app:hmda-groups}) and a resampling of
Diabetes by patient. Three scored new records: HMDA with single-token labels (above), the 24B model's FP64 check
on ACSEmployment (Runs, above) and a 4B training sample for the ACSEmployment base-rate threshold
(Appendix~\ref{app:posthoc}). The Diabetes resampling
addresses repeated patients: the pilot's bootstrap drew encounters, though 3{,}266 of the 13{,}998
test patients have more than one (9{,}206 of the 19{,}938 encounters). Drawing patients widened the Diabetes intervals
by a factor of 1.0 to 1.7, and it moved one pass-fail verdict at stake, the logistic regression's equalized-odds
gap by sex, into the open column of Table~\ref{tab:twoway}, which read 6 of 16 against 0 of 44 with encounters. It
moved one of the 24B model's as well, whose split reads 2 of 6 against 0 of 14 by patient; Section~\ref{sec:results-24b}
reports the 2 of 5 against 0 of 15 of the encounter-level intervals in use when its rule was fixed. No verdict that
a contract changed moved. The other reported Diabetes intervals are the patient-level ones.
\item The distinct-verdict counts of Table~\ref{tab:senses} and the split by sampling interval of
Table~\ref{tab:twoway} were computed after the first draft, from existing results files. The earlier counts, one per contract and attribute, are superseded.
\item The robustness checks of Appendix~\ref{app:robustness} were run post hoc, after the first
draft, from committed outputs: a chance level for cross-GPU flip overlap, paired intervals for the changed verdicts, a
null that keeps the contracts' overlap, $B = 10{,}000$ and BCa intervals, per-GPU flip slopes, the 24B model's
distances in logits, exact tests for sparse Diabetes cells and an MLP retrained with a new seed.
\item The checks of Appendix~\ref{app:split} were run post hoc, after the first draft, from
committed outputs and the raw ACS file: the 24B employment change along a chain of its contracts, the ACS audits with
survey weights, and the Diabetes verdicts at fixed capacity.
\end{itemize}

\paragraph{Models the rounding cannot reach.}\label{app:unreached} A one-output layer is a matrix-vector product, where neither NVIDIA's
TF32 nor the relaxed path engages, so logistic regression never flipped. Trees compare float32 values and sum their
leaves in FP32, beyond any Tier 1 contract.

\paragraph{A prospective test of the per-record model.}\label{app:prospective} The model of
Section~\ref{sec:mechanism-rounding} reconstructs flip sets it was built after. To test it on models it had not seen,
an MLP was retrained per task with a seed never used before (20261013), with the architecture, data, optimizer and
epochs of the original. The emulation, unchanged from the code that produced Section~\ref{sec:mechanism}, predicted
each new model's flips against FP64 under TF32, the M5's relaxed default and strict FP32, at the thresholds the
original rules give the new models (0.5 on the ACS tasks; the HMDA and Diabetes cutoffs recomputed from their FP64
training scores) and at the 199 thresholds of the grid. The predictions were committed to version control before any
GPU or the M5 scored the new models: the CPU and GPU runs record later start times, and the M5, whose metadata records
none, wrote its outputs after the commit by their file times. The M5's MLX build was rebuilt from the original commit and first reproduced the
original run's ACSEmployment MLP margins byte for byte. Under the relaxed default the M5 flipped exactly the
predicted records on every task, 2, 13, 6 and 2 against FP64, and every grid threshold matched; strict FP32 on the
M5 and on the CPU was predicted to flip nothing and flipped nothing. TF32 on an RTX 4090 and on an H100 flipped exactly
the predicted records at every prespecified threshold, 4, 6, 0 and 1, and matched at 183 to 199 of the 199 grid
thresholds per task.
On an RTX A6000, a GPU absent from the original study, it matched on three tasks; on ACSEmployment it flipped 5
records against 6 predicted, sharing 4, and matched at 24 of 199 grid thresholds. A post hoc comparison of layer patterns fits that measurement with
TF32 rounding in the second hidden layer only, a residual of 0.036 of the perturbation's RMS against 3.5 for the
predicted pattern. Which layers TF32 reaches depends on the kernels a GPU dispatches, which is why the protocol keeps
kernel libraries in the record (Section~\ref{sec:protocol-fit}). Each run repeated byte for byte.

\paragraph{Range against precision: an exploratory FP16 arm.}\label{app:fp16}
BF16 keeps FP32's exponent range, so every BF16 flip in this study is a rounding effect. An exploratory arm, designed
after every other result had been read, asked whether range moves a decision once a format has less of it. FP16 has
three more fraction bits than BF16, but its largest finite value is 65504, values below $2^{-14}$ lose precision as
subnormals, and values at or below $2^{-25}$ flush to zero. Table~\ref{tab:app-fp16-contract} gives the arm's
contract in the form of Table~\ref{tab:contracts}.

\begin{table}[ht]
\caption{The exploratory FP16 arm's contract, in the columns of Table~\ref{tab:contracts}.}
\label{tab:app-fp16-contract}
\footnotesize
\begin{tabular}{@{}l c >{\raggedright\arraybackslash}p{0.31\textwidth} l >{\raggedright\arraybackslash}p{0.23\textwidth}@{}}
\toprule
Contract & Tier & What changes in the arithmetic & Checked by & Hardware \\
\midrule
FP16 (exploratory) & 2 & inputs, every layer output and the tabular weights in FP16 (largest finite 65504) & canary, range canary & Apple M4 Pro CPU; RTX 4090 (LLM) \\
\bottomrule
\end{tabular}
\end{table}

The tabular models ran under FP16 on an M4 Pro CPU, twice,
bit for bit; its canary showed float16 products accumulated in a wider format and rounded once per output. Before
that run, the emulation of Section~\ref{sec:mechanism-rounding} predicted the flip sets of two formats with FP16's ten
fraction bits, one with FP16's range and one with FP32's, and the predictions were committed. FP16 flipped exactly the
predicted records in all eight cells under both readings: 40 decisions, against 450 under BF16
(Table~\ref{tab:app-fp16}). The two predicted formats flipped the same records everywhere. FP16's range acted on
17{,}141 stored values and 432 weights, flushing 3{,}877 values and overflowing none, and changed no decision.

The 4B model ran under FP16 and BF16 on RTX 4090 GPUs with the BF16 contract's set-up (Table~\ref{tab:app-fp16-llm}).
FP16 ran twice on one card, bit for bit, and flipped 80 decisions against BF16's 546, and its ACSIncome ties fell
from 122 to 18. No margin was lost to overflow, no layer output held an infinite or undefined value, and the largest
activation, 7{,}448, sat 8.8 times below FP16's limit.

Two emulations scored the 16{,}267 records within three logits
of the threshold, where every FP16 flip lay. Both had FP16's fraction bits, one with its range and one with FP32's.
They flipped 115 and 113 records but shared only 54. The expectation, written before the run, that range would
move at most 5\% of them was missed.

The direct range error was $6\times10^{-13}$ of the rounding error in squared
size. Yet the two margins differed by a median of 0.031 logits, as much as either differed
from FP64, and their errors correlated 0.20 to 0.38. In 36 layers a negligible perturbation changes how later
roundings fall. The records that flip near the threshold are therefore a property of the whole arithmetic path,
reproducible on that path and redrawn by each change tested, while the range moved how many flip by only 2. Native FP16 and the
emulation with its range shared flip sets with Jaccard indices of 0.19 to 0.30, one task below the expected floor of
0.2, so that expectation was missed as well.

After the rest of the arm had been read, a further run tested why the emulations flipped more records than native
FP16. They round every value that eager attention stores in FP16 (the attention scores, the softmax output and the
attention output), which SDPA's fused kernel keeps inside the kernel. With expectations written before the run,
native FP16 with eager attention scored the band and flipped 109 records, against 115 and 113 emulated, 80 under SDPA
in run 1 and 68 under SDPA on the eager run's band-only batches (Table~\ref{tab:app-fp16-llm}). Rounding where eager
attention stores values thus brought the count to the emulations', but not the same records: the eager flip set had
Jaccard indices of 0.22 with the emulation with FP16's range and 0.17 with the SDPA band run. No value overflowed,
and the largest activation was again 7{,}448. Of the run's 9 expectations, 8 held. The miss was the SDPA band run,
expected to repeat run 1 bit for bit. It changed both the batches and the card, and which of the two moved its
margins is not identified.

For the language models, rounding each reference's own label logits to BF16 leaves 132 and 177 ties at the ACSIncome
threshold, against 147 and 189 measured under BF16. Rounding them to FP16 leaves 10 and 26. The final rounding of
the logits thus sets how many records tie, though not which. Converting either checkpoint to FP16 overflows no weight.
Its range changes 0.02\% of the 4B model's weights and 0.14\% of the 24B model's, all of them small values held as
subnormals or flushed. Of the arm's 29 expectations, written before any run, 24 held. Besides the two above, the
misses concern how far FP16 re-rounding cut the 4B model's ties and flips and the 24B checkpoint's flushed share.

\begin{table}[ht]
\caption{Exploratory FP16 arm, tabular: decisions changed against FP64, run 1, primary reading (logit reading in
brackets). BF16 and FP16: measured on an M4 Pro CPU (BF16 matches the M5 Max CPU in every cell). E8M10 and E5M10:
emulated, committed before the FP16 run; E8M10 has FP16's fraction bits and FP32's range. Range: stored values whose
FP16 rounding differs from their E8M10 rounding; no value overflowed.}
\label{tab:app-fp16}
\small
\begin{tabular}{@{}l l r r r r r@{}}
\toprule
Task & Model & BF16 & FP16 & E8M10 & E5M10 & Range \\
\midrule
ACSIncome & LR & 39 (11) & 4 (2) & 4 (2) & 4 (2) & 0 \\
 & MLP & 15 (10) & 1 (3) & 1 (3) & 1 (3) & 2{,}240 \\
ACSEmployment & LR & 185 (231) & 6 (20) & 6 (20) & 6 (20) & 3 \\
 & MLP & 96 (60) & 5 (1) & 5 (1) & 5 (1) & 11{,}553 \\
HMDA & LR & 74 (19) & 18 (7) & 18 (7) & 18 (7) & 0 \\
 & MLP & 15 (15) & 1 (0) & 1 (0) & 1 (0) & 2{,}834 \\
Diabetes & LR & 18 (17) & 5 (5) & 5 (5) & 5 (5) & 0 \\
 & MLP & 8 (8) & 0 (0) & 0 (0) & 0 (0) & 511 \\
\bottomrule
\end{tabular}
\end{table}

\begin{table}[ht]
\caption{Exploratory FP16 arm, 4B model on RTX 4090 GPUs: decisions changed against FP64, run 1. Native FP16 and BF16
on all test records; E8M10 and E5M10 emulated on the records within three logits of the threshold (band), which hold
every native FP16 flip. Eager and SDPA band (addendum, post hoc): native FP16 with eager attention, and with SDPA on
the same band-only batches on another card, each scored on the band only. Shared: band records both emulations flip.
Range failures (a non-finite margin): none. Median $|\Delta m|$: E5M10 against E8M10, in logits.}
\label{tab:app-fp16-llm}
\small
\begin{tabular}{@{}l r r r r r r r r r@{}}
\toprule
Task & Band & BF16 & FP16 & Eager & SDPA band & E8M10 & E5M10 & Shared & Median $|\Delta m|$ \\
\midrule
ACSIncome & 6{,}062 & 184 & 28 & 41 & 21 & 41 & 44 & 19 & 0.031 \\
ACSEmployment & 1{,}306 & 13 & 3 & 6 & 2 & 3 & 3 & 1 & 0.031 \\
HMDA & 4{,}515 & 248 & 36 & 41 & 32 & 48 & 50 & 25 & 0.063 \\
Diabetes & 4{,}384 & 101 & 13 & 21 & 13 & 21 & 18 & 9 & 0.031 \\
\bottomrule
\end{tabular}
\end{table}

\paragraph{Expectations written before any run.} Table~\ref{tab:app-expect} sets each written expectation
against what was observed. Four expectations of flip rates or counts failed, all by being too high: three in the
study plan and one in the MI350X arm's plan. The kill criterion was the plan's condition for committing to the paper,
and it was not met on the tabular Tier 1 evidence alone.

\begin{table}[ht]
\caption{Expectations written before any run (the study plan) or before any scoring (the MI350X arm's plan and its
addendum) against what was observed, and the kill criterion per arm. Flip rates are shares of a dataset's decisions,
except the language model's TF32 row, pooled over the four datasets (at most 0.035\% per dataset). The plan's kill
criterion required all five items to hold before committing to the paper; item 1 asked for at least two model
families to flip under a realistic default, which the plan names as TF32, the relaxed default or BF16, and item 3,
a change in at least one of the six senses, is met by any sense-6 movement.}
\label{tab:app-expect}
\footnotesize
\begin{tabular}{@{}>{\raggedright\arraybackslash}p{0.2\linewidth} >{\raggedright\arraybackslash}p{0.3\linewidth}
>{\raggedright\arraybackslash}p{0.43\linewidth}@{}}
\toprule
Arm & Expected & Observed \\
\midrule
Tabular, relaxed default & 0.05\% to 0.1\% for LR and MLP & failed: LR 0; MLP 0.0026\% to 0.041\% \\
Tabular, TF32 & the order of the relaxed default & failed: LR 0; MLP at most 0.0077\% \\
Tabular, Tier 1 audit & most checks unchanged & held: only sense 6 changed \\
Language model, TF32 & 0.1\% to 1\% & failed: 0.020\% (H100), 0.019\% (RTX 4090) \\
Language model and deployment, audit & changes, if any, most likely in senses 2, 3 and 6 & partly: the language model's
Tier 1 changed only sense 6; Tier 2 contracts changed senses 1 to 4 and 6 \\
Deployment contracts & the largest, plausibly above 1\% for the language model & held: up to 4.123\% (FP8
serving, Diabetes) \\
MI350X, settings & defaults, strict and TF32 flags all IEEE-like & failed: the TF32 setting changes the products
\\
MI350X, language model TF32 setting & 40 to 200 flips & failed: 4 flips \\
MI350X, verdicts & BF16 and FP8 change some verdicts the H100's Tier 2 contracts changed & partly: FP8 changed
the H100's Diabetes ranking by sex and one other significance verdict; neither changed the H100's pass-fail or
significance verdicts \\
\midrule
\multicolumn{3}{@{}l}{Kill criterion: not met on tabular Tier 1 (item 1 failed: only the MLP flipped); met on Tier 1
once the language model is added} \\
\multicolumn{3}{@{}l}{(item 2 through the tabular relaxed default); met for the tabular arm once Tier 2 is included,
item 1 through BF16.} \\
\multicolumn{3}{@{}l}{The MI350X arm had no kill criterion.} \\
\bottomrule
\end{tabular}
\end{table}

\clearpage
\section{An Audit Record}\label{app:audit-record}
Table~\ref{tab:app-audit-record} is the record that the protocol of Section~\ref{sec:protocol} would have the auditor
keep for the 24B model's employment-status classification audit by sex: what was audited, under which declared contracts, the settings and
checks recorded with each contract's first run, the hash of each run's margins, run 1's gaps and point verdicts, and
each gap's stability interval and sampling interval. It is assembled after the fact from the run metadata. A change
to any entry in the stack column reopens the audit (Section~\ref{sec:protocol-fit}).

\begin{table}[ht]
\caption{The audit record for the 24B model's ACSEmployment audit by sex, as Table~\ref{tab:protocol} would have the auditor keep it. Margin hashes are the first eight hexadecimal digits of the SHA-256 of each run's margins. Gaps and verdicts are run 1's; vLLM BF16's run 3 gives 0.0518 and 0.0462, with the same verdicts.}
\label{tab:app-audit-record}
\footnotesize
\begin{tabular}{@{}p{0.2\linewidth} p{0.76\linewidth}@{}}
\toprule
Model & Mistral-Small-3.2-24B-Instruct-2506, revision \texttt{95a6d26c}, 24{,}011{,}361{,}280 parameters; BF16 weight fingerprint \texttt{b316a3ef} \\
Records & ACSEmployment test subset, 20{,}000 records; groups by recorded sex, each with at least 100 records and 30 positives and negatives \\
Decision & select when the margin of YES over NO at the first label token is at least $t_m = -5.782463073730469$, the quantile of 2{,}048 strict FP32 training margins that selects the training employment rate, 0.4566 \\
Tests & demographic-parity and equalized-odds gaps pass at 0.05 or less; 95\% percentile intervals from 1{,}000 bootstrap resamples of the records \\
\bottomrule
\end{tabular}

\vspace{4pt}
\begin{tabular}{@{}>{\raggedright\arraybackslash}p{0.12\linewidth} >{\raggedright\arraybackslash}p{0.26\linewidth} >{\raggedright\arraybackslash}p{0.15\linewidth} >{\raggedright\arraybackslash}p{0.15\linewidth} r r@{}}
\toprule
Contract & Stack & Settings and checks (run 1) & Runs, margin hash & DP gap, run 1 & EO gap, run 1 \\
\midrule
Strict FP32 (reference) & transformers 5.18.0, torch 2.11.0, CUDA 12.9; 2 H100; FP32; eager & TF32 off, precision highest; canary $4.2\times10^{-7}$ & 1, 2: \texttt{c941e02d} (identical) & 0.0567 (fail) & 0.0515 (fail) \\
TF32 & transformers 5.18.0, torch 2.11.0, CUDA 12.9; 2 H100; FP32; eager & TF32 on, precision high; canary $3.0\times10^{-4}$ & 1, 2: \texttt{d54a178d} (identical) & 0.0568 (fail) & 0.0516 (fail) \\
BF16 & transformers 5.18.0, torch 2.11.0, CUDA 12.9; 1 H100; BF16; SDPA & TF32 off; BF16 weights & 1, 2: \texttt{13dc5ecd} (identical) & 0.0554 (fail) & 0.0520 (fail) \\
vLLM BF16 & vLLM 0.25.0, 1 H100; BF16; KV cache in the model's type; run 3 logged attention FLASH\_ATTN and no linear-kernel selection (unquantized) & none reported by the engine & 1, 2: \texttt{5bdc7137}; 3: \texttt{98b963c6} (differ) & 0.0516 (fail) & 0.0464 (pass) \\
vLLM FP8 & vLLM 0.25.0, 1 H100; FP8 weights, quantized at load with one scale per tensor; activations with a scale per token; KV cache in the model's type; run 3 logged attention FLASH\_ATTN and FP8 products \texttt{CutlassFP8Scaled\-MMLinearKernel} & none reported by the engine & 1, 2, 3: \texttt{769cc63d} (identical) & 0.0417 (pass) & 0.0394 (pass) \\
\midrule
\multicolumn{4}{@{}l}{Stability interval over the Tier 1 contracts: robust} & [0.0567, 0.0568] & [0.0515, 0.0516] \\
\multicolumn{4}{@{}l}{Stability interval over all five contracts: contract-dependent} & [0.0417, 0.0568] & [0.0394, 0.0520] \\
\multicolumn{4}{@{}l}{Sampling interval of the reference} & [0.043, 0.070] & [0.033, 0.070] \\
\bottomrule
\end{tabular}
\end{table}

\paragraph{A cheaper way to rescore.} Rescoring every record under every contract
is the research protocol's cost. A cheaper heuristic, not validated here, rescores only the records whose
deployment score lies within the largest perturbation seen on a sample scored under both contracts; for the 4B
model's TF32 that was 0.57 in logit margin, although no flip lay beyond 0.0817 (Section~\ref{sec:results-margin}), and
a larger perturbation would escape it.

\end{document}